\documentclass[]{assets/matterlab}

\newif\ifdraftmode
\draftmodefalse  

\ifdraftmode
  \usepackage[draft]{microtype}
  \usepackage[draft]{graphicx}
\else
  \usepackage{microtype}
  \usepackage{graphicx}
\fi
\usepackage{float}
\usepackage{booktabs}
\usepackage{xurl}
\usepackage{hyperref}
\usepackage{titlesec}
\setcitestyle{numbers,comma}

\usepackage{amsmath}
\usepackage{amssymb}
\usepackage{mathtools}
\usepackage{amsthm}
\usepackage{bm}

\usepackage[noabbrev,nameinlink]{cleveref}

\usepackage{amsmath,amsfonts,bm}

\def\eqref#1{equation~\ref{#1}}

\def\1{\bm{1}}

\DeclareMathAlphabet{\mathsfit}{\encodingdefault}{\sfdefault}{m}{sl}
\SetMathAlphabet{\mathsfit}{bold}{\encodingdefault}{\sfdefault}{bx}{n}

\usepackage{orcidlink}
\usepackage{comment}
\usepackage[version=4]{mhchem}
\usepackage{longtable}
\usepackage{array}
\renewcommand{\arraystretch}{1.7}
\usepackage{multirow}           
\usepackage{amsfonts}           
\usepackage{nicefrac}
\usepackage{duckuments}         
\usepackage{thmtools,thm-restate}
\usepackage{enumitem}
\usepackage{xcolor,colortbl}
\usepackage{tikz}
\usepackage{tikz-cd}
\usepackage{caption}
\usepackage{subcaption}
\usepackage{listings}
\usepackage{gensymb}
\usepackage{textcomp} 
\usepackage{siunitx}
\usepackage{threeparttable}

\usepackage[hybrid]{markdown}
\markdownSetup{
  renderers = {
      interblockSeparator = {%
          \par
      },
  },
}
\usepackage[inkscapelatex=false]{svg}

\newcommand{\addressCHEM}{Department of Chemistry, University of Toronto,  80 St. George St., Toronto, ON M5S 3H6, Canada}
\newcommand{\addressAC}{Acceleration Consortium, 700 University Ave., Toronto, ON M7A 2S4, Canada}
\newcommand{\addressCS}{Department of Computer Science, University of Toronto, 40 St George St., Toronto, ON M5S 2E4, Canada}
\newcommand{\addressVECTOR}{Vector Institute for Artificial Intelligence, W1140-108 College St., Schwartz Reisman Innovation Campus, Toronto, ON M5G 0C6, Canada}
\newcommand{\addressMSE}{Department of Materials Science \& Engineering, University of Toronto, 184 College St., Toronto, ON M5S 3E4, Canada}
\newcommand{\addressCHEMENG}{Department of Chemical Engineering \& Applied Chemistry, University of Toronto, 200 College St., Toronto, ON M5S 3E5, Canada}

\newcommand{\addressCIFAR}{Canadian Institute for Advanced Research (CIFAR), 661 University Ave., Toronto,
ON M5G 1M1, Canada}
\newcommand{\addressNVIDIA}{NVIDIA, 431 King St W \#6th, Toronto, ON M5V 1K4, Canada}

\newcommand{\acknowSciNet}{Computations were performed on the Niagara supercomputer at the SciNet HPC Consortium. SciNet is funded by: the Canada Foundation for Innovation; the Government of Ontario; Ontario Research Fund - Research Excellence; and the University of Toronto.}

\newcommand{\acknowDARPA}{This work was supported by the Defense Advanced Research Projects Agency (DARPA) under Agreement No.~HR0011262E022.}

\usepackage{cinzel}
\newcommand{\elagente}{{\cinzel El Agente}}
\newcommand{\elagenteQ}{{\cinzel El Agente Q}}

\newcommand{\elagenteestructural}{{\cinzel El Agente Estructural}}
\newcommand{\estructural}{{\cinzel Estructural}}
\newcommand{\quntur}{{\cinzel El Agente Quntur}}

\usepackage{adjustbox}
\usepackage{wrapfig}
\usepackage{tabularx}
\newcolumntype{Y}{>{\raggedright\arraybackslash}X}

\usepackage{braket}
\tcbset{textmarker/.style={%
    enhanced,
    frame hidden,
    parbox=false,
    boxrule=0mm,
    boxsep=0mm,
    arc=2mm,          
    outer arc=2mm,    
    left=3mm,right=3mm,
    top=7pt,bottom=7pt,
    toptitle=1mm,bottomtitle=1mm,
    left skip=0.0cm
}}

\tcbset{textmarker2/.style={%
    enhanced,
    frame hidden,
    parbox=false,
    boxrule=0mm,
    boxsep=0mm,
    arc=2mm,          
    outer arc=2mm,    
    left=3mm,right=3mm,
    top=7pt,bottom=7pt,
    toptitle=1mm,bottomtitle=1mm,
    left skip=0.8cm
}}

\definecolor{hintblue}{RGB}{235,240,253}

\newtcolorbox{hintBox}{textmarker,
    colback=hintblue, fontupper=\small}

\newtcolorbox{hintBox2}{textmarker2,
    colback=blue!5!white}

\newtcolorbox{importantBox}{textmarker,
    colback=red!10!white}

\newtcolorbox{noteBox}{textmarker,
    colback=matterbg, fontupper=\small}

\usepackage{placeins}

\usepackage{xltabular}
\usepackage{makecell}

\title{El Agente Estructural: An Artificially Intelligent Molecular Editor}

\author[1,6,\dagger]{Changhyeok Choi}
\author[2,6,\dagger]{Yunheng Zou}
\author[1,6]{Marcel M\"uller}
\author [5]{Han Hao}
\author[1,6,9]{Yeonghun Kang}
\author[1,6]{Juan B. P\'erez-S\'anchez}
\author[1]{Ignacio Gustin}
\author [1]{Hanyong Xu}
\author[1,6]{Andrew Wang}
\author[2]{Aiwei Yin}
\author [1,2,6]{Mohammad Ghazi Vakili}
\author [2,5]{Chris Crebolder}
\author[1,2,3,4,5,6,7,8,*]{Al\'an Aspuru-Guzik}
\author[1,2,5,*]{Varinia Bernales}

\newcommand{\addressSKKU}{Department of Chemistry, Sungkyunkwan University, 2066 Seobu-ro, Suwon-si, 16419, Republic of Korea}

\affiliation[1]{\addressCHEM}
\affiliation[2]{\addressCS}
\affiliation[3]{\addressMSE}
\affiliation[4]{\addressCHEMENG}
\affiliation[5]{\addressAC}
\affiliation[6]{\addressVECTOR}
\affiliation[7]{\addressCIFAR}
\affiliation[8]{\addressNVIDIA}
\affiliation[9]{\addressSKKU}

\contribution[\dagger]{Contributed equally to this work.}

\abstract{

We present \elagenteestructural{}, a multimodal, natural-language–driven geometry-generation and manipulation agent for autonomous chemistry and molecular modelling.
Unlike molecular generation or editing via generative models, \estructural{} mimics how human experts directly manipulate molecular systems in three dimensions by integrating a comprehensive set of domain-informed tools and vision-language models.
This design enables precise control over atomic or functional group replacements, atomic connectivity, and stereochemistry without the need to rebuild extensive core molecular frameworks.
Through a series of representative case studies, we demonstrate that \estructural{} enables chemically meaningful geometry manipulation across a wide range of real-world scenarios.
These include site-selective functionalization, ligand binding, ligand exchange, stereochemically controlled structure construction, isomer interconversion, fragment-level structural analysis, image-guided generation of structures from schematic reaction mechanisms, and mechanism-driven geometry generation and modification.
These examples illustrate how multimodal reasoning, when combined with specialized geometry-aware tools, supports interactive and context-aware molecular modelling beyond structure generation.
Looking forward, the integration of \estructural{} into \quntur{}, an autonomous multi-agent quantum chemistry platform, could enhance its capabilities by adding sophisticated tools for the generation and editing of three-dimensional structures.
}

\date{\today}
\correspondence{\email{alan@aspuru.com}, and \email{varinia@bernales.org}}

\begin{document}

\maketitle



\newpage 
\section{Introduction}\label{sec1}
Computational chemistry has become an essential tool for understanding molecular structure, reactivity, and materials properties across chemistry, materials science, and catalysis~\cite{houk2017holy, norskov2009towards, huang2023central}. It is widely used in applications such as high-throughput screening~\cite{gomez2016design, greeley2006computational}, materials and catalyst design~\cite{norskov2009towards, seh2017combining, gu2020progress, greeley2009alloys}, and mechanistic studies~\cite{garza2018mechanism, peterson2010copper}, enabling the rational exploration of chemical space beyond what is experimentally accessible.

Recently, agentic systems that directly perform quantum chemistry calculations represent a significant step toward end-to-end automation. In our previous work, we introduced \elagenteQ{}~\cite{Zou2025}, a multi-agent system designed to autonomously execute quantum chemistry calculations. \elagenteQ{} enables users to specify computational tasks through natural-language dialogue and directly performs structure preparation, input generation, computation execution, automated error recovery, and result interpretation within a unified agentic framework. After our work, approaches for computational quantum chemistry workflows have been explored in systems such as AItomia~\cite{hu2025aitomiaintelligentassistantaidriven}, ChemGraph~\cite{pham2026chemgraph}, and DREAMS~\cite{dreams} for solid-state calculations. In our recent work, we described the capabilities of \quntur{}~\cite{ElAgenteQuntur}, a graduate-level computational chemistry agent.
The \elagente{} framework has expanded to systems such as {\cinzel El Agente Gr\'afico}~\cite{bai2026elagentegraficostructured}, which applies an execution-graph framework that lets LLM agents orchestrate scientific workflows with structured state and dynamic knowledge-graph as persistence memory, {\cinzel El Agente S\'olido}~\cite{kumar2026elagentesolidonew}, a hierarchical multi-agent framework for autonomous solid-state simulations, and {\cinzel El Agente Forjador}~\cite{zhang2026elagenteforjadortaskdriven}, a framework for autonomous generation of scientific computational tools, collectively demonstrating the scalability and adaptability of agentic AI systems for computational science.

Despite these advances, a major challenge remains in the reliable construction and manipulation of molecular geometries, particularly for complex molecular systems and structural modifications. Since chemical properties, including reactivity, electronic structure, and optical behaviour, are governed by three-dimensional molecular arrangements, the ability to construct and control molecular geometry is central to computational chemistry and to understanding structure–property relationships.

Existing approaches to molecular modelling typically rely on database retrieval, SMILES-based structure reconstruction using cheminformatics toolkits such as RDKit~\cite{rdkit} or Open Babel~\cite{obabel}, or generative models~\cite{GmezBombarelli2018,segler2018generating,gupta2018generative,popova2018deep}. These approaches have enabled major advances in data-driven chemical discovery and molecular generation. 

However, the methods described above impose severe limitations when investigating a chemical problem. Database retrieval is restricted to geometries already present in the database. SMILES~\cite{smiles_original_paper} is primarily designed for covalently bonded organic molecules and often fails to represent systems such as reaction intermediates, transition states, and transition-metal complexes. Generative models are similarly constrained by the distributions and representations present in their training data and often lack interpretability~\cite{krenn2022selfies}. More fundamentally, these approaches offer limited controllability. In many chemical investigations, the objective is not merely to obtain a plausible structure but to construct a geometry that reflects a specific mechanistic or structural hypothesis, such as a particular binding mode, coordination environment, stereochemical outcome, or transition-state (TS)-like distortion.

In contrast, computational chemists commonly manipulate structures directly using three-dimensional molecular editors such as Avogadro~\cite{avogadro}, GaussView~\cite{gv6}, IQmol~\cite{iqmol}, Molden~\cite{schaftenaar2000molden}, and Chemcraft~\cite{chemcraft}. These tools allow interactive editing of molecular geometries by modifying atomic coordinates, enabling chemists to construct stereochemically defined structures, reactant-bound complexes, and TS-like geometries while preserving the underlying molecular framework. However, this process remains largely manual and reasoning-intensive, which limits its scalability and poses a bottleneck for highly autonomous computational workflows.

To address this gap, we introduce \elagenteestructural{}, an agent designed for comprehensive molecular structure generation and manipulation through natural-language dialogue (\autoref{fig:schematic}). \estructural{} provides a set of 3D operations and structural analysis actions derived from the workflows used by computational chemists to construct and edit three-dimensional molecular structures. These operations act directly on atomic indices in \texttt{xyz} representations, enabling site-specific manipulation at the level of atoms or functional groups.
The closest related approaches in the literature are SynCraft~\cite{li2026guiding}, which introduces a simple set of targeted structural modifications to generated molecules to improve synthesizability, and AtomWorld~\cite{pmlr-v306-lv26b}, which includes a 3D geometry-editing benchmark for evaluating LLM agents' spatial understanding on crystalline materials.

During operation, \estructural{} analyzes and reasons about molecular structures both numerically and visually through inspection of the resulting geometries. The system supports a wide range of composable actions, including substitution, binding, stereochemical interconversion, direct control of interatomic distances, angles, and dihedral angles, and the construction or modification of organometallic structures. In addition, \estructural{} can interpret chemical schematics such as reaction mechanism diagrams and translate them into three-dimensional structures of intermediates or TS-like geometries.

Through a diverse set of case studies, we demonstrate how \estructural{} enables on-demand manipulation of molecular structures that are difficult to generate or edit using string-based approaches alone. These examples include site-selective functionalization, ligand binding to catalytic centres, stereochemically specific construction of transition-metal complexes, fragment-level structural analysis, and image-guided generation of reaction intermediates and TS-like geometries.

At the end of this paper, we discuss future development directions, including the integration of \estructural{} with \quntur{}~\cite{ElAgenteQuntur} to enable end-to-end exploration of broader chemical space within automated quantum chemistry pipelines.



\section{Results and Discussion}\label{sec2}
\subsection{System architecture}
\begin{figure}[t]
        \centering
        \includegraphics[width=1.0\textwidth]{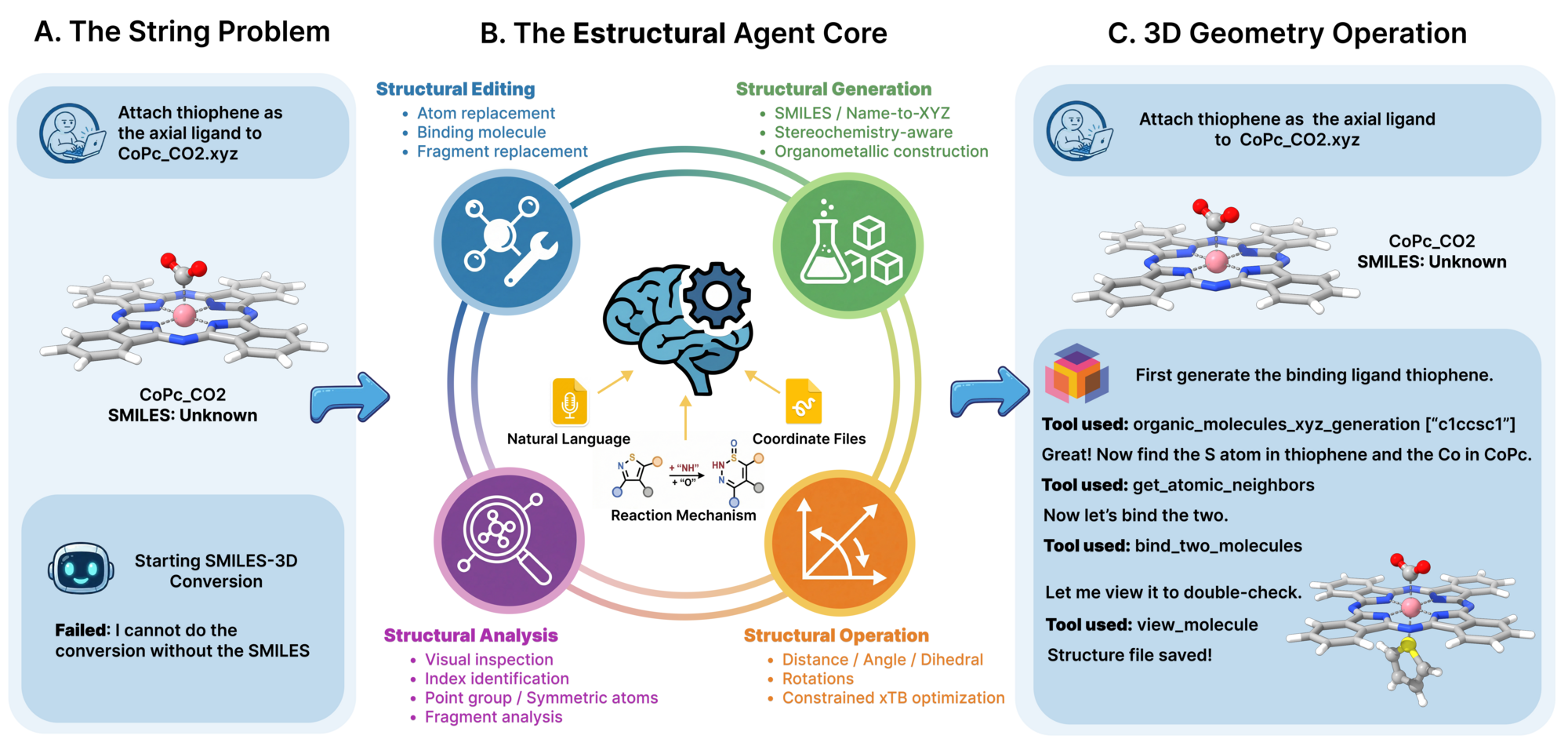}
        \caption{Schematic overview of the \estructural{} architecture and workflow. \estructural{} supports natural-language prompts, coordinate files, and reaction schematics as inputs, and manipulates molecular systems through four action categories: structural editing, structural generation, structural analysis, and geometric operations. The example illustrates attachment of a thiophene ligand to an organometallic complex: \estructural{} generates the ligand, identifies binding atoms, performs the coordination step, and validates the resulting structure. Atom colour codes: pink, Co; yellow, S; red, O; blue, N; grey, C; white, H. Selected illustrative elements, including the user, robot, and Agent Core icons, were generated using Gemini 3 (web-based access, 2026). These elements were subsequently incorporated into the figure by the authors.}
        \label{fig:schematic}
\end{figure}

\elagenteestructural{} builds upon the cognitive architecture~\cite{sumers2023cognitive} and graphical user interface (GUI) of \elagenteQ{}~\cite{Zou2025}. The system is centred on a single top-level agent, referred to as the \texttt{geometry\_operator}, which is responsible for high-level planning, reasoning, and orchestration of molecular-structure-related tasks (\autoref{fig:schematic}). Unless otherwise noted, the \texttt{geometry\_operator} was powered by GPT-5.4 with medium reasoning effort.

The \texttt{geometry\_operator} interprets user requests expressed in natural language and decomposes them into a sequence of geometry-aware actions, such as structure generation, editing, and structural analysis. Depending on the task, the agent dynamically selects between dedicated molecular-analysis and manipulation tools, visual inspection, and Python-based operations. These capabilities are used in a complementary manner. Most geometry manipulation tools use atomic indices to specify which atoms are involved in the requested structural change.

When reaction information is provided visually, for example through schematic reaction-mechanism diagrams, the \texttt{geometry\_operator} extracts structural and mechanistic cues from the images and returns a structured description of the reaction pathway and relevant geometric information, which is then used to guide subsequent structure-generation and editing steps.

\estructural{} also supports visual inspection of molecular structures by rendering \texttt{xyz} files at selected viewing angles and zoom levels. These rendered views can be used to obtain an overall spatial understanding of the structure or to support identification of relevant atoms or molecular subgroups when needed. 

This architecture enables flexible coordination between language-based reasoning, numerical structural analysis, visual interpretation, and molecular manipulation. To reduce costs and avoid contamination of the large--language model (LLM) and vision--language model (VLM) context from prior viewpoints, we employ an image-pruning context management strategy that removes prior molecular system views from memory upon the arrival of new views. This mechanism mimics how humans visually observe only the current snapshot of a scene. This approach saves a significant number of visual tokens as multi-step editing progresses.

The shared tool space includes structure-generation, atom- and fragment-level editing, symmetry and connectivity analysis, geometric manipulation, constrained optimization, and molecular visualization utilities, together with \texttt{bash\_kernel} for general-purpose shell commands and Python execution. Further details of the system architecture, individual tool implementations, and evaluation of the VLM are provided in Supporting Information Section~\ref{agents_tools}, Figures~\ref{fig:tools}--\ref{fig:mol_viz_tools}, and~\autoref{tab:agents_tools}.

We evaluated \estructural{} through a series of representative case studies reflecting realistic and practically relevant molecular modelling scenarios, including site-selective functionalization, ligand binding to organometallic complexes, stereoselective structure generation, stereoisomer interconversion, fragment analysis, and the generation of reaction intermediates or transition-state (TS)-like geometries from multimodal inputs. 

The case studies are presented as figures in a natural language user interface (NLUI) style to illustrate the interaction between the user and \estructural{}. In each figure, user-side speech bubbles show the exact prompts used, while agent-side bubbles summarize the reasoning steps and tool invocations required to complete the task. Molecular structures correspond either to user-provided inputs or to structures generated by the agent. Generated structures are labelled using the original filenames assigned by the agent, with the \texttt{xyz} extension omitted for readability. All structures were rendered using ChimeraX~\cite{pettersen2021ucsf,mengUCSFChimeraXTools2023}. 

We evaluated all case studies over five independent runs; per-run tool-call counts, runtimes, evaluation scores, and detailed scoring criteria for the principal case studies are provided in Supporting Information Section~\ref{SIsec:benchmark}, Tables~\ref{tab:per-spermidine}--\ref{tab:case7-opus45-endpoints}, and Figures~\ref{fig:SI_Case1_generic}--\ref{fig:SI_Case2_generic}.

Complete chat histories and generated structures for all case studies are available in our GitHub repository (see \hyperref[data_availability]{Data Availability}).

\subsubsection{Case study 1: Site-selective functionalization}\label{subsec:casestudy1}
\begin{figure}[!t]
        \centering        \includegraphics[width=1.0\textwidth]{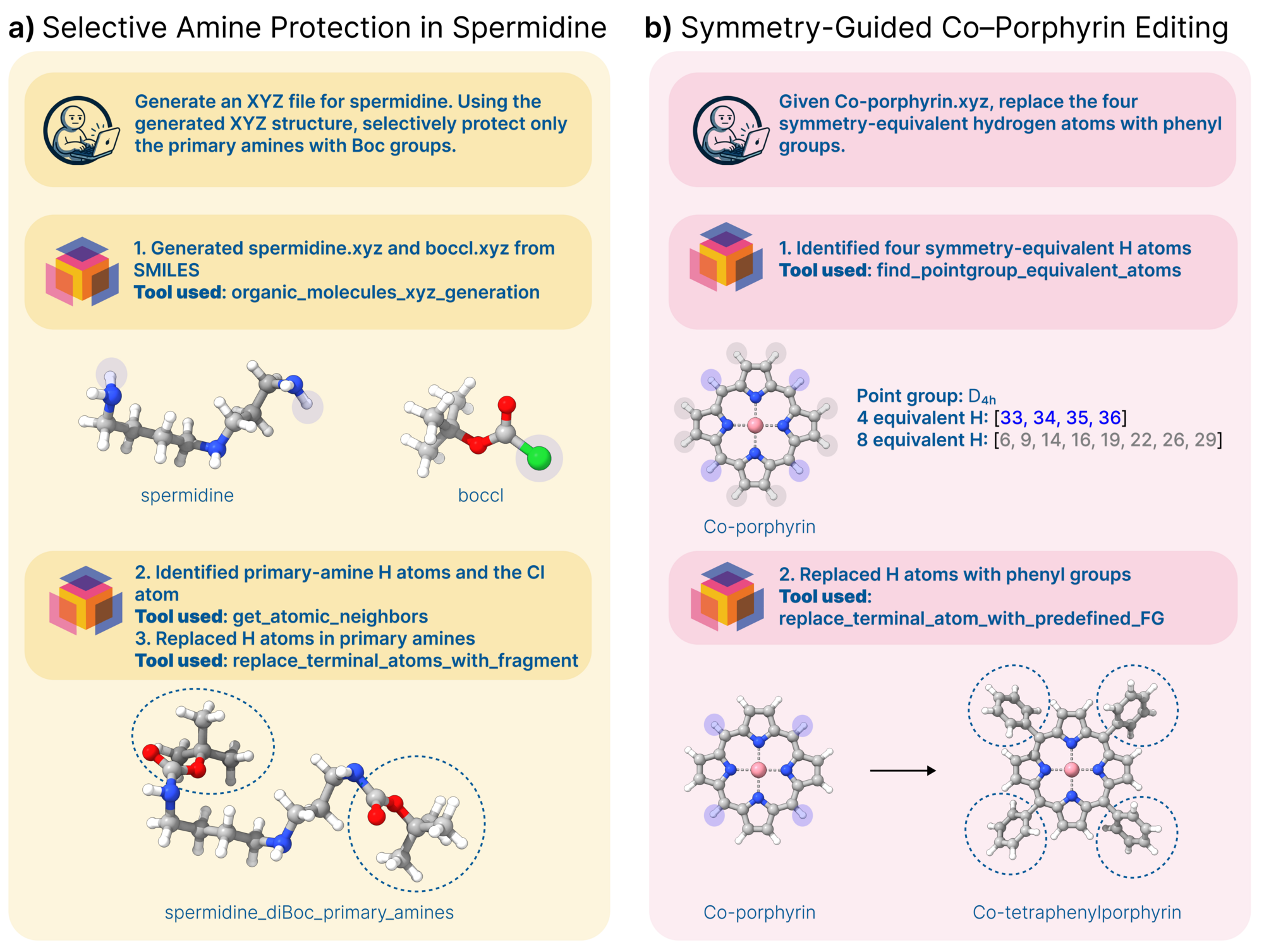}
        \caption{Representative examples of site-selective functionalization performed by \estructural{}. (a) Selective protection of primary amines in spermidine, where only terminal amine sites are functionalized with Boc groups while the secondary amine remains unchanged, (b) Symmetry-guided functionalization of a cobalt–porphyrin complex, in which four symmetry-equivalent hydrogen atoms are identified and selectively replaced with phenyl groups while preserving all other hydrogen atoms. The atomic sites selected for replacement and the newly introduced fragments are highlighted for clarity.
        Atom colour codes: pink, Co; lime green, Cl; red, O; blue, N; grey, C; white, H.}
        \label{fig:case_editing}
\end{figure}

In this case study, we demonstrate site-selective functionalization using \estructural{} across two representative molecular systems, highlighting both local chemical selectivity and symmetry-guided editing (\autoref{fig:case_editing}). These tasks require not only identifying appropriate atomic sites but also orchestrating multiple analysis and editing tools.

In the first example, \estructural{} is tasked with selectively protecting only the primary amines of spermidine, which contains two primary amines and one secondary amine, using tert-butoxycarbonyl (Boc) protecting groups while leaving the secondary amine unchanged (\autoref{fig:case_editing}a and~\autoref{tab:per-spermidine}). 
  
The user specifies the overall objective, namely generation of spermidine followed by selective Boc protection of the primary amines, but does not provide the target atomic indices, the editing sequence, or the tools required to accomplish the transformation. Since neither spermidine nor the Boc group is included in the predefined fragment library, \estructural{} needs to generate \texttt{xyz} structures for both components. 

Starting from a text prompt, \estructural{} generates \texttt{xyz} structures for spermidine and Boc chloride from SMILES representations. Structural analysis tools are then applied to identify the relevant connection indices in each \texttt{xyz} structure. \estructural{} successfully distinguishes hydrogen atoms bound to primary and secondary amine nitrogen atoms and identifies the chlorine atom of Boc chloride as the appropriate connection site based on local chemical environments. Following identification of the correct replacement sites, a structure editing tool selectively replaces hydrogen atoms on the primary amines with Boc groups, while preserving the remainder of the molecular geometry.

The second example illustrates symmetry-driven site-selective functionalization of a cobalt–porphyrin complex (\autoref{fig:case_editing}b and~\autoref{tab:per-Copor}). Symmetrically substituted metal–porphyrin complexes, particularly those functionalized at meso hydrogen positions, have been widely reported in catalytic applications~\cite{functionalized_porphyrin}. This case study probes \estructural{} 's ability to perform symmetry-aware functionalization tasks.
    
The user specifies the input structure, the four symmetry-equivalent hydrogen atoms as the substitution target, and phenyl as the desired substituent, but does not provide the corresponding atomic indices. The agent therefore has to identify which hydrogen atoms satisfy this symmetry criterion and determine the indices to be modified.

Given the provided \texttt{xyz} file, \estructural{} therefore performs point-group analysis to identify sets of symmetry-equivalent hydrogen atoms. The molecule is classified as belonging to the D$_{4h}$ point group, revealing two symmetry classes: four equivalent meso hydrogen atoms and eight equivalent $\beta$ hydrogen atoms. \estructural{} then identifies the four-membered symmetry class as the requested target and selectively replaces these meso hydrogen atoms with phenyl groups, while leaving all other hydrogen atoms unmodified.

This enables symmetry-preserving functionalization without the need to manually specify individual atomic indices. In this case, the phenyl group is available in the predefined fragment library, allowing \estructural{} to directly use the existing fragment rather than generating an additional \texttt{xyz} structure. 

These examples demonstrate how \estructural{} combines structural analysis, on-demand structure generation, and editing to perform site-selective and symmetry-aware molecular modifications.

The two examples above were designed to separately demonstrate molecular generation, site-selective functionalization, and symmetry-aware editing. To test whether these outcomes depended on procedural guidance, we repeated representative tasks using more generic prompts. In both the spermidine and Co-porphyrin cases, \estructural{} successfully inferred the required workflow and generated the expected structures (Figure~\ref{fig:SI_Case1_generic} and Tables~\ref{tab:per-spermidine_generic}--\ref{tab:per-Copor_generic}). Additional robustness tests using perturbed Co-porphyrin structures and porphyrins with different protonation states also yielded the expected functionalized products (Figures~\ref{SI:distortion}--\ref{SI:varied_protonation}).

\subsubsection{Case study 2: Binding of reaction intermediates and ligands}\label{subsec:casestudy2}
\begin{figure}[t]
        \centering        \includegraphics[width=1.0\textwidth]{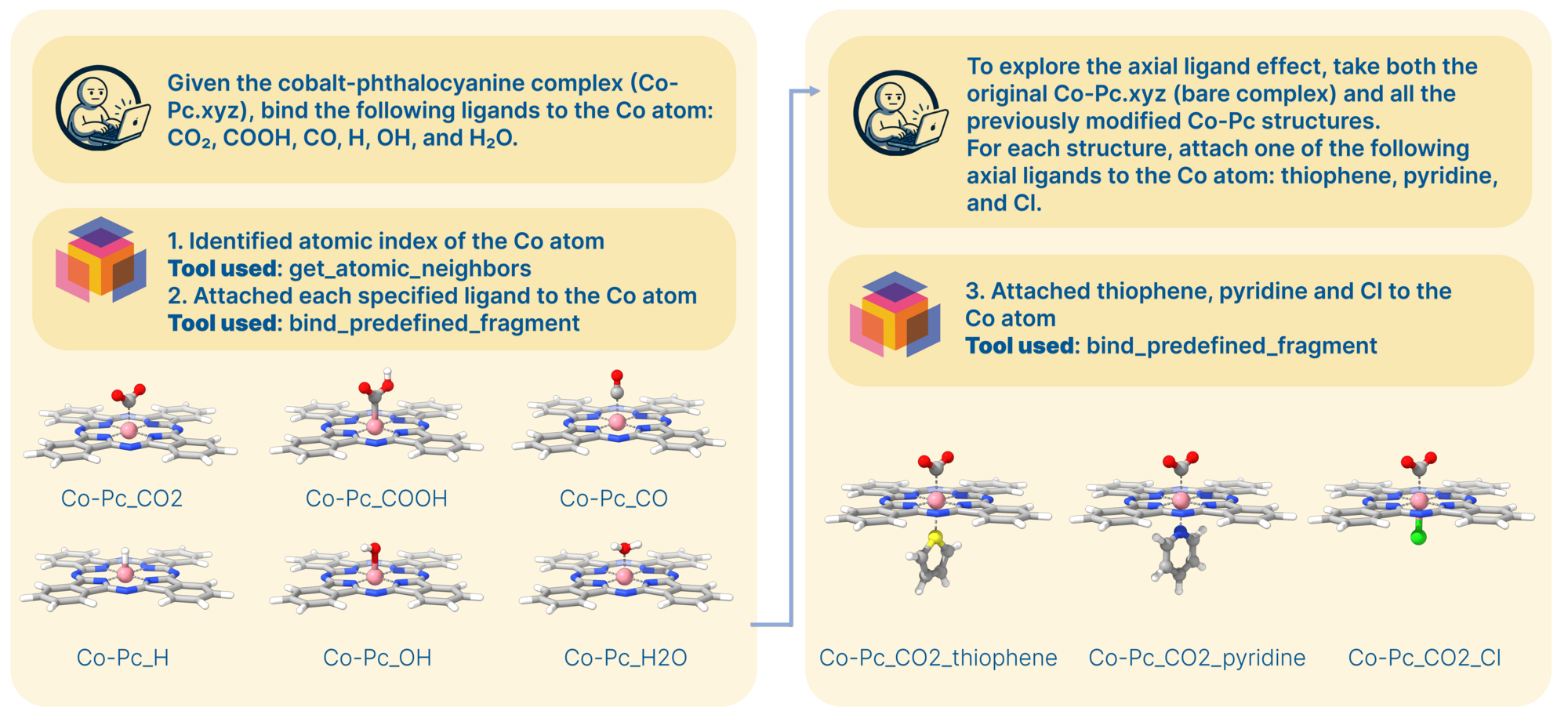}  \caption{Binding of reaction intermediates and axial ligands to a cobalt–phthalocyanine (Co-Pc) complex. The generated \texttt{xyz} structures are shown. In this example, sequential prompts are used first to bind reaction intermediates to the cobalt centre and, subsequently, to introduce axial ligands. 
        The Co centre, together with the bound intermediates (CO$_2$, COOH, CO, H, OH, and H$_2$O), and axial ligands (thiophene, pyridine, or Cl) are highlighted using a ball-and-stick representation, while the remaining atoms are rendered as sticks for clarity. 
        For axial-ligand–bound structures, only the CO$_2$-bound complexes are shown here. The full set of axial-ligand–coordinated structures is provided in~\autoref{fig:SI_CoPc}. 
        Atom colour codes: pink, Co; lime green, Cl; yellow, S; red, O; blue, N; grey, C; white, H.}
        \label{fig:case_binding}
\end{figure}
In this case study, we evaluate the ability of \estructural{} to bind reaction intermediates and ligands to a transition-metal catalyst, a common operation in computational catalysis (\autoref{fig:case_binding}, \autoref{fig:SI_CoPc} and~\autoref{tab:per-Case2}). The target system is a cobalt--phthalocyanine (Co--Pc) complex, which has been widely studied in electrochemical CO$_2$ reduction (CO$_2$RR)~\cite{CoPc_acidic, CoPc_CO2RR_Chem, CoPc_solvated, CoPc_strained} and oxygen reduction reactions (ORR)~\cite{CoPc_H2O2}. 

The user specifies the ligand identities, but not the Co atom index or the binding procedure. \estructural{} identifies the Co atom index and attaches CO$_2$, COOH, CO, H, OH, and H$_2$O to generate the corresponding intermediate-bound structures.   

Starting from the provided \texttt{Co–Pc.xyz} structure, \estructural{} employs fragment binding tools to attach each fragment to the cobalt atom, producing a series of Co–Pc structures bound with the corresponding reaction intermediates.

In the second step, the agent takes both the original Co--Pc structure and the structures generated in the first step, and attaches thiophene, pyridine, or chloride as an additional axial ligand. This requires \estructural{} to reuse its own intermediate outputs and apply the requested modification across the complete set of structures.

These examples illustrate how \estructural{} can efficiently construct a variety of catalyst–intermediate complexes for organometallic catalysts. Such structures are commonly required for modelling catalytic reaction mechanisms and energy profiles.

The two-step workflow above was used to demonstrate sequential reuse of previously generated structures. In a more direct setting, the same task can be requested in a single combinatorial prompt by specifying the allowed ligands for the two axial positions. \estructural{} then identifies the Co index, enumerates the requested ligand combinations, and generates the corresponding complexes without requiring the user to specify the intermediate sequence (\autoref{fig:SI_Case2_generic} and~\autoref{tab:per-Case2_generic}).

\subsubsection{Case study 3: Organometallic complex construction with stereochemical and ligand variations}\label{subsec:casestudy3}

\begin{figure}[!t]
        \centering        \includegraphics[width=1.0\textwidth]{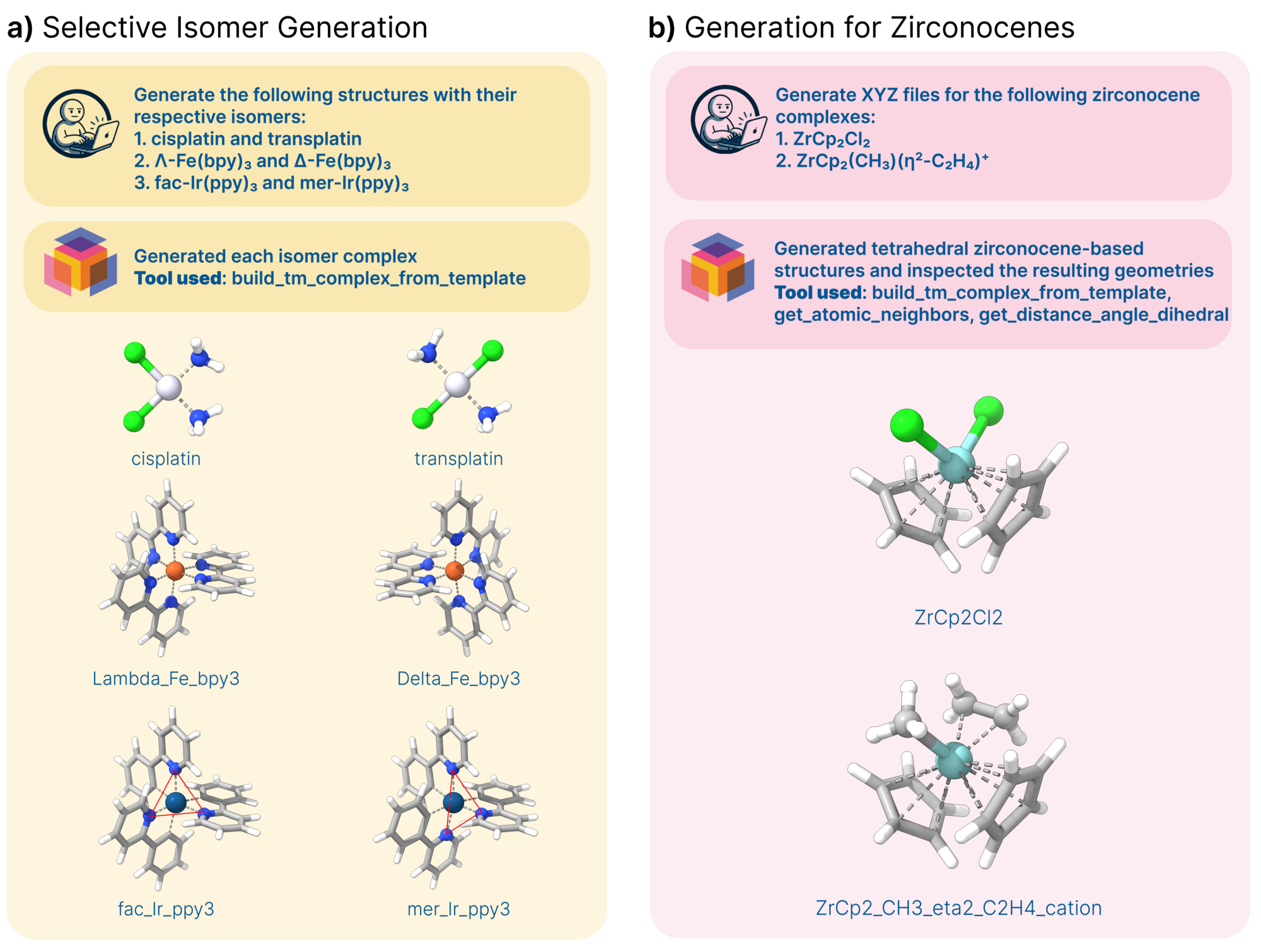}
        \caption{Generation of organometallic complexes with specified stereochemistry and ligand composition. (a) Generation of stereoisomeric pairs, including cisplatin/transplatin, $\Lambda$- and $\Delta$-Fe(bpy)$_3$, and \textit{fac}- and \textit{mer}-Ir(ppy)$_3$. For Fe(bpy)$_3$ and Ir(ppy)$_3$, the transition metal centres and the coordinating N atoms that define the stereochemistry are highlighted in ball-and-stick representation, while the remaining atoms are shown as sticks for clarity. For \textit{fac}- and \textit{mer}-Ir(ppy)$_3$, a red triangle is added to indicate the geometric arrangement of the coordinating N atoms to Ir.
        (b) Construction of zirconocene-based complexes with varying ligand sets, including ZrCp$_2$Cl$_2$ and ZrCp$_2$(CH$_3$)($\eta^2$-C$_2$H$_4$)$^{+}$. Cyclopentadienyl (Cp) rings are shown in stick models for clarity.
        Atom colour codes: dark blue, Ir; seashell, Pt; sky blue, Zr; dark orange, Fe; lime green, Cl; red, O; blue, N; grey, C; white, H.}
        \label{fig:case_Inorg}
\end{figure}
In this case study, we demonstrate the ability of \estructural{} to generate organometallic complexes with specified symmetry, stereochemistry, and ligand coordination (\autoref{fig:case_Inorg}a and Tables~\ref{tab:per-Case3A}--\ref{tab:per-Case3A_rmsd}). Generating such structures is a non-trivial task, as SMILES-based representations offer limited support for coordination environments, metal–ligand bonding, and stereochemical variants commonly encountered in organometallic chemistry. Even when organometallic structures can be encoded, reliably generating specific stereoisomers, such as \textit{cis}/\textit{trans}, \textit{fac}/\textit{mer}, or $\Delta/\Lambda$ enantiomers, remains challenging within conventional string-based molecular representations. Addressing these limitations is essential for studying structure–property relationships and reactivity trends in coordination chemistry and catalysis.

The user specifies the target complexes and their desired isomeric forms, while \estructural{} determines how the ligands should be placed and connected to generate the requested 3D structures. \estructural{} therefore selects the appropriate coordination templates and ligand placements to generate cisplatin and transplatin, $\Delta$- and $\Lambda$-Fe(bpy)$_3$ (bpy = 2,2'-bipyridine), and \textit{fac}- and \textit{mer}-Ir(ppy)$_3$ (ppy = 2-phenylpyridine). These systems represent common classes of geometric and chiral isomerism in coordination compounds.  

By selecting predefined coordination templates and specifying ligand identities and binding positions, \estructural{} constructs the corresponding \texttt{xyz} structures directly, ensuring the correct relative arrangement of ligands around the metal centre.

In addition to stereochemical variants, ligand substitution and coordination-mode variations are demonstrated using zirconocene-based complexes (\autoref{fig:case_Inorg}b and~\autoref{tab:per-Case3B}). 
The user specifies only the target complexes, while the coordination arrangement and ligand placement are determined by \estructural{}. 
Starting from a tetrahedral coordination template, \estructural{} generates ZrCp$_2$Cl$_2$ (Cp = cyclopentadienyl) and ZrCp$_2$(CH$_3$)($\eta^2$-C$_2$H$_4$)$^{+}$.

\begin{figure}[t]
        \centering        \includegraphics[width=1.0\textwidth]{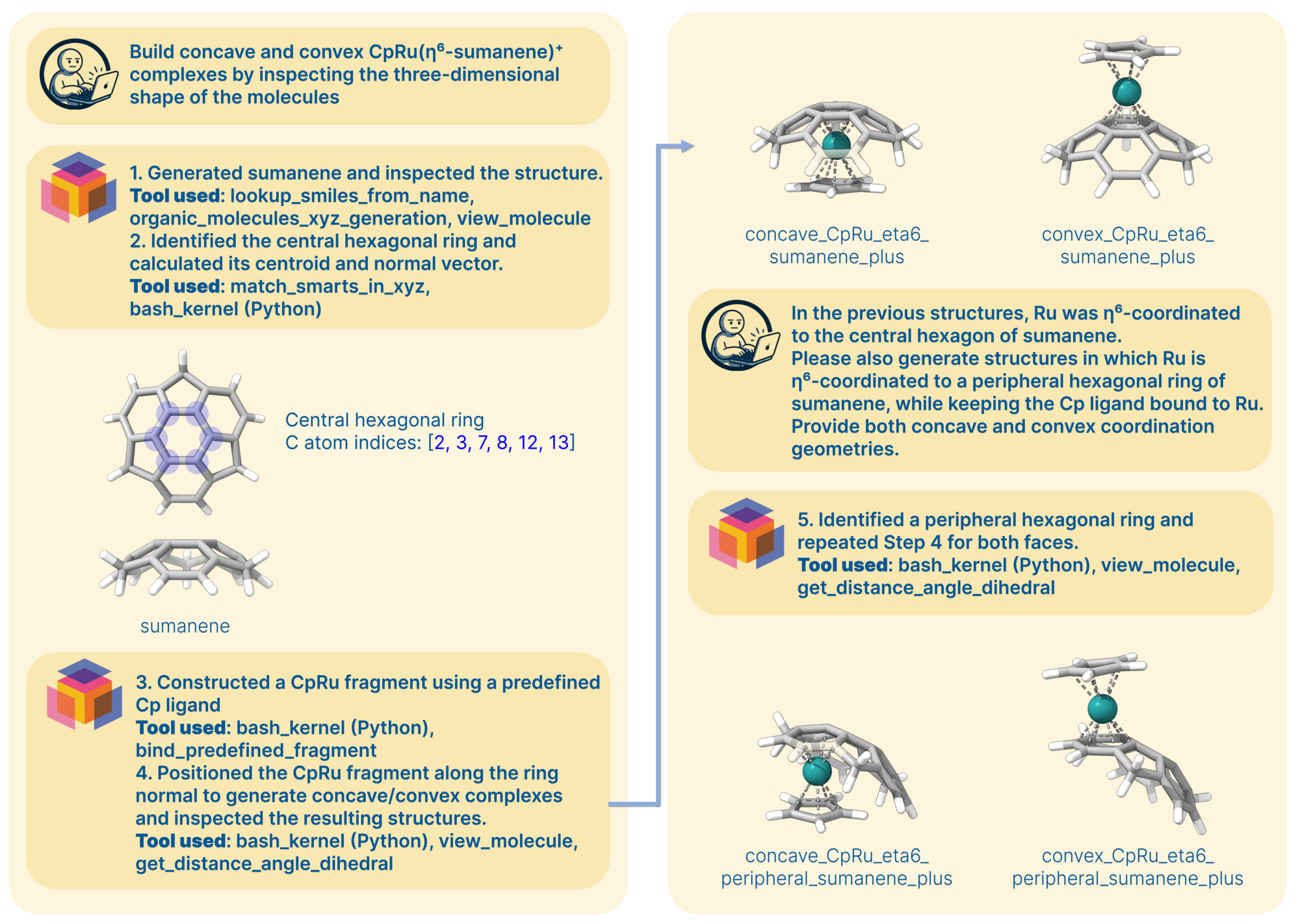}
        \caption{Stepwise construction of concave and convex CpRu($\eta^{6}$-sumanene)$^{+}$ complexes by \estructural{}. Initial construction of concave and convex CpRu($\eta^{6}$-sumanene)$^{+}$ complexes with Ru coordinated to the central hexagonal ring of sumanene, and extension of the construction in response to a follow-up prompt, generating concave and convex structures in which Ru is $\eta^{6}$-coordinated to a peripheral hexagonal ring while retaining the Cp ligand. Atom colour codes: teal, Ru; grey, C; white, H.}
        \label{fig:case_sumanene}
\end{figure}

To further test whether \estructural{} can generate organo-metallic structures through geometric reasoning rather than relying on predefined coordination templates, we examined the construction of concave and convex CpRu($\eta^{6}$-sumanene)$^{+}$ complexes (\autoref{fig:case_sumanene} and~\autoref{tab:per-Case3C}). 
The user specified only that both concave and convex structures should be generated, while the $\eta^{6}$ coordination site, the corresponding atomic indices, and the construction workflow were left to \estructural{}.

The agent first generates a sumanene structure and identifies a six-membered carbon ring using structural analysis and pattern matching. It then calculates the centroid and normal vector of the selected ring and constructs a CpRu fragment using a predefined Cp ligand. The CpRu fragment is subsequently oriented and positioned along the ring normal to generate the concave and convex coordination geometries. Visual and geometric inspections are also performed to examine the resulting structures.

Because the initial prompt did not specify which six-membered ring should coordinate Ru, some runs selected the central ring. This assignment differs from the experimentally reported coordination motif, in which Ru binds to a peripheral hexagonal ring~\cite{amaya2010dynamically}. A follow-up prompt specifying peripheral-ring coordination led \estructural{} to identify the corresponding carbon atoms, reposition the CpRu fragment, and generate the desired concave and convex structures. This example illustrates both the ability of \estructural{} to combine structural analysis with geometric operations and the need for additional user guidance when multiple chemically plausible coordination sites are present.

\subsubsection{Case study 4: Fragment-level replacement in organometallic complexes}\label{subsec:casestudy4}
\begin{figure}[!t]
        \centering        
        \includegraphics[width=1.0\textwidth]{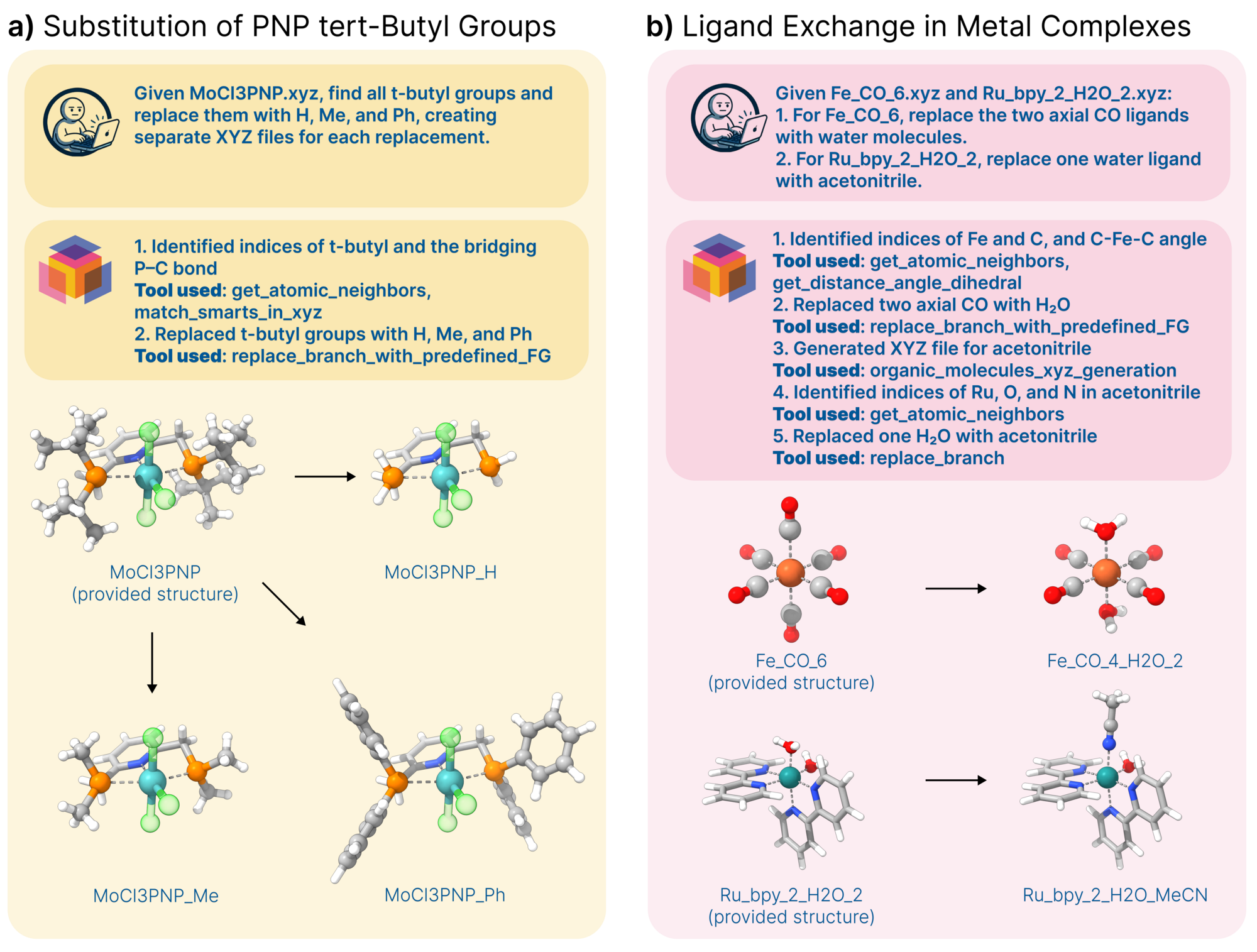}
        \caption{Fragment-level replacement and ligand exchange in organometallic complexes. (a) Functionalization of a literature-reported MoCl$_3$–PNP complex through systematic replacement of tert-butyl groups on a polydentate PNP ligand with hydrogen, methyl, and phenyl substituents.
        (b) Ligand exchange in transition-metal complexes, including replacement of axial CO ligands with water in an octahedral Fe(CO)$_6$ and substitution of a coordinated water molecule with acetonitrile in a Ru(bpy)$_2$(H$_2$O)$_2$ complex. In all cases, fragment-level replacements are performed while preserving the underlying coordination framework. 
        For the MoCl$_3$PNP and Ru(bpy)$_2$(H$_2$O)$_2$ complexes, the substituted groups and newly introduced ligands are highlighted using a ball-and-stick representation, while the remaining parts of the molecular framework are rendered as sticks for clarity.  
        Atom colour codes: green, Ru; light cyan, Mo; dark orange, Fe; lime green, Cl; orange, P; red, O; blue, N; grey, C; white, H.}        
        \label{fig:ligand_exchange}
\end{figure}

In this case study, we demonstrate fragment-level replacement operations in organometallic complexes using \estructural{} (\autoref{fig:ligand_exchange}a and~\autoref{tab:per-Case4A}). Constructing organometallic systems from scratch is often challenging, particularly for complexes featuring polydentate ligands and intricate coordination environments. In such cases, modifying an existing core structure can provide a more practical strategy.

The first example uses a literature-reported MoCl$_3$--PNP complex containing a tridentate ligand, which has been reported as an efficient catalyst for dinitrogen reduction to ammonia~\cite{Mo-PNP}. The initial \texttt{xyz} structure is taken directly from the literature source.

To explore substituent effects without rebuilding the PNP ligand and metal complex from scratch, the user requests replacement of all tert-butyl groups with H, Me, and Ph.

The user specifies the input structure, the group to be replaced, and the desired substituents, while \estructural{} identifies the corresponding tert-butyl groups and atomic indices and selects the required editing operations.

Starting from the literature-derived structure, \estructural{} systematically replaces each tert-butyl group with H, Me, or Ph to generate separate modified complexes while preserving the original metal coordination environment.

The second example focuses on ligand-exchange reactions, highlighting \estructural{}'s ability to replace not only terminal atoms but also entire ligand fragments. The user requests ligand exchange operations on two given complexes, Fe(CO)$_6$ and Ru(bpy)$_2$(H$_2$O)$_2$ (\autoref{fig:ligand_exchange}b and~\autoref{tab:per-Case4B}).

The input structures, ligands to be exchanged, and target coordination positions are specified by the user, while \estructural{} identifies the corresponding atomic indices and determines the required editing operations.

For Fe(CO)$_6$, \estructural{} identifies the six carbon atoms coordinated to the Fe using the \texttt{get\_atomic\_neighbors} tool and analyzes the C--Fe--C angles to determine their relative positions. A trans pair with a C--Fe--C angle approximately 180$^\circ$ is selected as the two axial CO ligands and replaced with water molecules while preserving the remaining coordination environment.

For Ru(bpy)$_2$(H$_2$O)$_2$, one coordinated water ligand is replaced with acetonitrile. Since acetonitrile is not included in the predefined fragment library, \estructural{} generates its \texttt{xyz} file on demand and uses it for the ligand-exchange operation. This shows that ligand exchange is not limited to predefined fragments and can accommodate newly generated ligand species.

Overall, this case study illustrates how \estructural{} performs targeted fragment replacement and ligand exchange on existing coordination structures without reconstructing the complexes from scratch.

\subsubsection{Case study 5: Molecular fragment operations and fragment analysis}\label{subsec:casestudy5}
\begin{figure}[ht]
        \centering
        \includegraphics[width=1.0\textwidth]{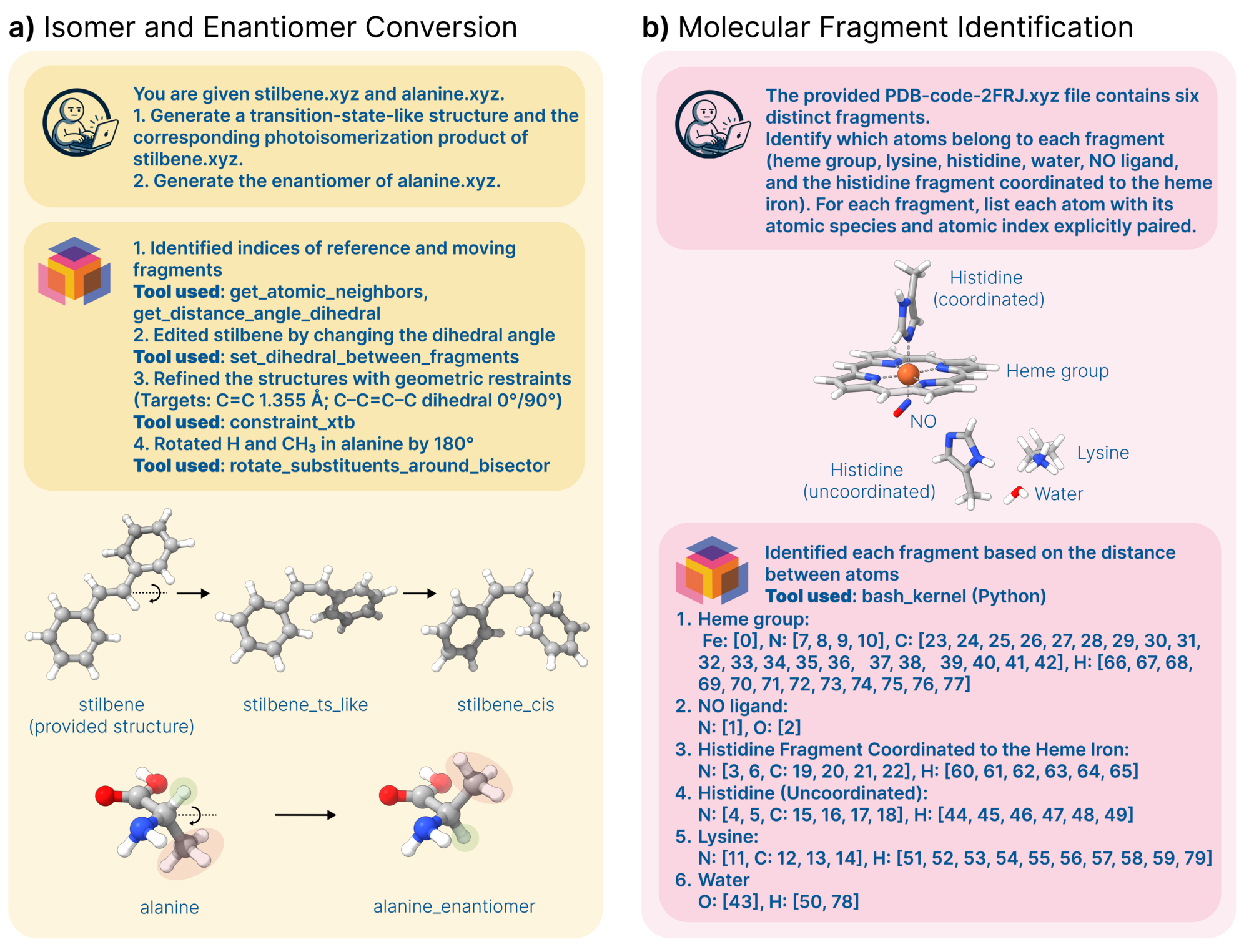}
        \caption{Molecular fragment operations and fragment analysis enabled by \estructural{}. (a) Conversion between isomeric forms through geometric operations, including the generation of \textit{cis}-stilbene and a photoisomerization TS-like geometry from \textit{trans}-stilbene, as well as enantiomer generation from L-alanine. (b) Fragment-level structural analysis of a multi-component system derived from a PDB structure, where distinct fragments, including a heme group, coordinated NO, coordinated and uncoordinated histidine, lysine, and H$_2$O, are identified, and atom indices are assigned explicitly for each fragment. Atom colour codes: dark orange, Fe; red, O; blue, N; grey, C; white, H.}
        \label{fig:operation_analysis}
\end{figure}

In this case study, we demonstrate \estructural{}'s ability to manipulate and identify molecular fragments using geometric operations and structural analysis tools. These capabilities complement structure generation and editing by enabling controlled isomer interconversion and fragment-aware analysis of complex molecular systems (\autoref{fig:operation_analysis}).

We first consider isomer interconversion in organic molecules, focusing on two representative cases: \textit{cis}/\textit{trans} isomerization of an alkene and enantiomer interconversion of an amino acid (\autoref{fig:operation_analysis}a and~\autoref{tab:per-Case5A}).

The user specifies the input structures and desired transformations, while \estructural{} determines the structural operations needed to generate the requested isomeric forms.

Alkene-centred \textit{cis}/\textit{trans} isomerization is frequently observed in photoexcited states and underlies the function of molecular photoswitches~\cite{kuramochi2024ultrafast,villaron2020stiff}. Starting from the provided \textit{trans}-stilbene structure, \estructural{} identifies the relevant molecular fragments and modifies the C--C=C--C dihedral angle describing the relative orientation of the two phenyl groups. By setting this dihedral angle to $0^\circ$ and $90^\circ$, the agent generates the \textit{cis}-stilbene isomer and a TS-like geometry associated with photoisomerization, respectively.
The resulting structures were further refined using \texttt{constraint\_xtb}, with the C=C bond length restrained toward 1.355~\AA{} and the C--C=C--C dihedral angle restrained toward $0^\circ$ and $90^\circ$ for the cis isomer and TS-like structure, respectively.

For enantiomer interconversion, we considered conversion of the provided L-alanine structure to D-alanine. \estructural{} identifies the stereogenic carbon and applies a targeted geometric rotation to rearrange the substituents around the stereogenic centre, yielding the opposite enantiomer. Such stereochemical conversions are relevant in biomolecular modelling and ligand preparation, where the correct enantiomer is often required.

The second example highlights fragment-level structural analysis for large molecular systems (\autoref{fig:operation_analysis}b and~\autoref{tab:Case5B}). For such systems, fragment information can be useful for constrained-geometry optimization or multilevel calculations such as QM/MM. Here, \estructural{} is tasked with identifying the atomic indices of six fragments in a PDB-derived structure adapted from an example in the ORCA 6.0 manual (Section 6.3.7)~\cite{orca6, ORCA_manual}.

The user specifies the identities of the target fragments, while \estructural{} determines the corresponding atomic indices and the procedure needed to separate them.

The agent analyzes the molecular connectivity using a distance-based graph and partitions the structure into the requested fragments, including the heme group, coordinated and uncoordinated histidine residues, lysine, a nitric oxide ligand, and water molecules. It then reports the atomic species and index mappings for each fragment.

\subsubsection{Case study 6: Multimodal reaction mechanism--guided structure generation}\label{subsec:casestudy6}

\begin{figure}[ht]
        \centering
        \includegraphics[width=1.0\textwidth]{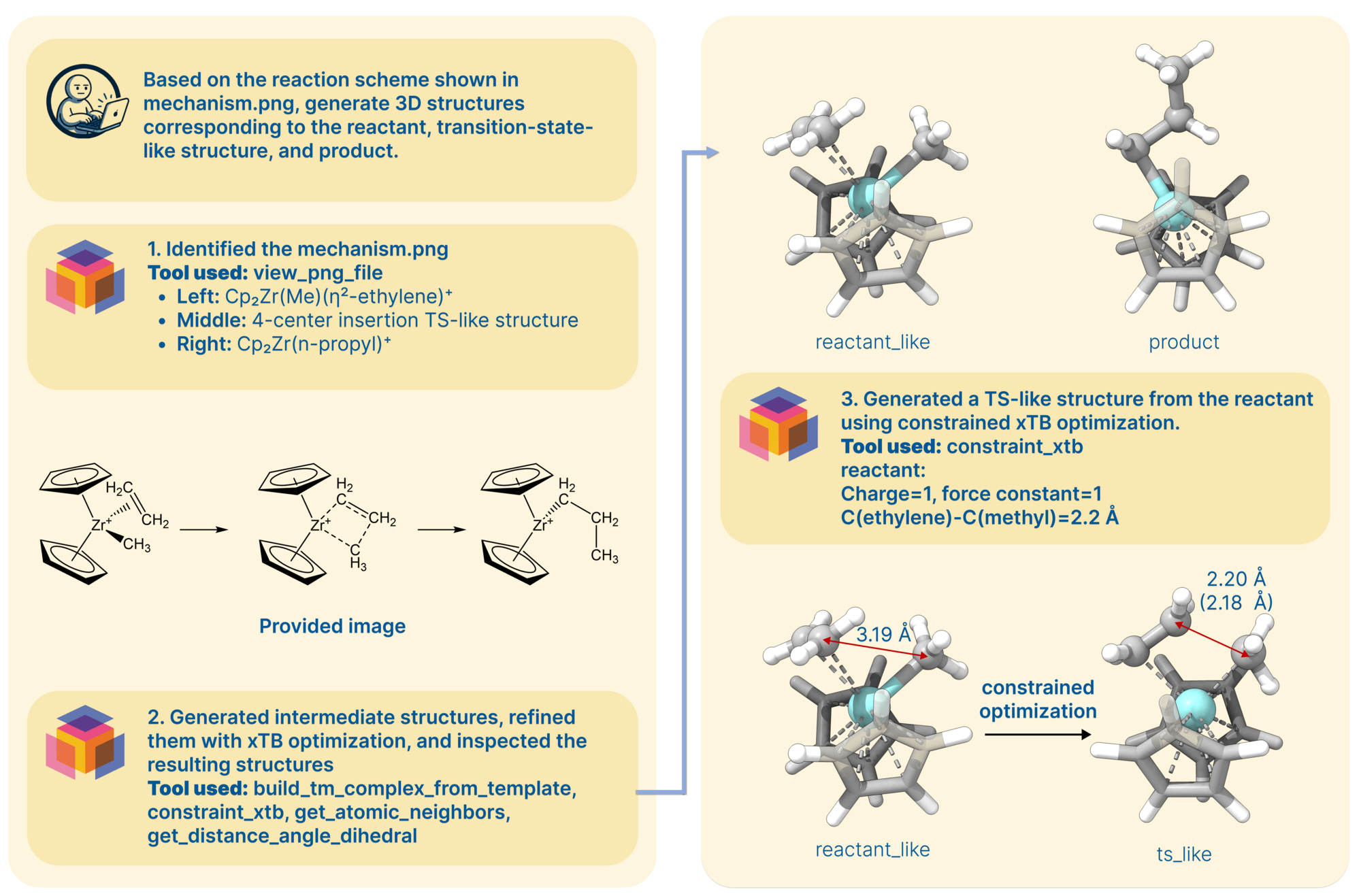}
        \caption{Multimodal reaction mechanism-guided generation of intermediates and TS-like structures. Starting from a user prompt and a provided reaction mechanism image, \estructural{} identifies the reaction intermediates and TS motif, generates three-dimensional structures for each intermediate, and constructs TS-like geometry using the \texttt{constraint\_xtb} tool. The highlighted C(ethylene)--C(methyl) distances are shown for the xTB-refined reactant-like structure (3.19~\AA) and the TS-like structure (2.20~\AA). 
        The full set of distance constraints and the force constant are provided in~\autoref{tab:gpt54_constraints}. The value in parentheses (2.18~\AA) denotes the corresponding reference distance obtained from an independent DFT transition-state optimization. Atom colour codes: sky blue, Zr; grey, C; white, H.}
        \label{fig:adv_case_ethylene}
\end{figure}

Unlike ordinary molecular specifications, reaction mechanisms involving intermediates and transition states (TSs) are often communicated through schematic images rather than narrative text. This is because key features of reactive structures, such as distorted bond lengths, angles, and partially formed or broken bonds, are difficult to describe precisely using text or string-based formats such as SMILES. By contrast, mechanism diagrams can convey these geometric and connectivity changes more directly.

Motivated by this common practice, this case study evaluates the ability of \estructural{} to perform multimodal reasoning by generating three-dimensional molecular structures from a reaction mechanism schematic (\autoref{fig:adv_case_ethylene} and Tables~\ref{tab:gpt54_constraints}--\ref{tab:Case6}). The example is the migratory insertion of ethylene into a metal--carbon bond, a key elementary step in ethylene polymerization~\cite{alt2000metallocene_review,musaev1997_Ni_diimine}. 
Here, the user provides only the reaction image and requests the reactant, TS-like structure, and product, while the identities, three-dimensional arrangements, and structure-generation workflow are determined by \estructural{}.

From the reaction scheme, \estructural{} identifies the zirconocene methyl--ethylene complex as the reactant, the propyl-bound complex as the product, and the insertion geometry as the TS-like structure. 

The agent captures the main molecular identities, coordination environment, and reaction-centre geometry shown in the schematic, but does not necessarily reproduce all fine conformational details. For example, the staggered relative orientation of the two Cp rings shown in the input image was not explicitly reproduced in the generated structure.

To generate the TS-like structure, the agent additionally identifies the reaction centre, determines which atoms should be constrained, selects physically reasonable constraint values, and assigns the molecular charge and spin multiplicity for the refinement. Using \texttt{constraint\_xtb}, \estructural{} generates a representative TS-like structure with the forming C--C distance constrained to 2.20~\AA ~(\autoref{tab:gpt54_constraints}), close to the reference DFT TS value of 2.18~\AA\ obtained at the B3LYP-D4/def2-SVP level in ORCA 6.0.~\cite{beckeDensityfunctionalThermochemistryIII1993,leeDevelopmentColleSalvettiCorrelationenergy1988,stephensInitioCalculationVibrational1994,caldeweyher2017extension_d4correction,caldeweyherGenerallyApplicableAtomiccharge2019,weigendBalancedBasisSets2005,orca6}. Subsequent TS optimizations initiated from the generated structures converged successfully and yielded a single imaginary frequency near $-300$~cm$^{-1}$ (\autoref{tab:Case6}). 
The constrained optimization is therefore used to construct a chemically reasonable TS-like starting geometry. Validation of the transition state is performed subsequently by quantum-chemical TS optimization and frequency analysis, while complete reaction-pathway validation remains outside the scope of the present workflow.

This example shows that \estructural{} can combine image interpretation, structural reasoning, and constrained geometry refinement to construct chemically meaningful reactive structures from schematic reaction information.

\subsubsection{Case study 7: Mechanism reasoning-driven geometry construction}\label{subsec:casestudy7}
\begin{figure}[!ht]
        \centering
        \includegraphics[width=1.0\textwidth]{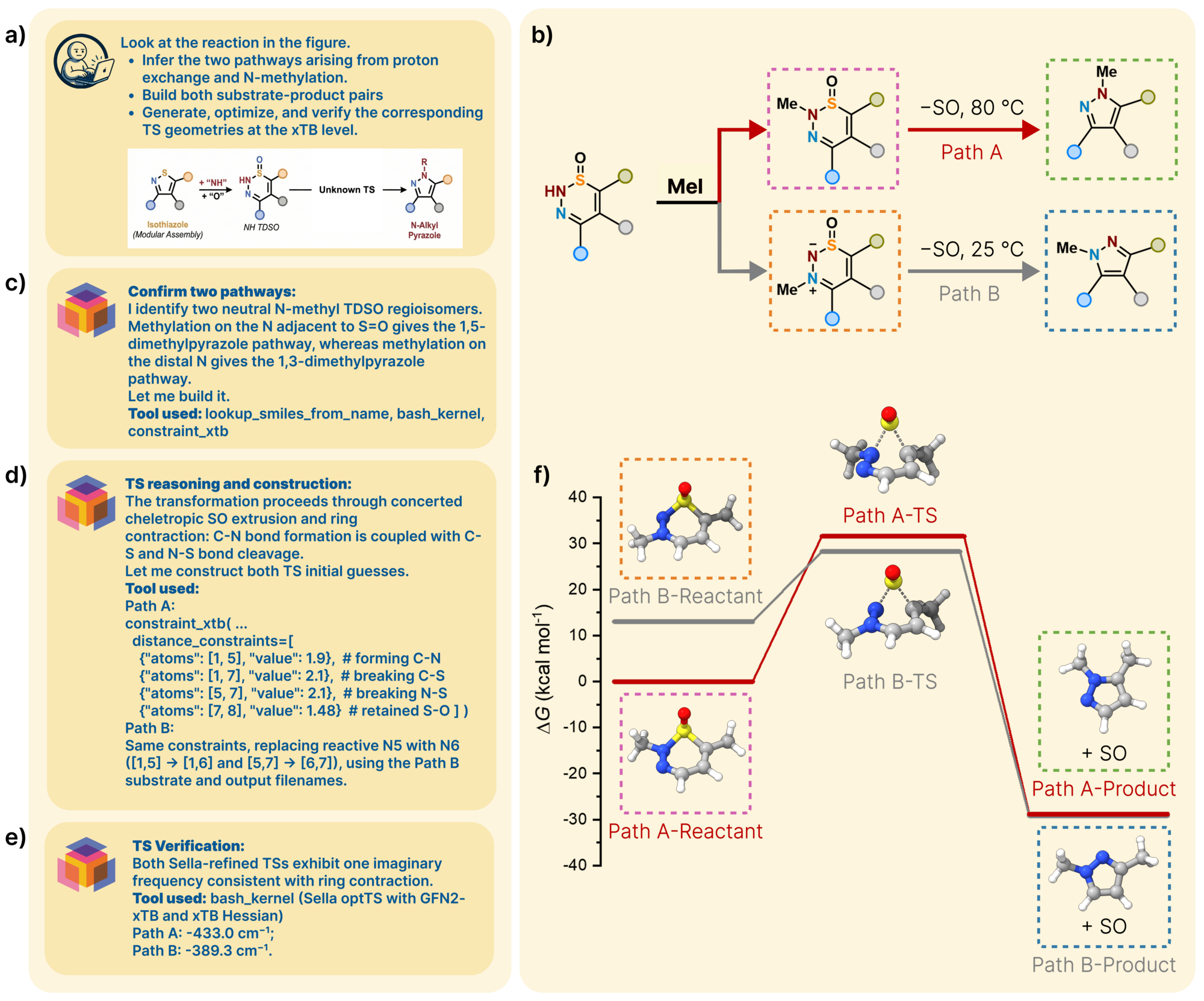}
        \caption{Mechanism reasoning-driven geometry construction for a single-atom skeletal-editing reaction. (a) Overview of the single multimodal request and supplied reaction diagram; the exact prompt used in the benchmark is given in Supporting Information Section~\ref{SIsec:case7}. (b) The two N-methyl TDSO substrates and corresponding pyrazole products arising from proton exchange before N-alkylation. (c) The agent's assignment and construction of the two substrate--product pairs. (d) Inference of concerted cheletropic SO extrusion and ring contraction, followed by construction of TS initial guesses using xTB distance constraints. (e) Sella optTS refinement at the GFN2-xTB level and frequency verification of the two first-order saddle points. No clarification or corrective user intervention was permitted during the trajectory shown in (a)--(e). (f) Free energy diagram calculated at the B3LYP-D4/def2-SVP level of theory using structures generated by \estructural{} as initial geometries. Elongated S–C and S–N bonds undergoing cleavage in the transition states are represented by dashed lines. Atom colour codes: yellow, S; red, O; blue, N; grey, C; white, H.}
        \label{fig:adv_case_NHDO}
\end{figure}

A common challenge in reaction discovery campaigns in organic and organometallic
chemistry is rationalizing reaction mechanisms to build an in-depth understanding and
facilitate knowledge-driven experimental design, where computations can play a pivotal
role.

Constructing a computational reaction profile requires plausible geometries for all
relevant ground and transition states. This process becomes laborious when several
reaction sites, substituent patterns, or mechanistic alternatives must be considered,
and it depends strongly on the operator's understanding of both reaction mechanisms
and computational chemistry. Taking a recent single-atom skeletal-editing reaction as
an example~\cite{fanourakis_strategic_2025}, we illustrate the potential of
\estructural{} for (1) following structures and substituent patterns through a reaction
pathway communicated through text and images, (2) proposing a chemically plausible TS
formation mechanism and reconstructing the corresponding three-dimensional structures
as \texttt{xyz} files, and (3) generating alternative pathways through systematic
structural modifications.

Specifically, we supplied the reaction diagram containing a masked transition state
together with a single prompt (\autoref{fig:adv_case_NHDO}a). The prompt specified the
substitution pattern and the proton exchange preceding N-alkylation, which produces two
neutral, compositionally identical starting structures and two possible pathways
(\autoref{fig:adv_case_NHDO}b). These chemical conditions were necessary to define the
target problem, but no substrate or product structures, atom indices, bond-change
assignments, geometric constraints, or intermediate instructions were provided. The
agent was asked to infer both substrate--product pairs and construct the corresponding
TS geometries; after this initial request, the complete trajectory proceeded without
human intervention.

From the reaction image, \estructural{} identified the six-membered TDSO scaffold,
tracked the substituent pattern into the five-membered pyrazole, and assigned the two
neutral N-methyl TDSO regioisomers to the corresponding 1,5- and
1,3-dimethylpyrazole products (\autoref{fig:adv_case_NHDO}b, c). It then inferred a
concerted ``SO extrusion + ring contraction'' mechanism involving simultaneous C--N
bond formation and S--C/S--N bond cleavage. This chemical hypothesis was translated
into geometric constraints to construct a TS initial guess for each pathway, followed
by Sella~\cite{hermes2022sella} optTS refinement at the GFN2-xTB level and frequency analysis
(\autoref{fig:adv_case_NHDO}d, e).

In a representative successful run, both substrate--product pairs were assigned
correctly, and the two Sella-refined TS structures each exhibited one significant
imaginary frequency ($-433.0$ and $-389.3$~cm$^{-1}$). The associated modes were
consistent with C--N bond formation coupled to S--C and S--N bond cleavage.

To assess reproducibility, we repeated the complete task five times using the same
single prompt. The mean result-oriented score was $61.0 \pm 37.0$ out of 100; however, only one of the five runs produced a fully correct result. The
observed failures included an incorrect product regioisomer, poorly constructed TS guesses that in one case led to an unintended proton-transfer saddle, a frequency-verification step that misread its own output, and one trajectory that modelled a different reaction. These results highlight substantial run-to-run variation in the chemical reasoning of GPT-5.4. We also tested Claude-Opus-4.5 on this case study with an end-point only variant that requested only the two substrates and two products, effectively testing whether it could correctly construct
the corresponding geometries from the provided image. In all five runs, the Opus-4.5 model failed to identify the correct structure from the given image and constructed wrong geometries, showcasing that accurate visual perception of chemical structures remains a prerequisite for this workflow. Full per-run results and
failure analyses are provided in Supporting Information Section~\ref{SIsec:case7}--\ref{SIsec:case7_endpoint} and Tables~\ref{tab:case7-rubric}--\ref{tab:case7-opus45-endpoints}.

In the representative run (Run~1), \estructural{} reported the relative energies of both reaction pathways at the GFN2-xTB level. The agent-generated structures were subsequently used as initial geometries for DFT geometry and TS optimizations at the B3LYP-D4/def2-SVP level, all of which converged successfully. The resulting DFT Gibbs free energies were used to construct the reaction energy profiles shown in \autoref{fig:adv_case_NHDO}f. The corresponding GFN2-xTB and DFT energetics are discussed in Supporting Information Section~\ref{Case7_refinement} (\autoref{tab:case7-energies}).

Overall, this case study demonstrates how \estructural{} can support mechanistic
investigation of competing reaction pathways by translating a chemical hypothesis into
concrete structural manipulations. The successful single-prompt trajectory shows that
extensive stepwise guidance is not required when the reaction image and chemical
context are interpreted correctly, substantially reducing the dependence on
human-guided construction. Nevertheless, the variation across repeated runs shows that
this remains an emergent capability rather than evidence of robust performance across
arbitrary reaction systems. Visual perception and chemical reasoning are no longer
prohibitive in every instance, but they remain important sources of failure. Thus, the
current system can operate autonomously for well-specified tasks, whereas ambiguous
reaction images and high-stakes TS assignments still warrant expert verification.
The demonstrated workflow should therefore be understood as mechanism-guided geometry
construction and frequency-based assessment of candidate TS geometries, rather than general
autonomous discovery or exhaustive validation of a reaction network.

\section{Discussion}\label{discussion}
In this section, we summarize the overall benchmark performance and discuss common failure modes associated with configurational ambiguity. We further examine the respective roles of deterministic tools and agent reasoning through ablation tests, provide practical guidance for users, and discuss the remaining limitations and future directions.

The overall benchmark results are summarized in~\autoref{tab:all_case_scores}. Across Cases 1--6, most tasks were completed consistently over five independent runs without human intervention. A notable exception was the CpRu($\eta^{6}$-sumanene)$^{+}$ case, where four of the five runs initially selected the central hexagonal ring rather than the experimentally reported peripheral coordination site and therefore required a follow-up instruction (\autoref{tab:per-Case3C}). This suggests that one of the main remaining challenges is the selection among multiple plausible molecular configurations rather than the execution of individual editing operations. When the user request does not uniquely specify the intended configuration, the agent should either select the most chemically appropriate structure, generate reasonable alternatives, or ask the user for clarification. Consistent with this interpretation, when the prompt was changed to request every unique complex, including both central and peripheral coordination and both concave and convex geometries, \estructural{} generated all four configurations correctly in all five runs (\autoref{fig:SI_sumanene_generic} and~\autoref{tab:sumanene_every}).

A similar configurational ambiguity emerged when we extended the ligand-binding task in Case Study 2 from the planar Co--Pc complex to a nonplanar Pb--Pc complex, where the two sides of the macrocycle are no longer geometrically equivalent (\autoref{fig:SI_PbPc}). The first ligand was consistently placed on the convex side of the macrocycle, a chemically reasonable placement consistent with previously reported axial coordination motifs for out-of-plane metal centres~\cite{jung2025spin}. When a second axial ligand was added, the binding tool initially selected a sterically accessible convex direction according to its geometry-placement rules. In three of five runs, the agent retained the resulting same-side configuration, whereas in two runs, it further repositioned the newly added ligand to place the two axial ligands on opposite sides of Pb. These results indicate that target-atom identification and individual editing operations are generally reliable, while tasks with several plausible configurations can still depend on the agent's higher-level structural interpretation. In such cases, additional user guidance may be needed to specify the desired configuration, to request generation of all plausible configurations, or to verify the generated structure.

We further examined the relative contributions of the deterministic tools and the underlying language model through tool-ablation tests (Figures~\ref{fig:SI_ablation_editing}--\ref{fig:SI_ablation_TS} and Tables~\ref{tab:ablation_CoPor}--\ref{tab:ablation_Case2}; see Supporting Information Section~\ref{SIsec:ablation} for details). For tasks such as atomic-index identification, structure editing, ligand binding, organometallic structure generation, and TS-like geometry construction, the dedicated deterministic tools had a major influence on reliability, geometric consistency, and execution efficiency. However, some tasks also depended strongly on the reasoning capability of the underlying LLM. In the multimodal reaction-mechanism case, for example, the agent had to identify the reacting atoms, select appropriate geometric constraints, and determine suitable constraint values before invoking the deterministic tools. Tests with different underlying LLMs therefore showed model-dependent performance in these reasoning-intensive steps (Tables~\ref{tab:LLMs}--\ref{tab:model_constraints_comparison} and \autoref{fig:SI_ethylene}). Overall, reliable performance arises from the combination of LLM-based chemical reasoning and deterministic geometry tools, with their relative importance depending on the task.

To clarify where \estructural{} provides added value relative to existing approaches, we compared its role with SMILES-based workflows, conventional molecular editors and other manual workflows, and custom scripted workflows. SMILES-based workflows are efficient for connectivity-level transformations but are less suited to specific three-dimensional conformations, coordination geometries, or TS-like structures. Conventional molecular editors and related manual workflows can handle such structures but require the user to identify the relevant atoms, determine the appropriate geometric manipulation, and select suitable structural parameters based on chemical knowledge and judgement. Custom scripted workflows can automate these operations for predefined targets or templates, but their applicability is often limited when the molecular symmetry, substituent, coordination environment, or broader structural context differs from that for which the workflow was originally designed. In contrast, \estructural{} can combine different analysis and editing methods and adjust its workflow to the input. For example, when exact symmetry was lost in perturbed Co-porphyrin structures, the agent used local connectivity analysis to identify the target sites (Supporting Information Section~\ref{SIsec_distortion}). This flexibility complements conventional molecular editors and scripted workflows by reducing the amount of task-specific setup and manual geometric reasoning required from the user. As a result, \estructural{} can serve as an interactive
molecular-editing assistant for researchers with different
levels of computational chemistry expertise. Routine tasks can often be automated, while ambiguous or mechanistically sensitive cases may still require expert verification or additional guidance.

\begin{figure}[t!]
        \centering        \includegraphics[width=1.0\textwidth]{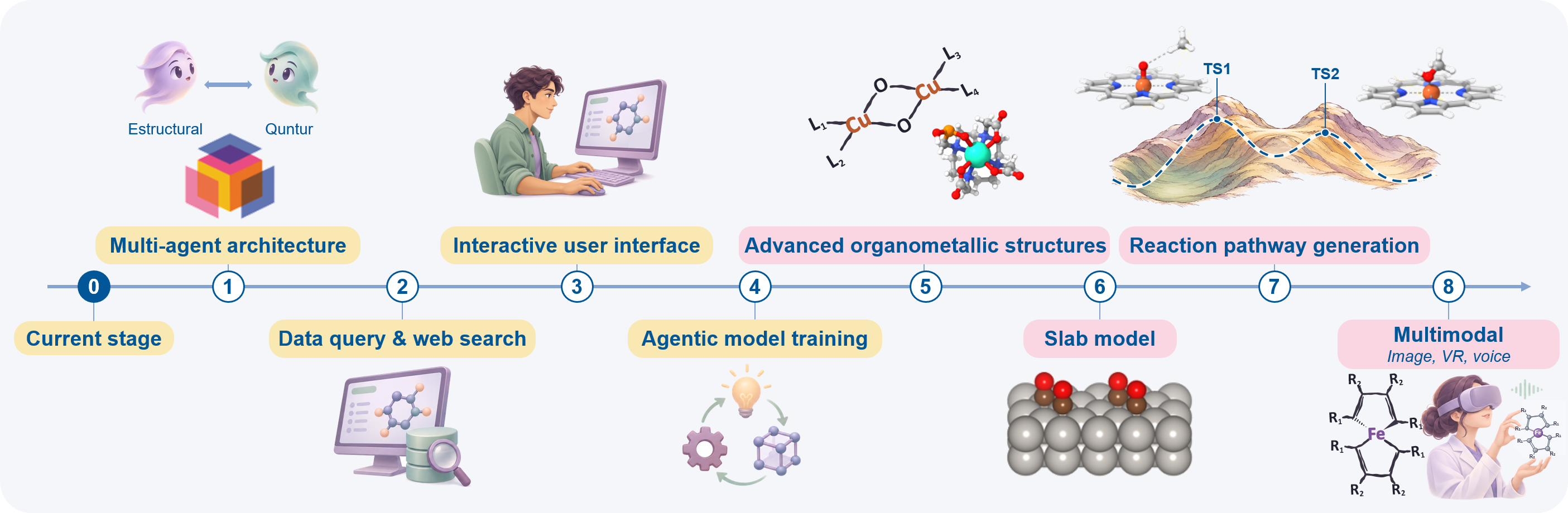}
        \caption{Conceptual roadmap for future development of \estructural{}.}
        \label{fig:roadmap}
\end{figure}

\subsection{Limitations and future directions}\label{limit_roadmap}
To provide practical guidance for users,~\autoref{tab:user_scope} summarizes representative applications of \estructural{}, current scope and limitations, typical failure modes, and recommended user guidance or checks across the major task categories examined in this work. These practical limitations also motivate several directions for further development.

The corresponding future directions are summarized in \autoref{fig:roadmap}. One important step is the integration of \estructural{} with multi-agent quantum chemistry workflows such as \quntur{}, enabling seamless coupling between geometry generation and electronic-structure calculations. Future versions may also incorporate data-driven structure retrieval from databases, enabling automated retrieval of
existing structures followed by targeted modifications, building on the structure-editing capabilities demonstrated in \autoref{fig:ligand_exchange}.

Additional directions include the development of interactive graphical interfaces for direct molecular editing, improved model training for molecular perception and geometric reasoning, and expanded capabilities for complex organometallic systems. Extending the framework to slab models, solid-state systems, and systematic reaction-pathway generation also represents an important long-term goal. Together, these developments aim to broaden the applicability of \estructural{} across more complex computational chemistry workflows. More detailed descriptions of each stage of the roadmap are provided in Supporting Information Section~\ref{SIsec:roadmap}.

\section{Conclusion}
In this work, we introduce \elagenteestructural{}, an agentic AI system that provides a natural-language–driven interface for comprehensive molecular-structure operations, including structure generation, editing, geometric operations, and analysis. By operating directly on three-dimensional Cartesian coordinates, \estructural{} enables modification of molecular structures following chemical intent while preserving their core geometry, beyond \textit{de novo} generation.
This capability is essential for modelling a broader spectrum of chemically relevant structures, including reaction intermediates, TS-like geometries, and coordination complexes that are difficult to access using conventional string-based approaches.

Through a series of realistic case studies, we demonstrated that combining large language model reasoning with structure-manipulation tools, explicitly designed to mirror how computational chemists manipulate structures in practice, enables users to solve practically relevant structure-manipulation tasks in computational chemistry, such as site-selective functionalization, fragment binding, and stereochemically controlled editing.
As a result, \estructural{} offers a practical and reliable pathway for preparing input geometries within automated and semi-autonomous computational chemistry workflows.

Continued development of \estructural{}, together with its integration into \quntur{}, is expected to facilitate increasingly autonomous exploration of chemical space spanning molecular, organometallic, and catalysis.
We anticipate that such agentic approaches will play an important role in facilitating more robust and reliable automated computational workflows, ultimately accelerating discovery and advancing our understanding across a wide range of chemical systems.

\section*{Data Availability}\label{data_availability}
The data supporting the findings of this study are available in the Supporting Information of this article. All data for the case studies are published at \url{https://github.com/the-matter-lab/ElAgenteEstructuralCaseStudies}.




\section*{Acknowledgments}
We gratefully acknowledge the longstanding contributions of the Matter Lab’s current
and past group members (\href{https://matter.toronto.edu}{matter.toronto.edu}), and in particular from El Agente team (\href{https://elagente.ca}{elagente.ca}).
\acknowDARPA\ 
This work was also supported by the AI2050 program of Schmidt Sciences.
C.C. acknowledges support from the DOE grant with the University of Minnesota award \#A006801504 and the Basic Science Research Program through the National Research Foundation of Korea (NRF) funded by the Ministry of Education (RS-2025-02634334).
M.M. was partially supported through a collaborative partnership with Merck KGaA, Darmstadt, Germany.
Y.K. was supported by the CIFAR AI Safety Catalyst Award (Catalyst Fund Project \#CF26-AI-001).
J.B.P.S. acknowledges funding of this project by the Natural Sciences and Engineering Research Council of Canada (NSERC) Alliance Grant \#ALLRP587593-23 (Quantamole). A.W. would like to acknowledge support from the Canada Graduate Scholarships - Doctoral (CGS-D) program.
A.A.-G. thanks Anders G. Fr{\o}seth, for his generous support. A.A.-G. and V.B. acknowledge the generous support of Natural Resources Canada and the Canada 150 Research Chairs program and the University of Toronto’s Acceleration Consortium, which receives funding from the CFREF-2022-00042 Canada First Research Excellence Fund. 
\acknowSciNet\ 
Molecular graphics and analyses were performed with UCSF ChimeraX \cite{pettersen2021ucsf,mengUCSFChimeraXTools2023}, developed by the Resource for Biocomputing, Visualization, and Informatics at the University of California, San Francisco, with support from National Institutes of Health R01-GM129325 and the Office of Cyber Infrastructure and Computational Biology, National Institute of Allergy and Infectious Diseases.

\section*{AI Declaration}
Generative AI tools were used during the preparation of this manuscript. ChatGPT (OpenAI; GPT-5.4--6.0, accessed via a commercial subscription in 2026) was used to assist with grammar correction and language polishing. In addition, ChatGPT (OpenAI) and Gemini (Google; Gemini 3, 2026) were used to generate selected illustrative elements. These include the brain, robot, and tool icons in the central panel of \autoref{fig:schematic}, the user icon reused across multiple figures, and the user illustrations and selected icons in \autoref{fig:roadmap}. The final figures were manually assembled and edited by the authors.

\section*{Conflict of Interest}
The authors declare no conflict of interest.

\clearpage


{
\small

\bibliographystyle{assets/plainnat}
\bibliography{references}


}


\clearpage

\setcounter{page}{1}

\appendix
\renewcommand{\thefigure}{S\arabic{figure}}
\renewcommand{\thetable}{S\arabic{table}}
\setcounter{figure}{0}
\setcounter{table}{0}

{\Huge \textbf{Supporting Information}}

\section{Agents and Tools}\label{agents_tools}
\subsection{The working principle}
The key principle---and core innovation---behind \estructural{} is atomic index-centric geometry operation.
This strategy mirrors how chemists manipulate structures in molecular viewers: we click, drag, and adjust angles or torsions, but each interaction ultimately represents a geometric transformation applied to satisfy a spatial intent.
The difficulty for an agent is that this workflow relies on continuous, high-frequency hand–eye feedback: humans can "steer" a structure through many micro-adjustments, while an LLM operates in coarse, low-frequency steps.
Bridging that gap requires a representation that turns spatial intent into discrete, executable commands.

Atomic indices provide that bridge.
An index is an unambiguous handle that links a visually identified atom in a three-dimensional object to its underlying coordinates, allowing the agent to move from qualitative spatial understanding (\textit{closer}, \textit{aligned}, \textit{opened}) to concrete geometry.
From these indexed coordinates, the agent can construct geometric primitives---e.g., vectors from two reference points and planes from three reference points---where the reference points are often atom positions but can also be inferred points computed from coordinates (such as bond midpoints, centroids, or fitted ring planes).
These primitives then support higher-level relationships such as distances, angles, dihedral angles, and relative orientations.
Consequently, most \estructural{} tools take atomic indices as primary inputs, enabling precise, composable manipulation of molecular geometry without relying on continuous manual dragging.
Crucially, these indices must be derived from trusted sources (e.g., a three-dimensional molecular viewer or analysis tools) to ensure that the geometry manipulations are stable and unambiguous.

With atomic indices as stable anchors, \estructural{} can translate natural-language spatial intent into deterministic, stepwise geometry operations that are repeatable, verifiable, and scalable across complex structures.

\subsection{Tool design and interface}
\begin{figure}[htbp]
\begin{center}
\includegraphics[width=1.0\textwidth]{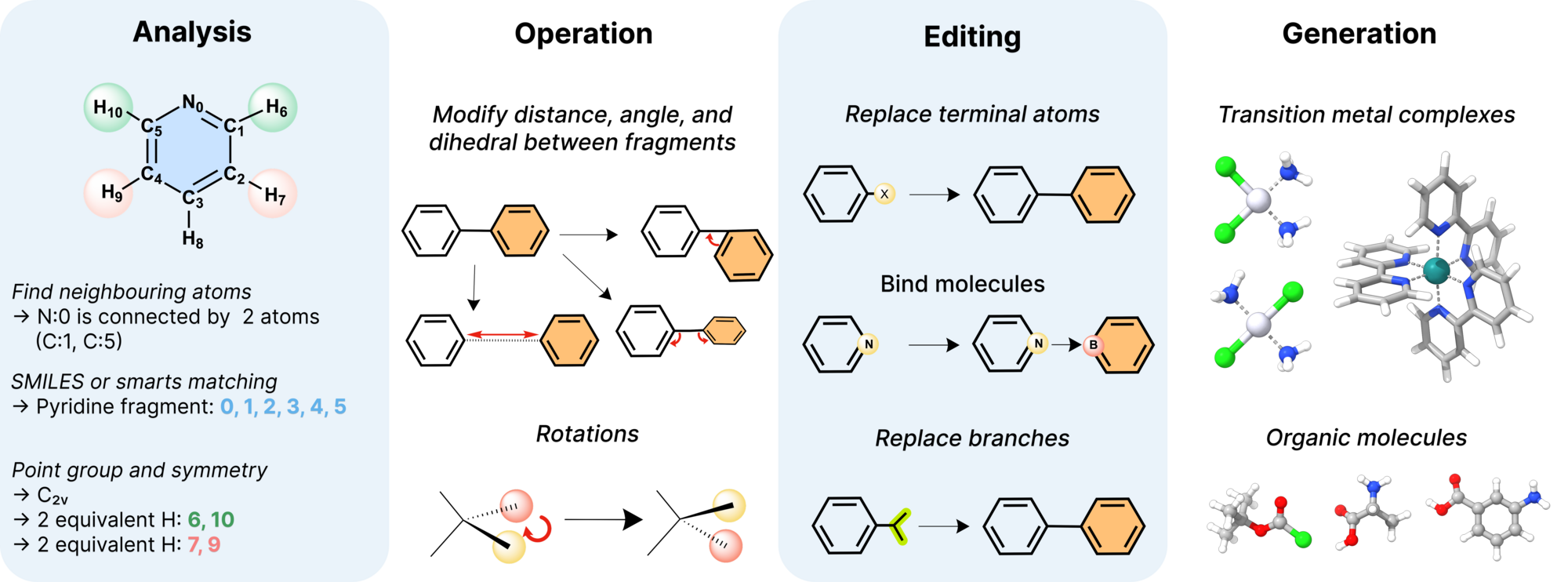}
\caption{Overview of the tools integrated into \estructural{}. The tools are categorized into structural analysis, geometric operation, structure editing, and structure generation, collectively enabling flexible construction, modification, and analysis of molecular geometries.
Atom colour codes: seashell, Pt; green, Ru; lime green, Cl; blue, N; red, O; grey, C; white, H.}
\label{fig:tools}
\end{center}
\end{figure}

To enable complex molecular manipulations, we introduce a set of specialized tools designed with domain-specific chemical knowledge of how computational chemists construct and manipulate three-dimensional molecular structures in practice (\autoref{fig:tools}). These tools are categorized into four groups based on their functionality: structural analysis, geometric operations, editing, and structure generation, formalizing common viewer-based actions, such as functional group substitution, fragment binding, and direct control of interatomic distances, angles, and dihedral angles, into programmable tools that can be invoked and composed by an LLM agent. 
The proposed tools are built upon widely adopted open-source Python libraries for molecular and materials modelling, including Atomic Simulation Environment (ASE)~\cite{ase-paper}, RDKit~\cite{rdkit}, Open Babel~\cite{obabel}, spglib~\cite{spglib}, and pymatgen~\cite{pymatgen}, which provide reliable representations and operations for atomic structures and chemical graphs.
Additionally, the geometry optimization tool uses the Fortran-based \texttt{xtb} package~\cite{bannwarthGFN2xTBAccurateBroadly2019,bannwarthExtendedTightbindingQuantum2021,spicherRobustAtomisticModeling2020,grimmeRobustAccurateTightBinding2017}.
Beyond rapid structural optimization, \texttt{xtb} is primarily employed to perform constrained geometry optimization, in which selected interatomic distances, angles, and dihedral angles are fixed to certain values. This enables more physically motivated geometric manipulations, which are essential for treating cases such as targeted reaction intermediates and transition-state-like initial structures (see Section~\ref{sec:geom_oper_tools} for details). All agents, including the LLM model and tools used in \estructural{}, are summarized in \autoref{tab:agents_tools}.

In addition to the chemistry-specific tools, \texttt{geometry\_operator} has access to \texttt{bash\_kernel}, a general shell-execution tool used to inspect and organize session files, execute Python scripts, and invoke locally installed computational programs when a dedicated wrapper is unavailable. Commands execute inside the session working directory with a 60-second timeout, and oversized command output is truncated before being returned to the model. The tool itself contributes no chemical heuristics; selection of commands, construction of input files, and interpretation of their outputs remain part of the agent's reasoning process.

\begingroup
\small
\begin{longtable}{@{}
  p{0.28\linewidth}
  p{0.67\linewidth}@{}}
\caption{Summary of agents and tools.} 
\label{tab:agents_tools} \\
\toprule
\textbf{Label / LLM model} & \textbf{Description} \\
\midrule
\endfirsthead

\multicolumn{2}{c}{{\bfseries \tablename\ \thetable{} -- continued from previous page}} \\
\toprule
\textbf{Label} & \textbf{Description} \\
\midrule
\endhead
\midrule
\endfoot
\bottomrule
\endlastfoot
\texttt{geometry\_operator} & 
Responsible for high-level structure manipulation tasks, including planning and executing actions such as geometry generation, editing, analysis, and other operations. GPT-5.4 was used as the default model unless otherwise specified. \\

\texttt{bash\_kernel} & Executes shell commands within the session working directory, enabling file operations and execution of Python scripts for molecular structure analysis and manipulation, as well as invocation of locally installed computational programs. Commands have a 60-second timeout, and oversized output is truncated before being returned to the agent. \\

\texttt{view\_png\_file} & View a PNG image file provided in the session working directory. \\

\texttt{view\_molecule} & Render the selected xyz file in 3D with adjustable view point. \\

\texttt{zoom\_molecule} & Render the selected xyz file in 3D with adjustable viewpoint and zoom level on a selected atom. \\

\texttt{find\_available\_fragments} & Search available predefined fragments. \\
\texttt{replace\_terminal\_atoms\_with\_fragment} & Connect two \texttt{xyz} structures at specified atomic indices. \\
\texttt{replace\_terminal\_atom\_with\_predefined\_FG} & Replace atomic indices in \texttt{xyz} file with a certain predefined fragment. \\
\texttt{bind\_two\_molecules} & Attach a certain \texttt{xyz} structure to specified atomic indices. \\
\texttt{bind\_predefined\_fragment} & Bind a predefined fragment to the provided atomic indices in the \texttt{xyz} file. \\
\texttt{bind\_atom\_ligands\_library} & Bind a predefined fragment to the provided element in the \texttt{xyz} file. \\
\texttt{replace\_branch\_with\_predefined\_FG} & Replace molecular branches connected to a specified atomic index with another branch from predefined fragments. \\
\texttt{replace\_branch} & Replace a molecular branch in one \texttt{xyz} with another \texttt{xyz} file. \\

\texttt{build\_tm\_complex\_from\_template} & Build transition metal complex from predefined templates. \\
\texttt{build\_predefined\_organometallic} & Build predefined organometallic complexes. \\
\texttt{organic\_molecules\_xyz\_generation} & Generates \texttt{xyz} files for organic molecules from a list of SMILES strings using RDKit for 3D structure generation. \\

\texttt{move\_atoms} & Move selected atomic indices by a specified vector. \\
\texttt{get\_distance\_angle\_dihedral} & Get distance, angle, and dihedral defined by selected atomic indices. \\
\texttt{set\_distance\_between\_fragments} & Set distance between two selected atomic indices (or involved fragments). \\
\texttt{set\_angle\_between\_fragments} & Set angle defined by three selected atomic indices (or involved fragments). \\
\texttt{set\_dihedral\_between\_fragments} & Set dihedral angle defined by four selected atomic indices (or involved fragments). \\
\texttt{remove\_atoms} & Remove selected atomic indices from the \texttt{xyz} file. \\
\texttt{rotate\_substituents\_around\_bisector} & Rotate a group of atoms around the connected atom bisector. \\
\texttt{insert\_atom\_at\_centroid} & Insert an atom at the centroid of selected atomic indices. \\

\texttt{get\_atomic\_neighbors} & Find neighbouring atoms for a given element symbol or atomic index. \\
\texttt{match\_smarts\_in\_xyz} & Match SMARTS patterns in the \texttt{xyz} file and return the corresponding atomic indices (without H atoms). \\
\texttt{get\_connected\_subgraph\_indices} & Get all atomic indices connected to a specified atomic index. \\
\texttt{find\_pointgroup\_equivalent\_atoms} & Find equivalent atoms in the molecule based on point group symmetry. \\
\texttt{constraint\_xtb} & Conducts a geometry optimization using xTB 6.5.1 with GFN\textit{n}-xTB or GFN-FF. Optionally, force constants on specific angles, torsions, and distances can be set. \\ 
\texttt{lookup\_smiles\_from\_name} & Lookup SMILES string from common chemical name using PubChem database. \\

\end{longtable}

\endgroup

\subsubsection{Structural analysis tools}\label{SI_Note:analysis}
\begin{figure}[htbp]
        \centering        \includegraphics[width=1.0\textwidth]{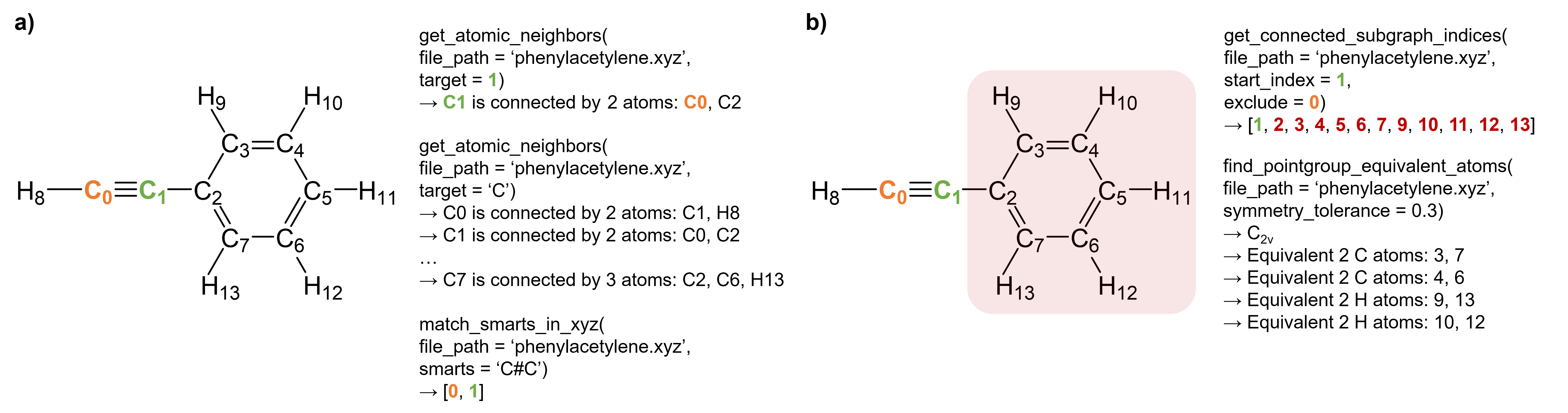}
        \caption{Overview of the analysis tools, their input arguments, and the returned results. For each example, the input molecular structure and specified arguments are shown together with the corresponding outputs produced by the analysis tools. The illustrated cases include (a) \texttt{get\_atomic\_neighbors}, \texttt{match\_smarts\_in\_xyz}, (b) \texttt{get\_connected\_subgraph\_indices} and \texttt{find\_pointgroup\_equivalent\_atoms}.
        }
        \label{fig:SI_analys}
\end{figure}

Structural analysis tools are designed to identify atomic indices and structural features within a target molecule that can subsequently be used for structure editing or geometric operations. These tools extract local and global structural information from \texttt{xyz} representations, enabling informed and site-selective manipulation of molecular geometries. The arguments of the structural analysis tools, along with representative results, are provided in \autoref{fig:SI_analys}.

\texttt{view\_molecule} and \texttt{zoom\_molecule} tools allow the agent to visually observe the molecule in 3D. The 3D view is rendered by PyMOL~\cite{PyMOL} with each atom labelled by its atomic index and element symbol. This allows the agent to flexibly review the molecular system and spatially identify points of interest or visually verify the resulting geometry after structure editing and geometry operations.

\texttt{bash\_kernel} tool allows the agent to execute shell commands, including routine file inspection and management (e.g., \texttt{sed}, \texttt{cp}, and \texttt{ls}), as well as custom Python scripts for molecular structure analysis and manipulation. These Python scripts enable flexible analysis and direct manipulation of molecular geometries.
In the workflow figures, Python-based structural operations are labelled as \texttt{bash\_kernel (Python)} to distinguish them from routine shell operations. As the figures summarize the principal structural analysis and manipulation steps rather than every tool invocation, routine file inspection and management operations are generally omitted for clarity.

The agent is instructed to preferentially use dedicated molecular tools when an appropriate tool is available. These tools provide predefined geometric operations and built-in checks for connectivity and steric clashes, whereas direct Python-based coordinate manipulation requires such safeguards to be implemented explicitly. Python-based operations are therefore used mainly as a fallback for operations not covered by the dedicated toolset.

\texttt{get\_distance\_angle\_dihedral} tool is used to calculate interatomic distances, angles, and dihedral angles.
Its interface follows the same conventions as the geometric operation tools (\autoref{fig:SI_operation}), allowing measurements to be defined between single atoms or between molecular fragments via centroid dummy atoms.

\texttt{get\_atomic\_neighbors} tool takes an \texttt{xyz} file and a \texttt{target} argument, which can be specified either as an atomic index (integer) or as an element symbol.  
If the target is provided as an atomic index, the tool returns a list of atoms directly bonded to the target atom, together with their element types and atomic indices.  
If the target is specified as an element symbol, the tool identifies all atoms of the given element in the structure and returns, for each such atom, its neighbouring atoms along with their element types and indices.

\texttt{match\_smarts\_in\_xyz} tool first converts the input \texttt{xyz} file into a MOL object, from which molecular connectivity is inferred. Based on this connectivity, the tool identifies atomic indices that match the input SMILES or SMARTS pattern.  
Hydrogen atoms are excluded from the matching results.

\texttt{get\_connected\_subgraph\_indices} tool operates using a pair of atomic indices: a starting index and an excluded index. It returns all atomic indices belonging to the branch originating from the starting index, excluding the direction toward the excluded index. Therefore, the \texttt{start\_index} and the \texttt{exclude} atom must be directly bonded to each other. The output is a list of atomic indices belonging to the identified subgraph. This functionality is used internally by both the branch-replacement and geometric-operation tools and enables the identification of complete fragment indices using only a single bond pair.

\texttt{find\_pointgroup\_equivalent\_atoms} is a wrapper around the point-group analysis implemented in \texttt{pymatgen} and takes an \texttt{xyz} file and a symmetry tolerance as input. In this work, the default symmetry tolerance is set to 0.3~\AA, the same default value used by \texttt{pymatgen}.  
Using this tolerance, the tool determines the molecular point group and returns lists of symmetrically equivalent atoms with their corresponding atomic indices.

Point-group detection is sensitive to both the numerical tolerance and the spatial orientation of atoms in the molecular geometry. To improve robustness, the tool internally performs additional point-group analyses using a looser tolerance (0.5~\AA, increased by 0.2~\AA~from the default) and by repeating the analysis after removing hydrogen atoms. If the point group detected using the default tolerance differs from that obtained using the looser tolerance or from the hydrogen-removed structure, the tool reports the discrepancy and recommends using a looser symmetry tolerance. 

    In addition, the symmetry tolerance is exposed as a tool argument, allowing the agent to actively adjust the tolerance and repeat the analysis when searching for a higher-symmetry interpretation of a distorted or near-symmetric structure.

\subsubsection{Geometric operation tools}\label{sec:geom_oper_tools}
\label{SI_Note:operation}
\begin{figure}[!h]
        \centering        \includegraphics[width=1.0\textwidth]{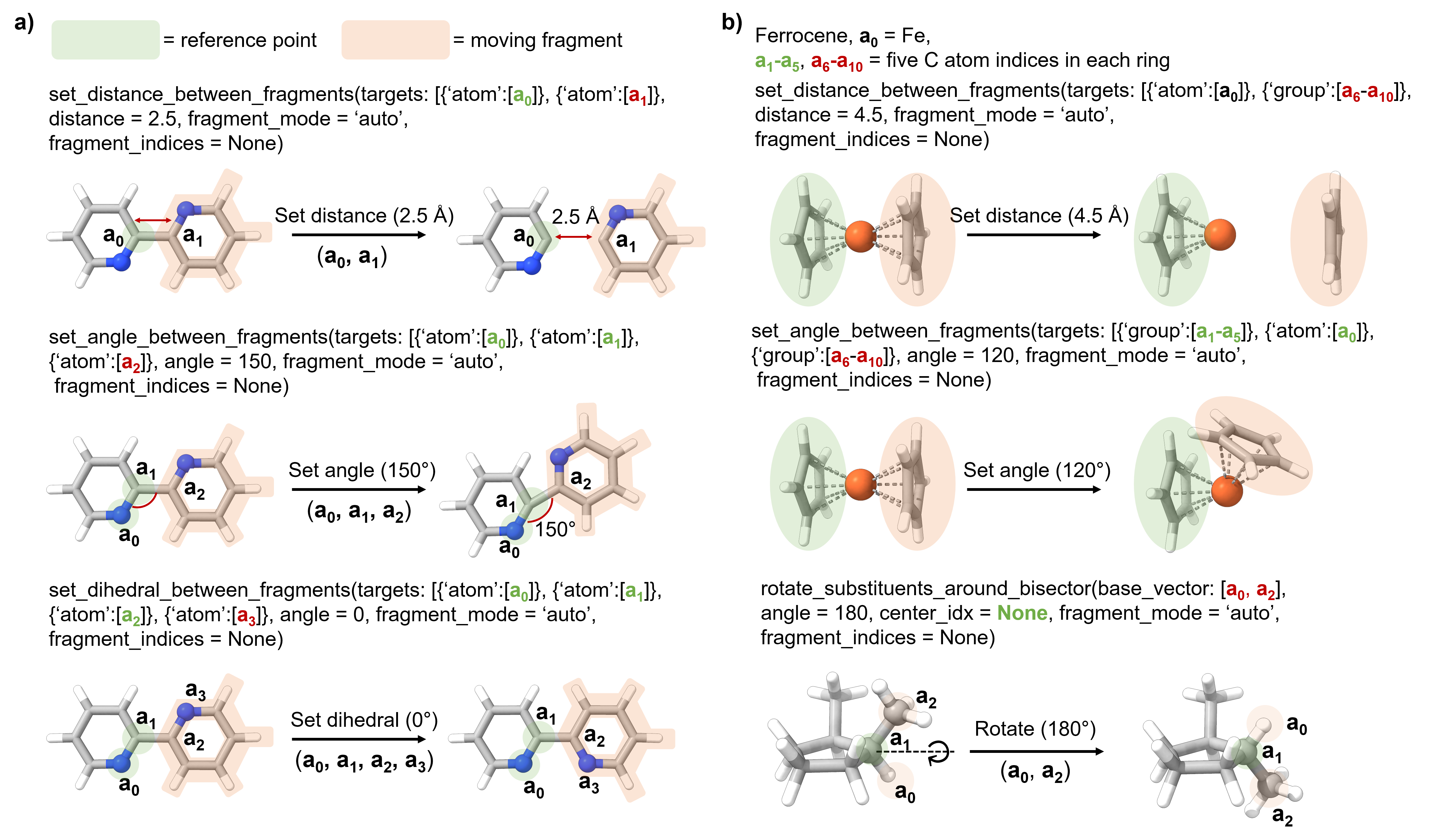}
        \caption{Overview of geometric operation tools in \estructural{}. (a) Atom-based operations, (b) group-based operations and rotation.         Each example shows the initial molecular structure, the applied geometric operation with its input parameters, and the resulting structure after the operation. The operations include distance, angle, dihedral control, and fragment rotation.}        \label{fig:SI_operation}
\end{figure}
Geometric operation tools are designed to directly manipulate molecular geometry by modifying interatomic distances, bond angles, dihedral angles, and fragment orientations through rotation.
These operations are applicable not only to individual atoms but also to connected molecular subgroups, enabling coordinated motion of entire fragments rather than isolated atomic displacements.

To modify the interatomic distance between a bonded atom pair $(i, j)$, the tool first identifies all atomic indices belonging to the branch extending from atom $j$, excluding the direction toward atom $i$, following the same branch-identification strategy used in the branch replacement tool. All atoms in the identified branch are then translated collectively to achieve the target distance between atoms $i$ and $j$. This procedure is similarly applied to bond-angle and dihedral-angle modifications, as well as to fragment rotations.


The rotation tool is conceptually related to dihedral angle manipulation but is particularly useful for generating alternative isomeric configurations. It takes a list of atomic indices defining a base vector, identifies the mutually connected atom shared by these indices, and computes unit vectors from each index to this shared atom. The fragment is then rotated by a specified angle about the resultant vector obtained from the vector sum of the two vectors. This operation enables efficient fragment swapping, facilitating conversion between distinct isomeric forms. Further details on the operational mechanisms and tool arguments are provided in the following sections and in \autoref{fig:SI_operation}.

\textbf{Target specification and dummy atom construction}

Geometric operations are specified using a \texttt{targets} dictionary.
For atom-level operations, a target is defined by a single atomic index.
For group-level operations, the target is defined as a list of atomic indices corresponding to a molecular fragment or functional group. In this case, a dummy atom is introduced at the centroid of the specified indices, and this centroid is used as the effective reference point for geometric control.

Distances, angles, and dihedral angles are then evaluated and manipulated using these atomic or centroid-based reference points, allowing consistent treatment of both individual atoms and extended groups within a unified framework. This design allows the tools to be naturally extended to geometric operations involving metal centres and polydentate ligands.

\textbf{Fragment selection modes}

To determine which atoms are displaced during a geometric operation, geometric operation tools support three fragment selection modes via the \texttt{fragment\_mode} parameter:

\texttt{fragment\_mode = `auto'}
The moving fragment is automatically identified based on molecular connectivity.
For example, when modifying the distance between atoms $a_0$ and $a_1$, the algorithm identifies all atoms belonging to the branch extending from $a_1$ in the direction opposite to $a_0$. This is achieved using a connectivity-based subgroup detection procedure (via \texttt{get\_connected\_subgraph\_indices}). The entire detected fragment is then displaced collectively.

\texttt{fragment\_mode = `manual'}
The user explicitly specifies the indices of atoms to be moved via \texttt{fragment\_indices}. This mode provides full control when automatic fragment detection is ambiguous or when custom selections are required.

\texttt{fragment\_mode = `none'}
Only the explicitly specified atom or group indices are displaced, without any automatic expansion to connected atoms.

\textbf{Rotation operations}

Rotation operations are defined by specifying a set of \texttt{base\_vector\_indices}, which determine the rotation axis and reference direction. Let these indices correspond to atoms $a_1, a_2, \ldots, a_n$. The rotation centre, denoted as $a_0$, is identified as the atom that is commonly bonded to all atoms in \texttt{base\_vector\_indices}. If not explicitly provided via \texttt{center\_idx}, this atom is automatically inferred from molecular connectivity.

For each base atom $a_i$, a unit vector $\hat{v}_i = (a_i - a_0)/|a_i - a_0|$ is constructed. These unit vectors are summed and normalized to define the effective rotation axis. Atoms in the selected fragment are then rotated by the target angle around this axis, with the same \texttt{fragment\_mode} and \texttt{fragment\_indices} logic applied as in distance and angle operations.
These operation tools generalize common manual manipulations performed in molecular viewers into programmatic actions that can be invoked by the agent.

\textbf{Constrained geometry optimization}

The constrained geometry optimization tool (\texttt{constraint\_xtb}) employs the semiempirical quantum-mechanical GFN\textit{n}-xTB methods~\cite{bannwarthGFN2xTBAccurateBroadly2019,bannwarthExtendedTightbindingQuantum2021,grimmeRobustAccurateTightBinding2017}, which are designed for accurate descriptions of geometries, vibrational frequencies, and non-covalent interactions, as well as the related GFN force field from the same family of methods~\cite{spicherRobustAtomisticModeling2020}.
In addition to unconstrained energy minimization with respect to nuclear coordinates for the relaxation of generated structures, harmonic-constraining potentials can be applied to selected distances, bond angles, and dihedral angles.
These additional potentials enable optimizations toward TS-like geometries or the enforcement of specific binding motifs, such as the attachment of ligands to the metal centre or the stabilization of metal--organic complexes in predefined configurations.
Each constraining potential is defined by the atomic indices it acts on (two for
distances, three for angles, and four for dihedral angles) and a force constant expressed in atomic units (\si{\hartree\per\bohr\squared}), corresponding to a harmonic potential in Cartesian space.
Further details on the technical background are provided in the official \texttt{xtb} documentation (see \href{https://xtb-docs.readthedocs.io/en/latest/xcontrol.html}{xtb-docs.readthedocs.io/en/latest/xcontrol.html}).
From the perspective of the agent, constrained optimization is exposed as a lightweight Python tool that takes a molecular structure, the relevant atomic indices, and the force constant as input, and returns the correspondingly relaxed, constraint-consistent structure.

Unless otherwise specified, \texttt{constraint\_xtb} uses GFN2-xTB with four CPU
cores and invokes the standard xTB geometry optimizer through \texttt{--opt}; the
default xTB convergence thresholds and maximum number of optimization cycles are
therefore retained. When no constraints are specified, the tool performs an unconstrained geometry optimization to refine the input structure. No implicit-solvent model, finite-temperature correction, or
external field is applied. The tool supports total molecular charge and the number
of unpaired electrons through the xTB \texttt{--chrg} and \texttt{--uhf}
arguments, respectively. When these arguments are omitted, the xTB defaults are
used, corresponding to the neutral, zero-unpaired-electron case for the closed-shell
systems considered here. Non-neutral or open-shell settings are selected from the
intended molecular composition, formal charge, ligand charges, and chemically
expected electronic state rather than inferred from the \texttt{xyz} coordinates alone.
Case-specific charge and spin assignments for the reported coordination complexes
are stated in the corresponding benchmark subsections.

Constraints are limited to coordinates needed to retain the intended coordination
motif or approximate reaction coordinate while all other internal coordinates are
allowed to relax. Target values are selected from the starting geometry and the
chemically expected bond-forming, bond-breaking, or metal--ligand distances and are
subsequently checked in the optimized structure. The default working force constant
is 0.5 atomic units; it may be increased to 1.0 when the constrained coordinate is
not retained to within approximately 5\% of its target.


In addition to the operations described above, \estructural{} provides several auxiliary geometric manipulation tools, including \texttt{move\_atoms}, \texttt{remove\_atoms}, and \texttt{insert\_atom\_at\_centroid}. 
\texttt{move\_atoms} tool translates specified atomic indices by a user-defined Cartesian displacement vector $(x,y,z)$. 
\texttt{remove\_atoms} tool deletes selected atoms from the structure based on their indices. 
\texttt{insert\_atom\_at\_centroid} tool introduces a new atom of a specified element at the geometric centroid of a user-defined set of atomic indices.

\subsubsection{Structure editing tools}\label{SI_Note:editing}
\begin{figure}[!t]
\begin{center}
\includegraphics[width=1.0\textwidth]{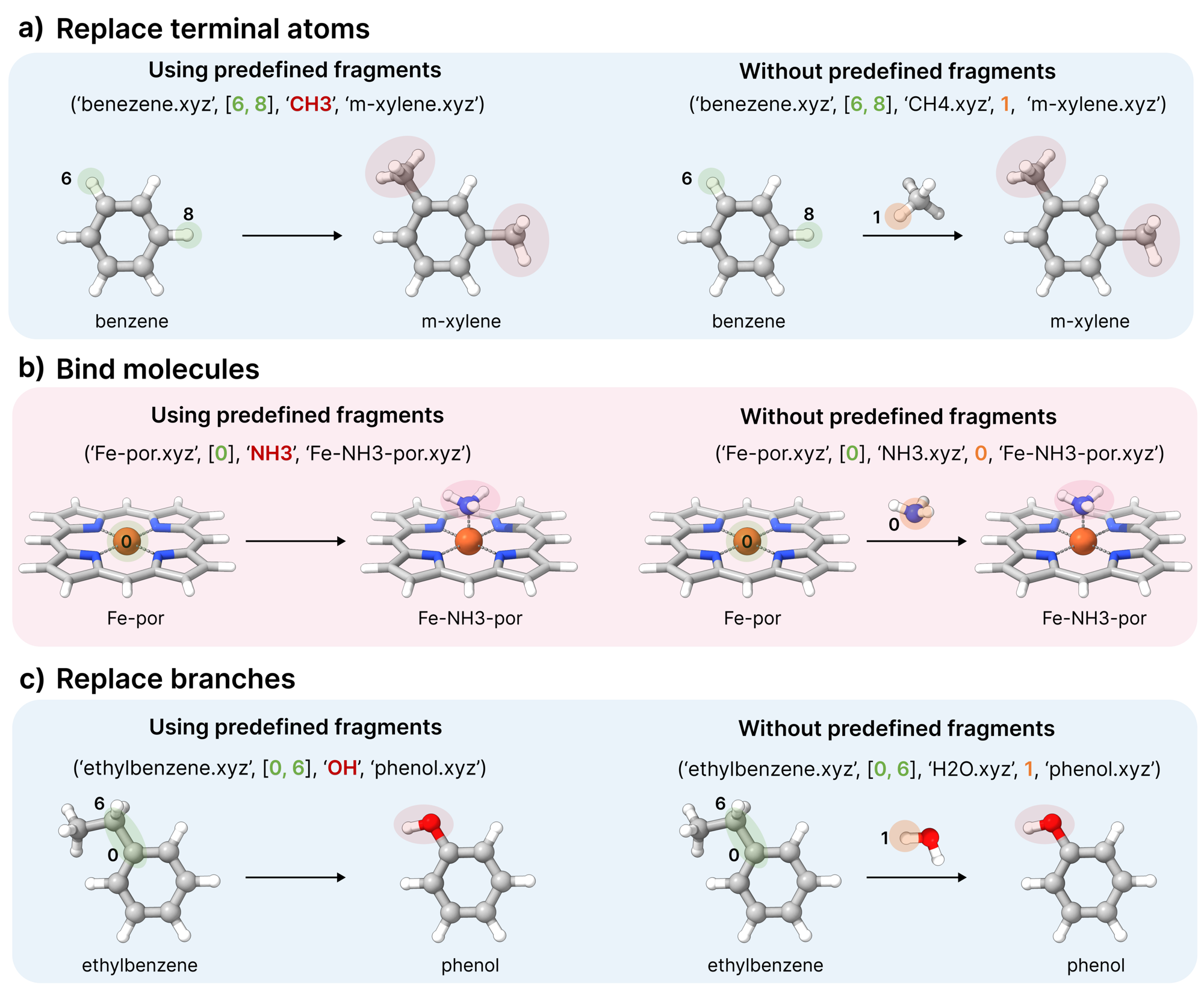}
\caption{Schematic illustration of structure editing tools, including (a) terminal atom replacement, (b) molecule binding, and (c) branch replacement. These tools enable functionalization, fragment attachment, and subgroup substitution by modifying selected atomic sites while preserving the rest of the molecular geometry. Representative arguments are shown in each example, including the input \texttt{xyz} structure filename, target atomic indices for replacement or binding, fragment specifications, and the name of the newly generated output structure. 
Atom colour codes: dark orange, Fe; red, O; blue, N; grey, C; white, H.}
\label{fig:editing_tools}
\end{center}
\end{figure} 

The structure editing tools support three primary operations: replacement of terminal atoms, binding of molecular fragments to designated atomic sites, and substitution of molecular branches (subgroups) within an \texttt{xyz} structure (\autoref{fig:editing_tools}).
Across all editing operations, the Cartesian coordinates (or internal coordinates) of atoms outside the replaced region remain unchanged, thereby preserving the original conformation of the molecular core. For all structure editing tools, an additional sanitization step is applied to prevent unphysical atomic overlaps between the core structure and newly attached fragments. If steric clashes are detected at the interface, the attached fragment is iteratively rotated about the newly formed bond until overlaps are resolved, ensuring a physically reasonable initial geometry.

\begin{figure}[!t]
        \centering        \includegraphics[width=1.0\textwidth]{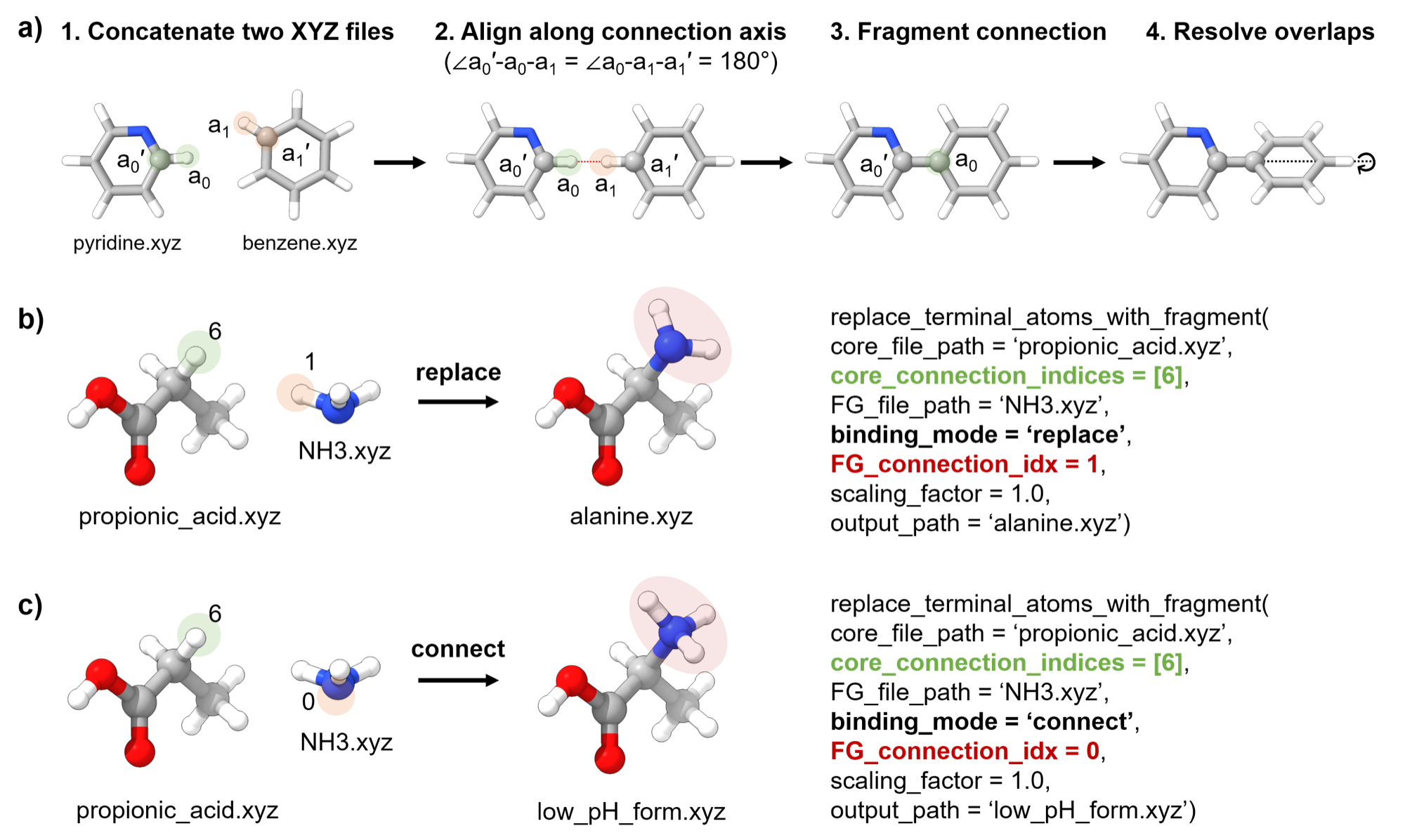}
        \caption{Detailed mechanism and arguments of replace\_terminal\_atoms\_with\_fragment tool. (a) Schematic workflow for the replacement tool. The initial molecular structure, the applied replacement with its input parameters, and the resulting structure after the operation for (b) binding\_mode = `replace', and (c) binding\_mode = `connect'.
        Here, $a_0$ and $a_1$ denote the terminal atoms selected as substitution sites in the two structures. Since both are terminal, each is bonded to a single neighbouring atom, denoted as $a_0'$ and $a_1'$, respectively.
        Atom colour codes: red, O; blue, N; grey, C; white, H.}
        \label{fig:SI_editing}
\end{figure}

The terminal atom replacement tool is conceptually analogous to functionalization operations in molecular visualization software, in which a functional group is selected and substituted at a chosen terminal atom.
This tool takes as input a core \texttt{xyz} file to be edited, the atomic indices to replace, and a substituent.
To streamline common use cases, we provide a predefined library of frequently used functional groups, allowing fragments to be specified by name alone. If the requested substituent is not available in the library, the agent first generates the corresponding fragment as a standalone \texttt{xyz} structure and then attaches it to the specified atomic site, with the connection atom explicitly defined. This design enables highly flexible molecular functionalization, ranging from commonly used substituents to custom-designed fragments. The detailed mechanism and full argument specifications of the terminal atom replacement tool are described in \autoref{fig:SI_editing}.

\begin{figure}[!t]
        \centering        \includegraphics[width=1.0\textwidth]{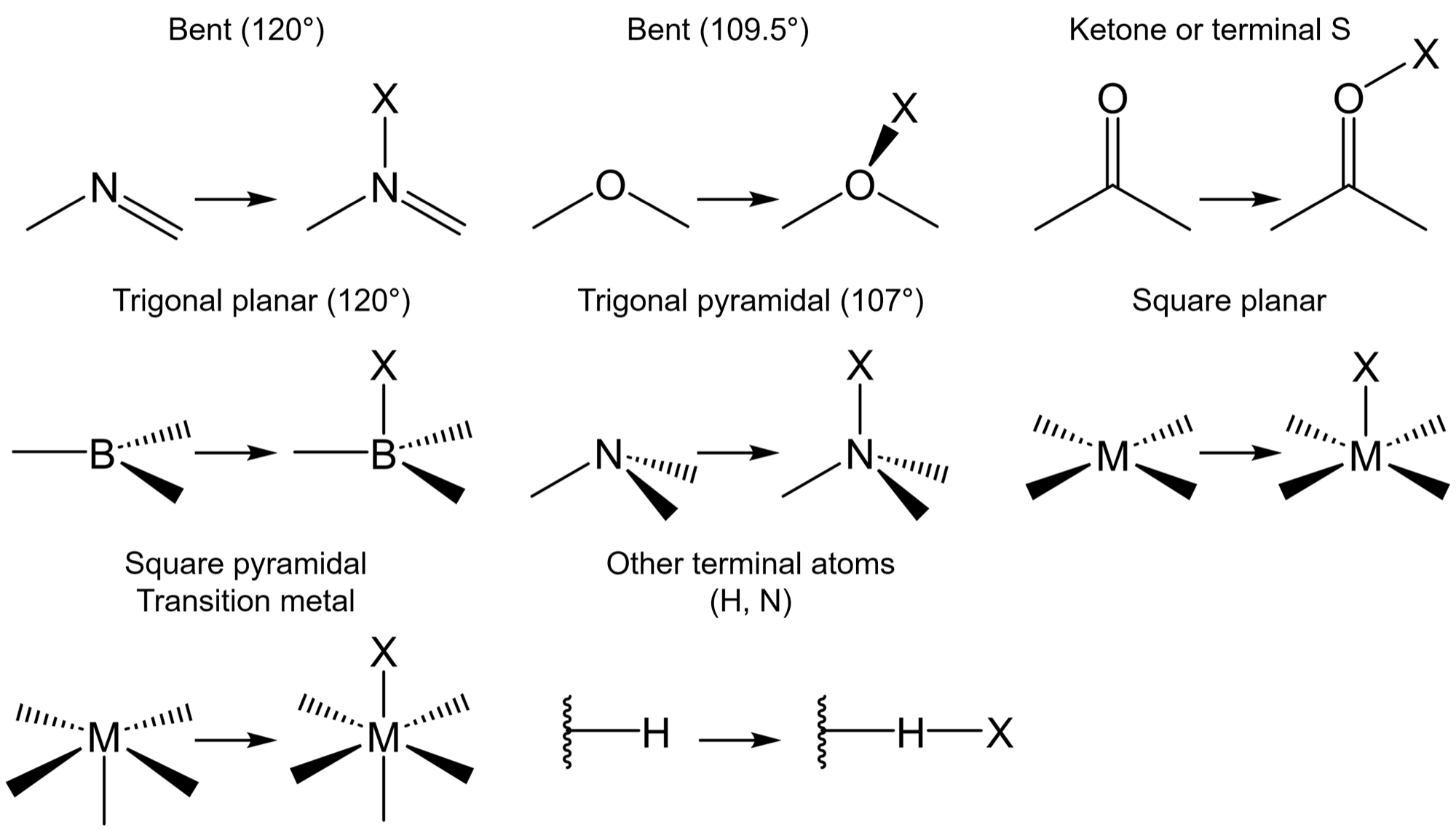}
        \caption{VSEPR-guided placement of new bonds in the binding tool. A dummy atom (X) is introduced to indicate the direction of the newly created bond, with its position determined according to VSEPR-based idealized geometries for different central atoms and coordination environments.
        }
        \label{fig:SI_dummy}
\end{figure}

\begin{figure}[!t]
        \centering        \includegraphics[width=1.0\textwidth]{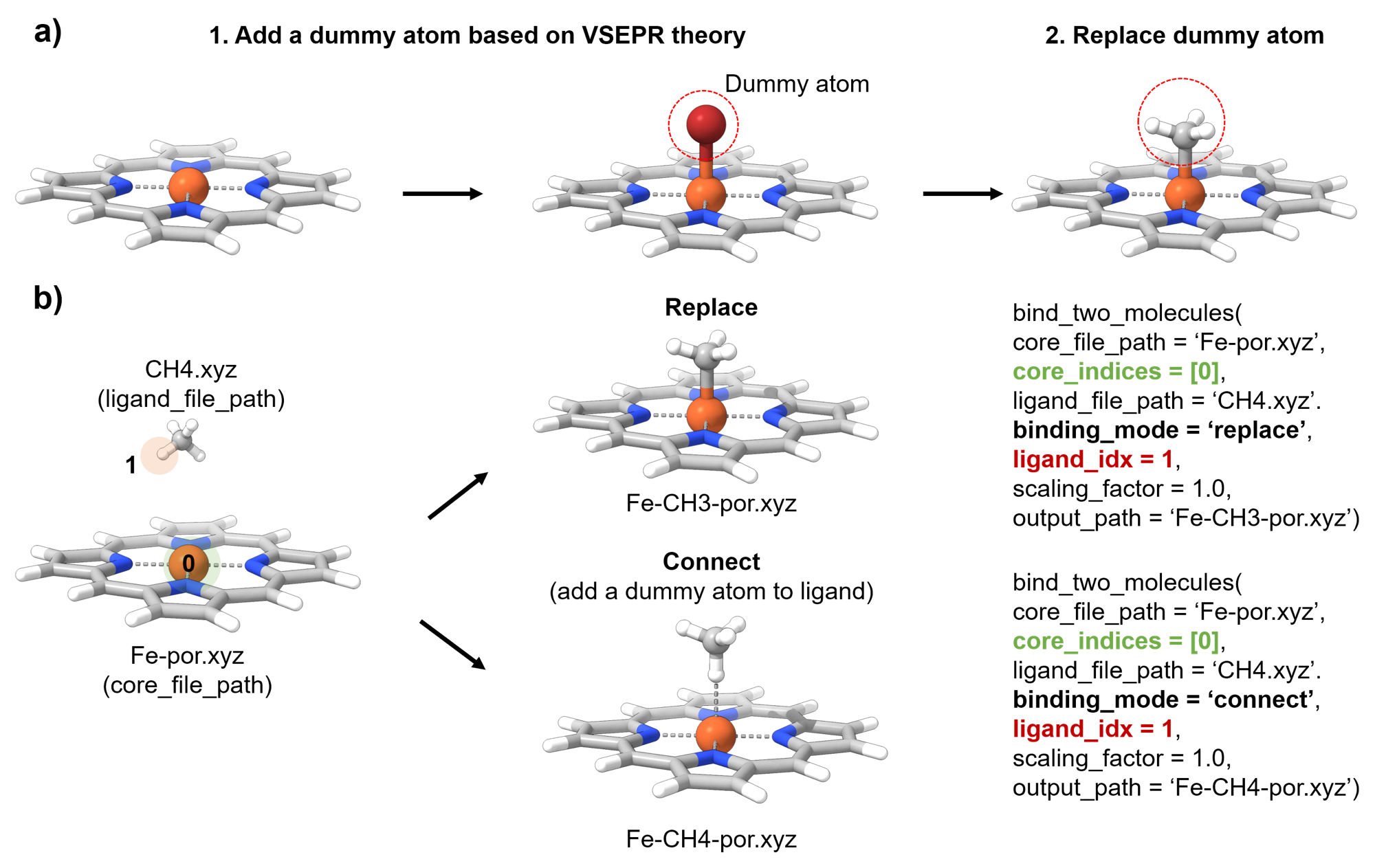}
        \caption{Detailed mechanism and arguments of bind\_two\_molecules tool. (a) Schematic workflow of the binding tool. The initial molecular structure, the applied replacement with its input parameters, and the resulting structure after the operation for (b) binding\_mode = `replace', (c)
        binding\_mode = `connect'.
        Atom colour codes: dark orange, Fe; blue, N; grey, C; white, H.}
        \label{fig:SI_bind_molecules}
\end{figure}

The molecule binding tool attaches an additional molecular fragment, provided as an \texttt{xyz} file, to specified atomic indices within a core molecular structure. Compared to terminal atom replacement, this operation is technically more challenging. While terminal atom replacement can be performed by preserving the internal coordinates of each fragment, fragment binding requires the explicit definition of new inter-fragment bonds with appropriate bond directions.
To address this challenge, the tool introduces a pair of dummy atoms, each attached to a designated connection atom in the core structure and the incoming fragment, respectively. These dummy atoms are positioned according to VSEPR-based geometrical considerations, thereby defining both the bonding direction and the initial orientation of the new bond. The binding operation is then carried out by applying the same replacement procedure used in the terminal atom replacement tool, with the dummy atoms serving as temporary connection points.
The molecule binding tool supports both predefined fragments from the functional group library and arbitrary \texttt{xyz} fragments. Detailed descriptions of the binding mechanism and the positioning strategy for dummy atoms are provided in \autoref{fig:SI_dummy} and \autoref{fig:SI_bind_molecules}.

The branch replacement tool substitutes a molecular subgroup (e.g., alkyl chains or ring systems) with an alternative fragment provided as an \texttt{xyz} file. The operation is defined using bond pairs specified by atomic index tuples $(i, j)$ in the core structure, where the branch remains connected to atom $i$ but originates from atom $j$.
Given the specified bond pair $(i, j)$, the tool first identifies all atomic indices belonging to the branch extending from atom $j$, excluding the direction toward atom $i$, using the \texttt{get\_connected\_subgraph\_indices} tool internally. These branch atoms are then removed from the core structure, with atom $j$ retained as the replacement point. Subsequently, analogous to the terminal atom replacement procedure, atom $j$ is replaced with the new branch fragment.
This tool is particularly useful for systematic subgroup substitution and ligand exchange reactions. In the context of ligand exchange, the operation can be performed efficiently by specifying the atomic index of the metal centre and the coordinating atom within the ligand, enabling rapid modification of coordination environments.

\textbf{Core--substituent attachment using terminal connection indices}

The input arguments include a core structure file (\texttt{core\_file\_path}) and a list of terminal connection indices,
\texttt{core\_connection\_indices} $= \{a_0\}$, which specifies the terminal atoms in the core to be replaced (typically H atoms).
For each specified terminal atom $a_0$, its directly bonded neighbour in the core is determined internally and denoted as $a_0'$.

A substituent fragment is provided separately. Based on the user prompt, the agent identifies
the terminal attachment atom in the substituent, which is denoted as $a_1$.
The unique bonded neighbour of $a_1$ within the substituent is then determined internally by the tool
and denoted as $a_1'$.

In the case of \texttt{binding\_mode = `connect'}, the substituent fragment is attached without removing
the fragment-side connection atom.
Instead, a dummy atom is introduced at the fragment connection site according to the local VSEPR
geometry (\autoref{fig:SI_dummy}), as discussed in detail in the following section.
The newly introduced dummy atom is treated as the terminal attachment atom $a_1$, while the originally
specified fragment connection index (\texttt{FG\_connection\_idx}) is treated as its bonded neighbour $a_1'$.

\textbf{Alignment to preserve the core bonding direction}

For each core connection site $a_0$, the substituent fragment is rigidly aligned such that the fragment is placed along the
original bond direction in the core. The alignment enforces linear geometries
\[
\angle a_0' - a_0 - a_1 = 180^\circ
\quad \text{and} \quad
\angle a_0 - a_1 - a_1' = 180^\circ,
\]
so that the substituent is oriented ``straight out'' from the core along the $a_0' - a_0$ direction.

\textbf{Translation to set the target bond length}

After angular alignment, the substituent fragment is translated along the attachment axis without additional rotation.
The translation distance is chosen such that
\[
d(a_0', a_1') = \left[ r(a_0') + r(a_1') \right] \times s,
\]
where $r$ is the covalent radius and $s$ is a \texttt{scaling\_factor} parameter (default: $s = 1.0$).

\textbf{Bond formation via terminal-atom deletion and index retention}

Finally, the terminal atoms $a_0$ (in the core) and $a_1$ (in the substituent) are removed, which results in the formation of a new covalent bond between $a_0'$ and $a_1'$ at the attachment site.
To enable robust handling of multiple substitution sites specified by the core connection index list, the atom index of $a_1'$ in the merged structure is reassigned to $a_0$ (i.e., the original terminal atom
index in the core).
With this convention, the connection site indices remain consistent throughout successive substitutions,
thereby preventing index-shift issues during multi-site structure modification.

\textbf{Post-attachment sanitization by rotational adjustment.}

For all structure editing tools, a geometry sanitization procedure is applied to eliminate unphysical atomic overlaps between the core structure and newly attached fragments. After fragment attachment, interatomic distances are evaluated between atoms in the core structure and atoms in the newly added fragment, excluding the pair of atoms forming the newly created bond. Atom pairs within the original core or within the attached  fragment are not considered in this procedure. Let $i$ denote an atom index in the core structure and $j$ an atom index in the attached fragment. A steric clash is defined when the distance $d(i,j)$ satisfies
\[
d(i,j) < 1.5 \times [r(i) + r(j)],
\]
where $r(i)$ and $r(j)$ are the covalent radii of atoms $i$ and $j$, respectively. 

For H--H pairs, a separate minimum-distance criterion of 2.0~\AA\ is applied; that is, an H--H pair is considered clashing when $d(i,j) < 2.0$~\AA.

If any such clashes are detected, the attached fragment is iteratively rotated about the newly formed bond in increments of 5$^\circ$. After each rotation, the interatomic distances are re-evaluated until all $d(i,j)$ exceed the threshold defined above. This procedure preserves the internal geometry of both the core structure and the fragment while ensuring a sterically reasonable initial configuration for subsequent calculations.

\textbf{Binding tool via dummy-atom introduction}

The binding tool introduces a dummy atom (X) at the intended attachment direction of a specified atom in the core structure and subsequently uses this dummy atom as the terminal site for fragment binding (\autoref{fig:SI_dummy}). The placement of the dummy atom is determined from the local coordination environment around the selected central atom.

To determine the appropriate dummy-atom position, the local coordination geometry around the central atom is first analyzed and compared against a set of predefined ideal VSEPR geometries. For each candidate geometry with the corresponding coordination number, the angles between neighbouring atoms are computed and compared with the ideal angular pattern.

A similarity score is assigned to each candidate geometry based on the deviation between the observed and ideal angles. Let $\{\theta_k\}$ denote the set of observed angles and $\{\theta_k^{\mathrm{ideal}}\}$ the corresponding ideal angles. For each candidate geometry, the score is computed as

\[
\mathrm{score} = \frac{1}{N} \sum_k \exp\left(-\frac{(\theta_k - \theta_k^{\mathrm{ideal}})^2}{2\sigma^2}\right),
\]

where $N$ is the number of angles considered and $\sigma$ is a smoothing parameter controlling the angular tolerance. In this work, $\sigma$ is set to 5.7$^\circ$.

The candidate geometry with the highest score is selected when the score exceeds a threshold of 0.6. The missing coordination position defined by the selected ideal geometry is then used to place a dummy atom at a fixed distance of 1.0~\AA{} from the central atom. In this way, the dummy atom indicates the direction of the newly created bond.

If no candidate VSEPR geometry satisfies the scoring threshold, a fallback procedure is used. A grid-based angular search is performed around the central atom to identify sterically accessible directions with minimal overlap with the existing structure. The dummy atom is then placed in the least sterically hindered direction, again at a fixed distance of 1.0~\AA{} from the central atom. This fallback enables dummy-atom placement for irregular or highly distorted local coordination environments that do not closely match the predefined ideal geometries.

Once the dummy atom has been introduced, it is treated as a terminal attachment site, and the subsequent binding procedure follows the same workflow as the \texttt{replace\_terminal\_atoms\_with\_fragment} tool (\autoref{fig:SI_bind_molecules}). Specifically, the same procedures are used for fragment alignment, bond-length setting, selection of the binding mode (\texttt{replace} or \texttt{connect}), and use of predefined fragments. The binding tool can therefore be viewed as a two-step process: (i) determination and introduction of a terminal attachment site through local coordination analysis and dummy-atom placement, followed by (ii) standard fragment replacement or connection.

\textbf{Predefined fragments}

\begin{figure}[!t]
        \centering        \includegraphics[width=0.90\textwidth]{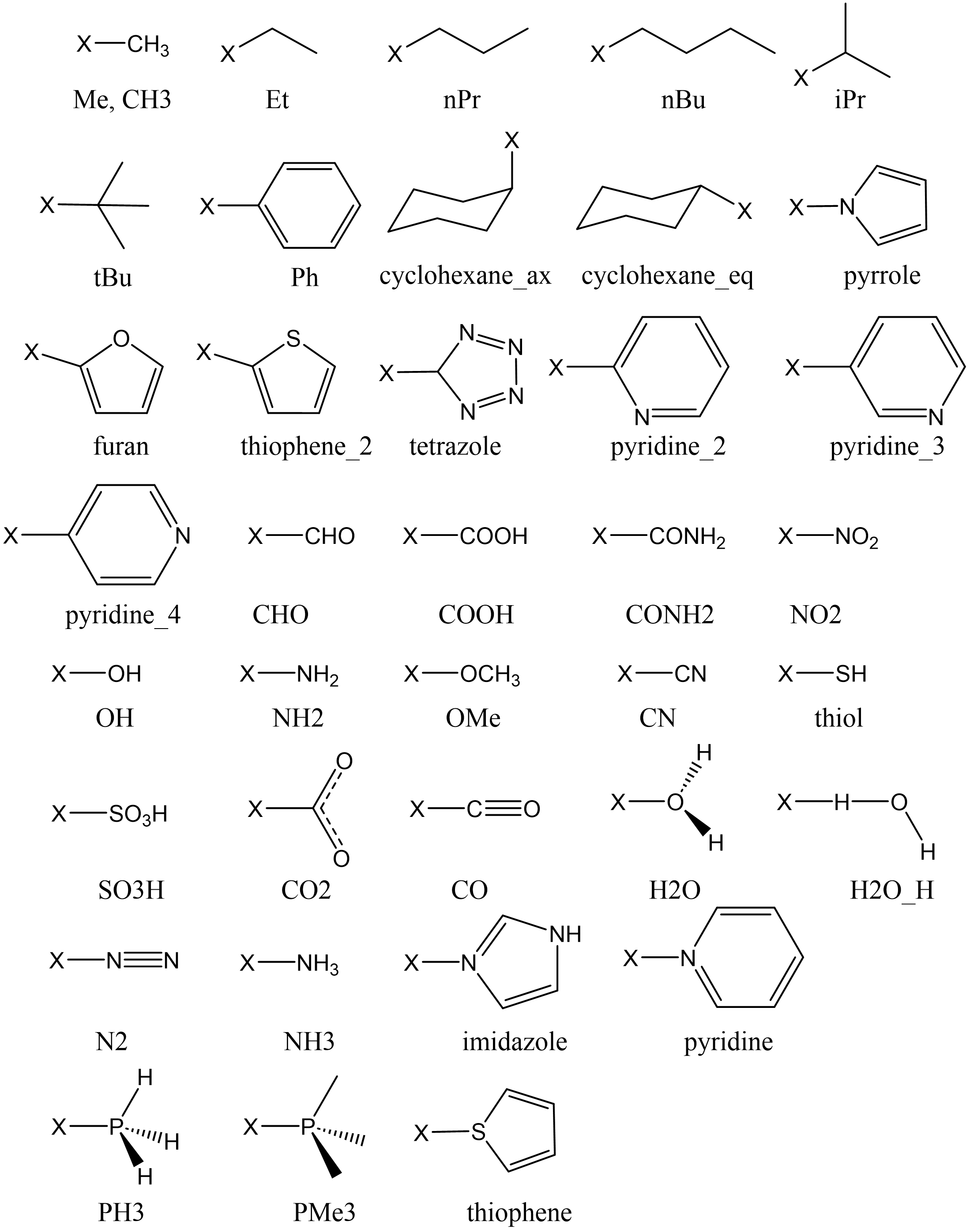}
        \caption{Predefined fragment library used for replacement and binding operations. Labels shown below each structure correspond to the fragment names as stored in the library. X denotes the connection site.}
        \label{fig:SI_predefined}
\end{figure}

For commonly used substituents, we provide a predefined fragment library
(\autoref{fig:SI_predefined}).
For fragments in this library, the terminal attachment atom and its bonded neighbour are predefined,
with $a_1$ and $a_1'$ fixed to indices 0 and 1, respectively.
As a result, when a predefined fragment is used, it is not necessary to generate a separate
substituent-specific \texttt{xyz} file.
Instead, the agent can simply specify the fragment name.

In this case, both \texttt{FG\_connection\_idx} and \texttt{binding\_mode} are implicitly defined by the
fragment template and therefore do not need to be provided.

\clearpage
\subsubsection{Structure generation tools}\label{SI_Note:generation}
Structure generation tools are designed to construct \texttt{xyz} representations of both organic and organometallic molecules. For generating organic molecules, we implement a conversion pipeline that converts SMILES strings to \texttt{xyz} coordinates using RDKit~\cite{rdkit} and Open Babel~\cite{obabel}. Leveraging the pretrained chemical knowledge of the LLM, common names and IUPAC names can be automatically converted into SMILES representations, which are then passed to the generation tool.
To improve robustness in cases where the LLM fails to generate a correct SMILES string, \estructural{} also supports name-based structure retrieval from PubChem database~\cite{kim2025pubchem} via the \texttt{pubchempy}. In such cases, common or systematic names are resolved to canonical SMILES directly from PubChem, providing a reliable fallback mechanism for generating organic molecules.

\begin{figure}[htbp]
\begin{center}
\includegraphics[width=1.0\textwidth]{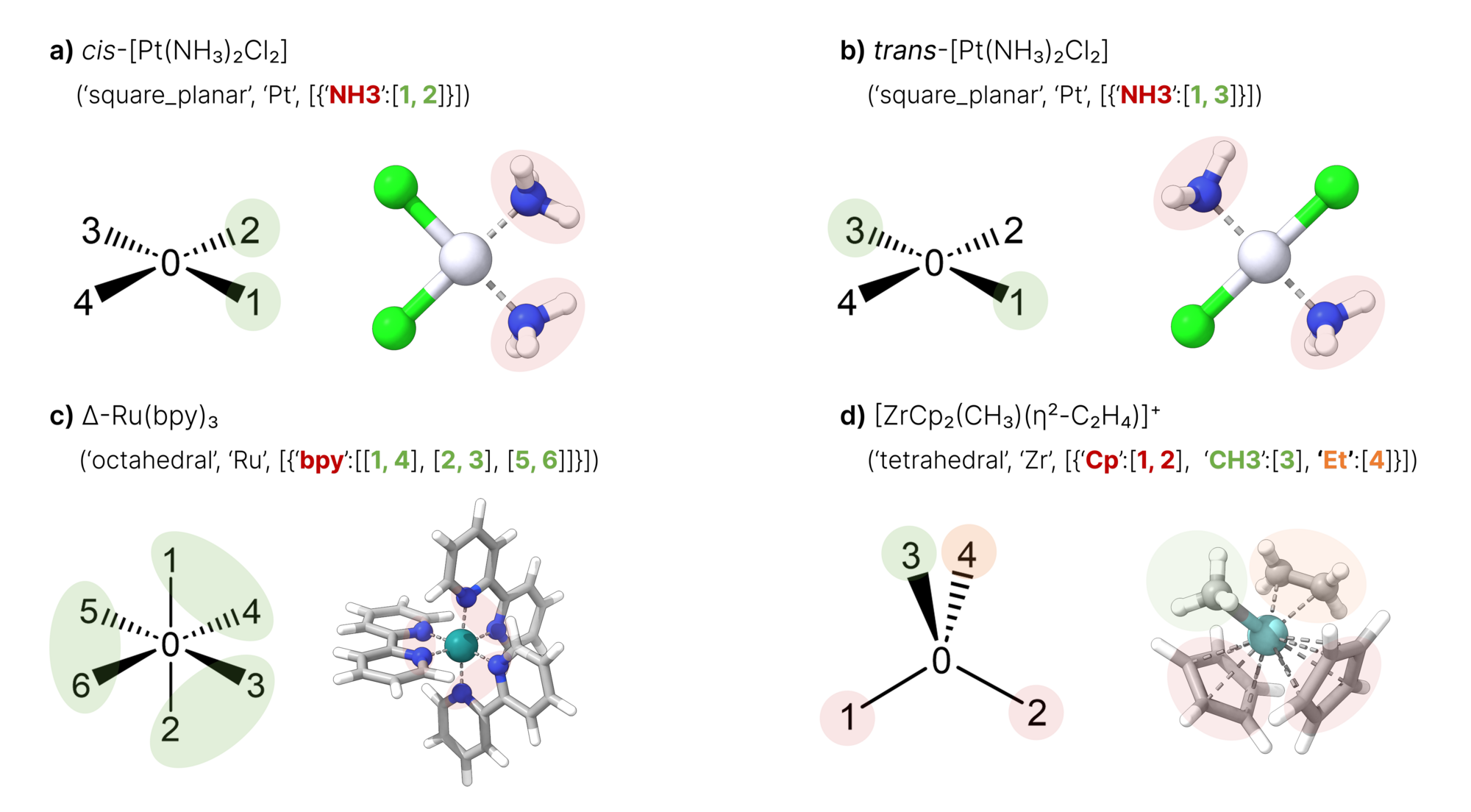}
\caption{Overview of structure generation tools for organometallic molecules. Organometallic structures are constructed using predefined coordination–symmetry templates, in which ligands are assigned to annotated atomic indices to generate stereochemically specific geometries. The representative input arguments and the corresponding generated structures for (a) square planar \textit{cis}-[Pt(NH$_3$)$_2$Cl$_2$] and (b) \textit{trans}-[Pt(NH$_3$)$_2$Cl$_2$], (c) $\Delta$-Ru(bpy)$_3$ (bpy = 2,2$'$-bipyridine), and (d) tetrahedral 
[ZrCp$_2$(CH$_3$)($\eta^2$-C$_2$H$_4$)]$^+$
(Cp = $\eta^5$-cyclopentadienyl). 
Atom colour codes: seashell, Pt; green, Ru; sky blue, Zr; lime green, Cl; blue, N; grey, C; white, H.}
\label{fig:inorg_builder}
\end{center}
\end{figure} 

For organometallic structures, we introduce a builder tool based on predefined coordination symmetry templates (\autoref{fig:inorg_builder}).
This tool requires specification of the central metal, coordination geometry, ligand identities, and their binding positions. Coordination geometries corresponding to coordination numbers from 2 to 6 are predefined.
In each coordination template, the central metal atom is assigned a fixed atomic index of 0, while surrounding atoms are initialized as chlorine atoms by default.
Each atomic index in the template is annotated with coordination-context information, such as \textit{cis} relationships in square-planar geometries or axial positions in octahedral geometries. After selecting the desired coordination template and replacing the central metal with the target element, ligands are introduced by substituting the corresponding atomic indices with ligand fragments. This procedure enables the generation of stereochemically well-defined organometallic structures. 

For more complex organometallic systems, such as those with high coordination numbers ($>6$) or multiple polydentate ligands, direct construction from templates may be less practical. In such cases, we provide an alternative tool (\texttt{build\_predefined\_organometallic}) that loads predefined organometallic structures, which can subsequently be modified using the structure editing tools.

\begin{figure}[!t]
        \centering        \includegraphics[width=1.0\textwidth]{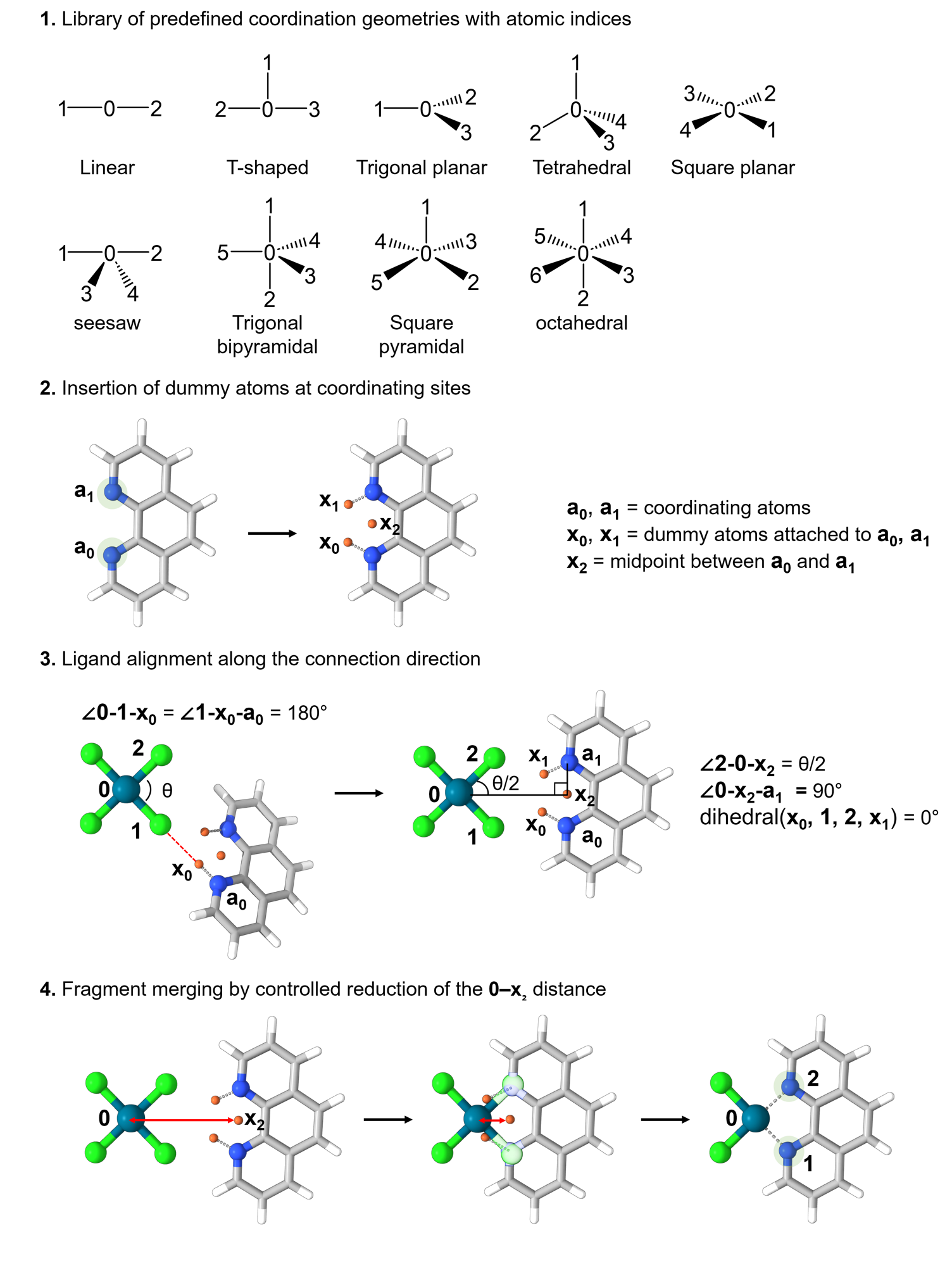}
        \caption{Detailed mechanism and arguments of the organometallic structures builder.
        Atom colour codes: green, Pd; lime green, Cl; blue, N; grey, C; white, H; orange, dummy atom.}
        \label{fig:SI_inorg}
\end{figure}

\begin{figure}[!t]
        \centering        \includegraphics[width=0.95\textwidth]{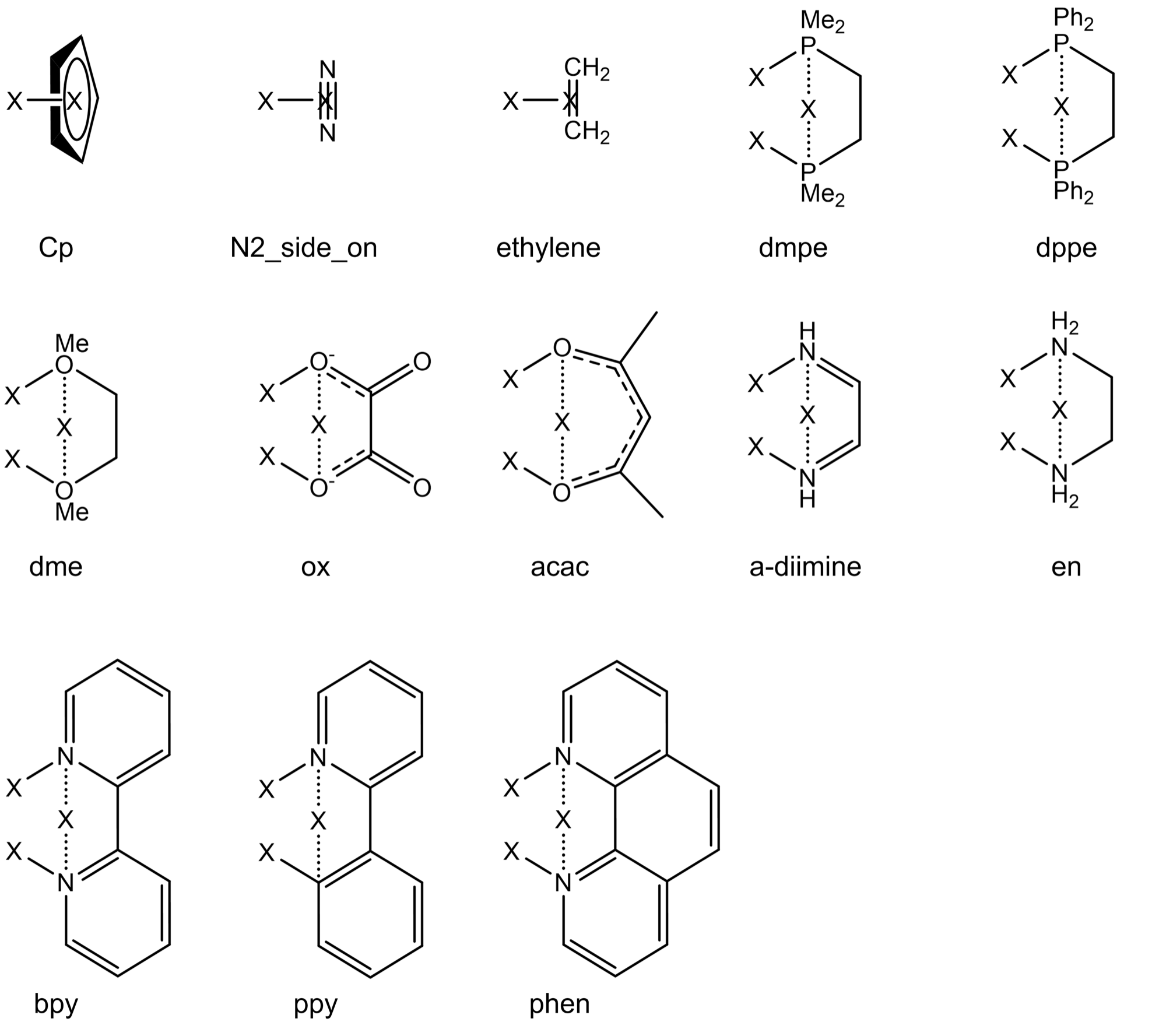}
        \caption{Predefined polydentate ligands library used for replacement and binding operations. Labels shown below each structure correspond to the fragment names as stored in the library. X denotes the connection site.}        \label{fig:SI_predefined_ligands}
\end{figure}

The full set of predefined templates and detailed construction mechanisms is provided in the following sections and \autoref{fig:SI_inorg}. The builder tool also supports a predefined library of commonly used bidentate and polydentate ligands (\autoref{fig:SI_predefined_ligands}).

\textbf{Transition-metal complex construction from coordination templates.}

The transition-metal complex construction tool builds metal--ligand structures starting from a
predefined coordination-geometry template.
In all templates, the central metal atom is fixed to atom index 0.
A library of predefined coordination geometries (e.g., square planar, octahedral, tetrahedral) is
provided, and the desired geometry is selected via the \texttt{coordination\_geometry} argument.

By default, all coordination sites in the selected template are occupied by monodentate placeholder
ligands (Cl atoms).
Therefore, explicitly specifying monodentate ligands such as Cl is optional.
When a metal element is provided via the \texttt{metal} argument, the element at atom index 0 is
reassigned accordingly.

\textbf{Ligand binding specification and polydentate attachment.}

Ligand attachment is controlled by the \texttt{ligand\_bindings} argument, which is provided as a list of dictionaries mapping ligand names to indexed coordination sites in the selected template. The coordination sites of each template are predefined and explicitly numbered. For polydentate ligands, multiple coordination indices are specified simultaneously and treated as a single binding event.

A predefined library of common polydentate ligands is provided (\autoref{fig:SI_predefined_ligands}). Each ligand template contains predefined binding atoms and dummy atoms that encode its preferred coordination geometry. The coordination sites of each metal template are explicitly indexed, and the tool description provides the site combinations associated with different coordination geometries and stereochemical arrangements. The predefined ligand library also specifies the identities and ordering of the ligand donor atoms. The agent reads these template and ligand definitions and constructs the corresponding \texttt{ligand\_bindings} argument for the requested complex and stereochemistry. Thus, the complete \texttt{ligand\_bindings} assignment is not retrieved from an external database during structure generation, but is assembled from the predefined coordination-site and ligand-binding information.

The ordering of the coordination indices can be chemically meaningful. For a bidentate ligand assigned to template positions \texttt{[1,2]}, the first predefined donor atom is placed at coordination site 1 and the second donor atom at site 2. Reversing the assignment to \texttt{[2,1]} can give an equivalent result when the two donor atoms are symmetry-equivalent, as in a symmetric ligand such as 1,10-phenanthroline (phen). For an unsymmetrical ligand such as 2-phenylpyridine (ppy), however, the donor atoms are chemically distinct, and reversing the index order changes their mapping to the coordination sites and can produce a different ligand orientation or stereoisomer. Likewise, exchanging ligands between symmetry-equivalent template positions can yield symmetry-equivalent structures, whereas permutations involving inequivalent positions can generate distinct coordination arrangements. The tool therefore preserves the coordination-site labels and predefined donor-atom ordering rather than assuming that arbitrary permutations are equivalent.

During polydentate attachment, the ligand is aligned to the coordination template using the predefined dummy atoms according to the mechanism shown in \autoref{fig:SI_inorg}, ensuring consistent chelation geometry. For standard monodentate substitutions not covered by the polydentate library, ligand attachment follows the general atom-replacement workflow described for the \texttt{replace} tool.

The structure shown in \autoref{fig:SI_inorg} was generated using the following tool invocation:
\begin{verbatim}
build_tm_complex_from_template(
    coordination_geometry = `square_planar',
    metal = `Pd',
    ligand_bindings = [{`phen': [1,2]}],
    output_path = `Pd_phen_Cl2.xyz'
)
\end{verbatim}

\clearpage
\subsection{Vision language model benchmark on spatial understanding}\label{sec3:VLM}
In order for \estructural{} to perform well on vision-based molecular tasks, we first evaluate a fundamental capability of vision--language models (VLMs): whether they can perceive molecular structures clearly. To quantify this capability, we design a simple benchmark: given a target atom from a molecule, the model must identify its neighbouring atoms that are bonded to it. Although this task can be solved easily by computing interatomic distances and applying empirical bond-length thresholds, the goal here is to assess performance under a vision-only setting.

Accordingly, the benchmarked VLM agent is restricted to using only the \texttt{view\_molecule} and \texttt{zoom\_molecule} tools. The system prompt and an example of tool usage are reported in Section~\ref{SI: VLM System Prompt}. We sample 300 molecules from the GeomConf dataset~\cite{axelrod2022geom} with sizes drawn from a uniform distribution, and benchmark frontier VLMs on this task, including Gemini-3-Pro, Gemini-3-Flash, Claude-Opus-4.5, Claude-4.5-Sonnet, GPT-5.2-Thinking-High, and GPT-5.4-Thinking-High.

\begin{figure}[htbp]
    \centering
    \includegraphics[width=1.0\linewidth]
    {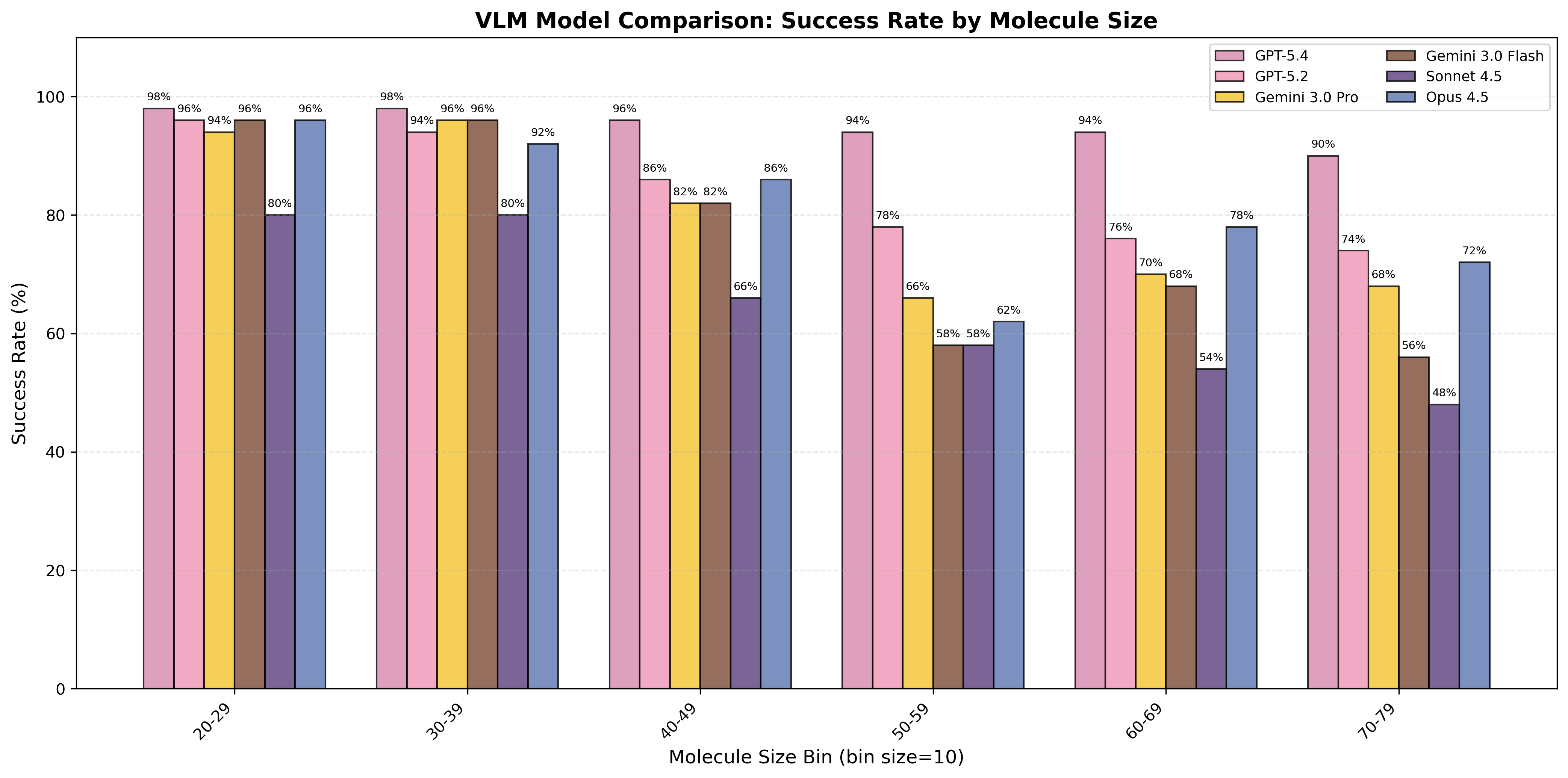}
    \caption{\textbf{VLM Benchmark Performance by Molecule Size.} 
    Comparing the success rates of six Vision Language Models (Gemini 3 Pro, Gemini 3 Flash, Sonnet 4.5, Opus 4.5, GPT 5.2, and GPT 5.4) across varying molecule sizes, measured by the number of atoms. Most tested models show decreasing accuracy as molecular size increases, whereas GPT 5.4 maintains a success rate of at least 90\% across all size bins.}
    \label{fig:VLM_Bench}
\end{figure}

As shown in \autoref{fig:VLM_Bench}, most models exhibit substantial
size-dependent degradation. In contrast, GPT-5.4 maintains 90--98\% accuracy
across all molecular-size bins. The marked improvement over GPT-5.2 provides
strong evidence that this visual-perception bottleneck is not intrinsic to
image-based molecular interfaces and can be substantially mitigated by advances
in the underlying VLM. Although this benchmark probes local bonding perception
rather than general chemical reasoning, the result suggests that molecular vision
capabilities may improve rapidly with continued model development.

The case studies therefore retain a dynamic strategy in which \estructural{}
selects between explicit structural analysis tools and vision-based inspection
depending on the task context. Structural analysis and visual reasoning play
complementary roles: connectivity- and index-based analysis provides robustness
for complex, multi-step manipulations, while visual inspection is invoked when it
offers additional intuition or confirmation. This hybrid design remains important
because strong performance on local-neighbour identification does not yet establish
reliable interpretation of complete reaction mechanisms or error-free multi-turn
geometry editing. Targeted systems such as Ether0~\cite{narayanan_training_2025}
and ChemVLM~\cite{li2025chemvlm} further indicate that chemistry-specific training
can improve combined visual and chemical reasoning.

\clearpage
\subsubsection{VLM benchmark system prompt}\label{SI: VLM System Prompt}
In this section, we list the VLM benchmark system prompt for all six Vision-Language Models (Gemini 3 Pro, Gemini 3 Flash, Sonnet 4.5, Opus 4.5, GPT 5.2, and GPT 5.4), and list an example for the use of \texttt{view\_molecule} and \texttt{zoom\_molecule} tool in the benchmark.

\textbf{System Prompt}

You're a helpful scientific agent.
Here is Your Current Role: you are a helpful chemistry vision agent. You will be tasked to identify correct neighbour atoms. You should use vision capability only, which you can view the molecule by using the given tools. You can rotate the molecule to inspect it in different angles. You can also zoom in to get a better view especially when view is too crowded. Do not overthink, this is a simple vision task.

\begin{figure}[!htbp]
        \centering        \includegraphics[width=1.0\textwidth]{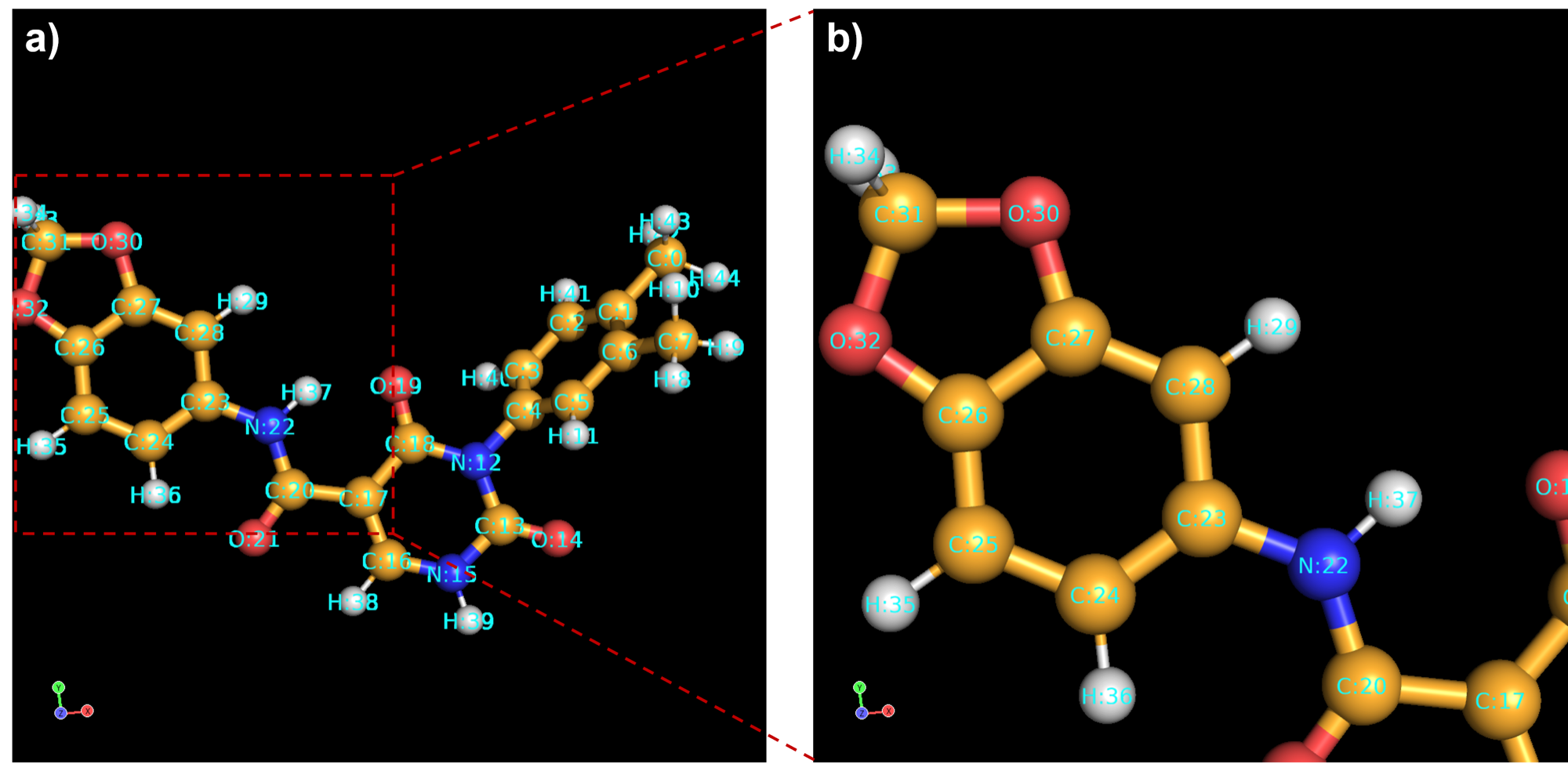}
        \caption{(a) Visualization using the \texttt{view\_molecule}. The tool displays the full molecule (global view) by resetting previous zoom levels. (b) Zoomed-in view of the molecular structure. This perspective allows for a detailed inspection of local atomic environments.
}
        \label{fig:mol_viz_tools}
\end{figure}

\textbf{Tool Used for geomconf\_1027}
The following pictures show an example of geomconf\_1027 using \texttt{view\_molecule} and \texttt{zoom\_molecule} to zoom in on the target region for a clearer view (\autoref{fig:mol_viz_tools}).

\clearpage

\section{Case study benchmark}\label{SIsec:benchmark}
To assess reproducibility, each case study was repeated for five independent iterations. For each iteration, the number of tool calls, runtime, and evaluation score were recorded and summarized in this section. Complete agent traces, including the detailed tool-call sequence and generated outputs, are provided separately in our group GitHub repository at \url{https://github.com/the-matter-lab/ElAgenteEstructuralCaseStudies}.

To score each case study, we evaluated the generated structures at multiple levels, including whether the requested structural task was completed, whether the resulting molecular connectivity and geometry remained physically reasonable, and whether relevant stereochemical, configurational, or other task-specific structural features were preserved. Because the case studies span organic molecules, coordination complexes, stereochemical transformations, and TS-like structures, we used complementary validation criteria rather than a single universal metric. The general validation procedures applied across the benchmark are described below, while additional task-specific criteria and scoring rubrics are provided in the corresponding case-study subsections.

To evaluate molecular identity and connectivity, generated \texttt{xyz} structures were converted into SMILES using distance-based bond perception implemented in RDKit. For structures containing only nonmetal atoms, the resulting canonical SMILES were compared with the expected molecular connectivity. Agreement with the target representation directly confirms that the intended molecular identity and connectivity were obtained. This analysis also provides a check for severe geometric corruption because substantial atomic overlaps can introduce spurious short-range connections, while sufficiently elongated bonds can result in loss of expected connectivity and therefore alter the resulting SMILES.

For transition-metal-containing structures, where conventional SMILES representations of the complete metal complex are less informative, the transition-metal atoms were removed, and the remaining disconnected components were treated as individual ligand fragments. A SMILES was generated independently for each fragment, and the resulting ligand SMILES were sorted to produce an order-independent list for each structure. The identities and multiplicities of these fragments were then compared with those expected for the target structure. This procedure was particularly useful for organometallic functionalization cases in which the metal coordination framework was retained while the requested transformation occurred within the ligand framework, such as meso functionalization of Co-porphyrin and substitution of the tert-butyl groups in MoCl$_3$PNP. Severe intra- or inter-ligand overlaps can also alter the perceived connectivity and thereby change the resulting ligand-fragment representation.

Steric clashes were evaluated independently of the SMILES-based connectivity analysis. The structure-editing and ligand-binding tools include post-editing sanitization that identifies steric conflicts between the original structure and newly attached fragments. Only atom pairs between the original core structure and the newly attached fragment were considered, excluding the pair forming the newly created bond. Atom pairs within the original core structure or within the newly attached fragment were not evaluated. A clash was defined when an interatomic distance was less than 1.5 times the sum of the corresponding covalent radii, with an additional H--H criterion of 2.0~\AA. Rotational adjustment was applied where possible to remove detected clashes (Section~\ref{SI_Note:editing}). For structures generated through these editing workflows, we confirmed that no clashes remained according to these criteria.

For tasks involving stereochemistry or specific coordination geometries, additional task-specific structural metrics were used. For organic stereochemical inversion, stereochemical SMILES were used to confirm inversion at the targeted stereocentre. For \textit{cis}/\textit{trans} coordination isomers, relevant dihedral angles or ligand--metal--ligand angles were evaluated, while \textit{fac}/\textit{mer} structures were distinguished using the characteristic pattern of coordination angles around the metal centre.

For Case~6, involving multimodal reaction-mechanism-guided structure generation, the generated TS-like structures were subjected to subsequent DFT transition-state optimization. Successful convergence to a transition state with the expected principal imaginary frequency was used as downstream validation of the generated TS-like starting geometry.
Unless otherwise specified, all DFT calculations performed for downstream structural validation across the case studies employed the B3LYP-D4/def2-SVP level of theory using ORCA 6.0~\cite{beckeDensityfunctionalThermochemistryIII1993,leeDevelopmentColleSalvettiCorrelationenergy1988,stephensInitioCalculationVibrational1994,caldeweyher2017extension_d4correction,caldeweyherGenerallyApplicableAtomiccharge2019,weigendBalancedBasisSets2005,orca6}.

For organometallic structures involving higher-hapticity ligands, such as $\eta^5$-Cp and $\eta^6$-sumanene, a single bond length or angle does not adequately describe the overall coordination geometry. We therefore additionally compared the generated structures with their corresponding GFN2-xTB-relaxed structures using heavy-atom RMSD. Heavy-atom RMSD values of approximately 0.5~\AA\ have previously been used in conformer-generation studies as a practical criterion for identifying geometrically redundant organic conformers~\cite{seidel2023high,fuentes2026anisotropic}. Here, 0.5~\AA\ was used only as a practical reference for the magnitude of structural relaxation rather than as a universal criterion for organometallic structural correctness. Structures below this value remain close to their GFN2-xTB-relaxed geometries and require only limited overall structural rearrangement during optimization. Because the goal of \estructural{} is to generate chemically reasonable three-dimensional starting structures rather than fully optimized geometries, we used GFN2-xTB relaxation as a practical low-cost reference to assess the extent of structural rearrangement required after generation. The resulting heavy-atom RMSD therefore reflects the magnitude of relaxation rather than agreement with a high-level DFT minimum. 

All case-study evaluations were scored on a 100-point scale according to predefined case-specific evaluation rubrics. The detailed scoring criteria, quantitative results, and any additional validation metrics used for each task are described in the corresponding subsections below.

\subsection{Case study 1: Site-selective functionalization}
\subsubsection{Selective amine protection in spermidine}
In this case study, the user prompt was: ``Generate an XYZ file for spermidine. Using the generated XYZ structure, selectively protect only the primary amines with Boc groups.'' 

Across all five runs, the agent first generated spermidine from its SMILES representation, then generated BocCl from SMILES, identified the primary amine sites, and sequentially replaced the corresponding N--H hydrogens using the structure-editing tool. The evaluation score was based on whether the final generated structure had the correct molecular connectivity corresponding to spermidine with all primary amines Boc-protected. Full credit was assigned when the generated SMILES exactly matched the expected product connectivity (\autoref{tab:per-spermidine}).

\renewcommand{\arraystretch}{1.05}
\begin{table}[H]
\centering
\caption{Per-iteration evaluation scores for ``Selective amine protection in spermidine''.}
\label{tab:per-spermidine}
\small

\begin{tabularx}{\linewidth}{@{}
>{\centering\arraybackslash}p{0.16\linewidth}
>{\centering\arraybackslash}p{0.20\linewidth}
>{\centering\arraybackslash}p{0.20\linewidth}
>{\centering\arraybackslash}p{0.20\linewidth}
@{}}
\toprule
\textbf{Iteration} &
\textbf{Number of tool calls} &
\textbf{Runtime (s)} &
\textbf{Evaluation score} \\
\midrule

Run 1 & 9 & 50 & 100 \\
Run 2 & 8 & 49 & 100 \\
Run 3 & 7 & 50 & 100 \\
Run 4 & 8 & 52 & 100 \\
Run 5 & 9 & 83 & 100 \\

\midrule
\textbf{Average} &
$\mathbf{8.2 \pm 0.8}$ &
$\mathbf{56.8 \pm 14.7}$ &
$\mathbf{100 \pm 0}$ \\
\bottomrule
\end{tabularx}
\end{table}

\subsubsection{Symmetry-guided Co-porphyrin Editing}
In this case study, the user prompt was: ``Given Co-porphyrin.xyz, replace the four symmetry-equivalent hydrogen atoms with phenyl groups.''  
Across all five runs, the agent performed point-group analysis of the Co-porphyrin structure to identify the four symmetry-equivalent meso hydrogen atoms and replaced each of them with a phenyl group. The resulting structures were validated by removing the Co atom and converting the remaining organic framework into SMILES, confirming formation of the tetraphenyl-substituted porphyrin connectivity. The dihedral angles between the porphyrin framework and the attached phenyl groups were approximately $50^\circ$, and no steric clashes were observed according to the post-attachment sanitization criteria.
The variation in the number of tool calls across iterations primarily arose from differences in the agent's structural verification steps. For example, in Run~1, after identifying the four symmetry-equivalent H atoms, \estructural{} additionally called \texttt{get\_atomic\_neighbors} for each identified H atom to confirm the C atom bonded to it before performing the replacement. In Runs~2--4, the agent used fewer \texttt{get\_atomic\_neighbors} calls, while Run~5 did not use this tool. These differences in neighbour inspection were the main contributor to the higher tool-call count in Run~1 (8 calls) compared with the other runs (4--5 calls), although \texttt{xyz}-file inspection using \texttt{bash\_kernel} also contributed to the variation. Thus, the variation in tool-call counts reflects differences in the extent of structural analysis and verification performed by the agent rather than failures or retries.
Full credit was assigned when the generated SMILES exactly matched the expected product connectivity (\autoref{tab:per-Copor}).

\renewcommand{\arraystretch}{1.05}
\begin{table}[H]
\centering
\caption{Per-iteration evaluation scores for ``Symmetry-guided Co-porphyrin Editing''.}
\label{tab:per-Copor}
\small

\begin{tabularx}{\linewidth}{@{}
>{\centering\arraybackslash}p{0.16\linewidth}
>{\centering\arraybackslash}p{0.20\linewidth}
>{\centering\arraybackslash}p{0.20\linewidth}
>{\centering\arraybackslash}p{0.20\linewidth}
@{}}
\toprule
\textbf{Iteration} &
\textbf{Number of tool calls} &
\textbf{Runtime (s)} &
\textbf{Evaluation score} \\
\midrule

Run 1 & 8 & 19 & 100 \\
Run 2 & 4 & 11 & 100 \\
Run 3 & 4 & 13 & 100 \\
Run 4 & 5 & 16 & 100 \\
Run 5 & 4 & 17 & 100 \\

\midrule
\textbf{Average} &
$\mathbf{5 \pm 1.7}$ &
$\mathbf{15.2 \pm 3.2}$ &
$\mathbf{100 \pm 0}$ \\
\bottomrule
\end{tabularx}
\end{table}

\begin{figure}[htbp!]
        \centering        \includegraphics[width=1.0\textwidth]{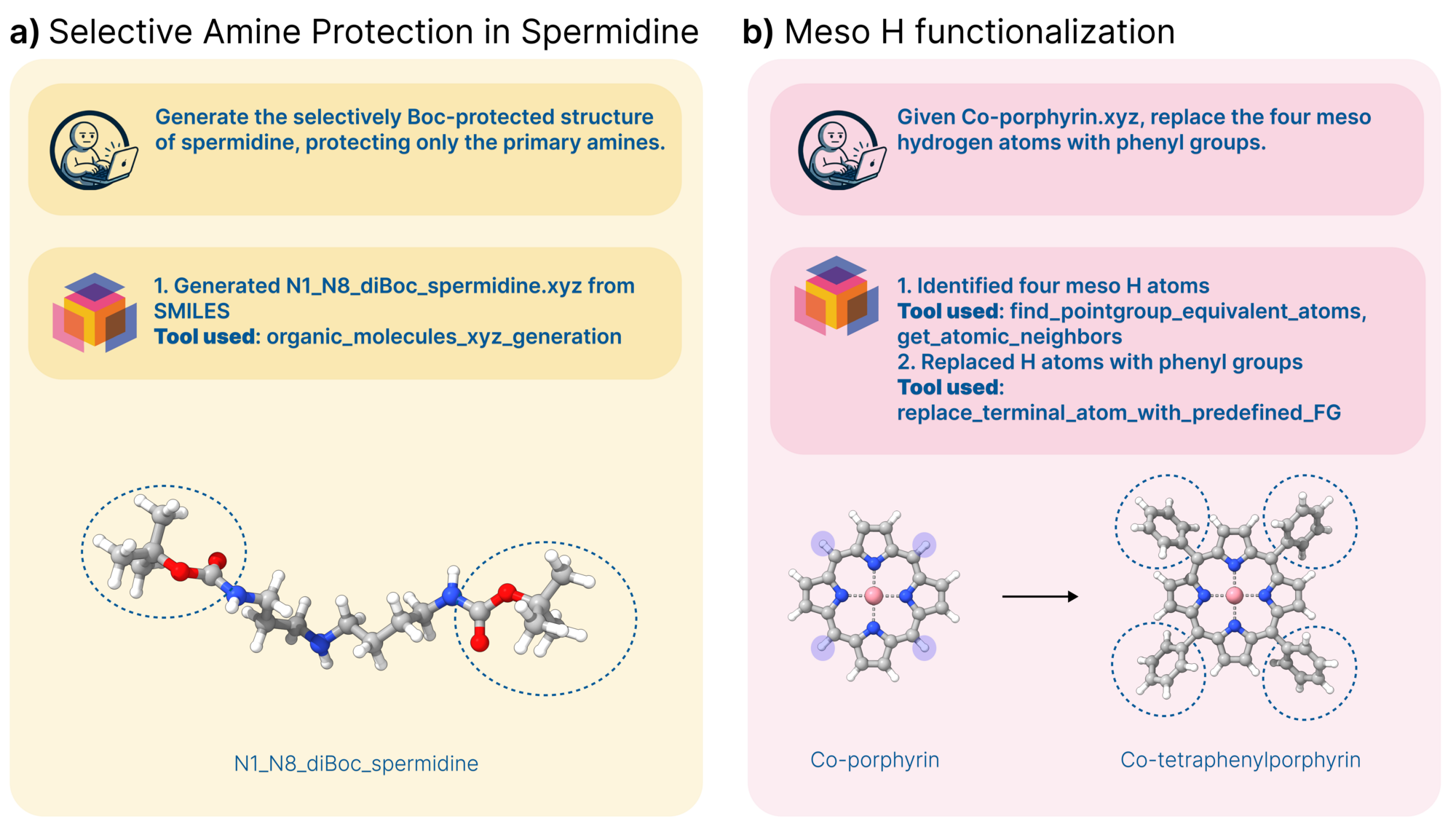}
        \caption{Representative results obtained using more generic prompts. (a) Selective Boc protection of the primary amines in spermidine. (b) Identification and phenyl substitution of the four meso hydrogen atoms in Co-porphyrin.
        Atom colour codes: pink, Co; red, O; blue, N; grey, C; white, H.
}
        \label{fig:SI_Case1_generic}
\end{figure}

To further evaluate the agent under more generic instructions, we repeated two representative case studies using prompts that provided less procedural guidance (\autoref{fig:SI_Case1_generic}). For spermidine, the prompt was: ``Generate the selectively Boc-protected structure of spermidine, protecting only the primary amines.'' In contrast to the standard prompt, no intermediate structure-generation step or editing procedure was specified, and the agent had to infer the appropriate route to the requested final product. In all five iterations, the agent recognized the target product and directly generated the selectively Boc-protected spermidine structure from its SMILES representation. The resulting structures showed the expected molecular connectivity and received full credit (\autoref{tab:per-spermidine_generic}).

For Co-porphyrin functionalization, ``Given Co-porphyrin.xyz, replace the four meso hydrogen atoms with phenyl groups.'' Here, the explicit hint that the target hydrogen atoms were symmetry-equivalent was removed and replaced by the chemically descriptive term meso. In all five iterations, the agent identified the four meso hydrogen atoms by combining point-group equivalence analysis with local connectivity inspection using the \texttt{find\_pointgroup\_equivalent\_atoms} and \texttt{get\_atomic\_neighbors} tools, and subsequently replaced the identified hydrogen atoms with phenyl groups. 
In the generic-prompt test (\autoref{tab:per-Copor_generic}), Run 1 required substantially fewer tool calls than Runs 2--5. In Run 1, \estructural{} efficiently identified the four target H atoms using symmetry analysis and proceeded with the substitutions, resulting in only five tool calls. In Runs 2--5, the agent performed more extensive structural analysis before substitution, repeatedly using \texttt{get\_atomic\_neighbors} to examine and confirm the atomic connectivity of candidate sites, in combination with symmetry analysis. These additional structural identification and verification steps account for the higher numbers of tool calls in these runs.
The generated structures exhibited the expected Co-tetraphenylporphyrin connectivity after excluding the Co centre from the SMILES-based validation, and all five iterations received full credit (\autoref{tab:per-Copor_generic}).

\renewcommand{\arraystretch}{1.05}
\begin{table}[H]
\centering
\caption{Per-iteration evaluation scores for ``Selective amine protection in spermidine (generic prompt)''.}
\label{tab:per-spermidine_generic}
\small

\begin{tabularx}{\linewidth}{@{}
>{\centering\arraybackslash}p{0.16\linewidth}
>{\centering\arraybackslash}p{0.20\linewidth}
>{\centering\arraybackslash}p{0.20\linewidth}
>{\centering\arraybackslash}p{0.20\linewidth}
@{}}
\toprule
\textbf{Iteration} &
\textbf{Number of tool calls} &
\textbf{Runtime (s)} &
\textbf{Evaluation score} \\
\midrule

Run 1 & 1 & 26 & 100 \\
Run 2 & 3 & 42 & 100 \\
Run 3 & 1 & 16 & 100 \\
Run 4 & 1 & 15 & 100 \\
Run 5 & 1 & 10 & 100 \\

\midrule
\textbf{Average} &
$\mathbf{1.4 \pm 0.9}$ &
$\mathbf{21.8 \pm 12.7}$ &
$\mathbf{100 \pm 0}$ \\
\bottomrule
\end{tabularx}
\end{table}

\renewcommand{\arraystretch}{1.05}
\begin{table}[H]
\centering
\caption{Per-iteration evaluation scores for ``Symmetry-guided Co-porphyrin Editing (generic prompt)''.}
\label{tab:per-Copor_generic}
\small

\begin{tabularx}{\linewidth}{@{}
>{\centering\arraybackslash}p{0.16\linewidth}
>{\centering\arraybackslash}p{0.20\linewidth}
>{\centering\arraybackslash}p{0.20\linewidth}
>{\centering\arraybackslash}p{0.20\linewidth}
@{}}
\toprule
\textbf{Iteration} &
\textbf{Number of tool calls} &
\textbf{Runtime (s)} &
\textbf{Evaluation score} \\
\midrule

Run 1 & 5 & 31 & 100 \\
Run 2 & 31 & 62 & 100 \\
Run 3 & 16 & 28 & 100 \\
Run 4 & 21 & 55 & 100 \\
Run 5 & 17 & 27 & 100 \\

\midrule
\textbf{Average} &
$\mathbf{18.0 \pm 9.4}$ &
$\mathbf{40.6 \pm 16.6}$ &
$\mathbf{100 \pm 0}$ \\
\bottomrule
\end{tabularx}
\end{table}

\subsubsection{Robustness to perturbed and nearly symmetric Co-porphyrin structures}\label{SIsec_distortion}
\begin{figure}[h!]
        \centering        \includegraphics[width=1.0\textwidth]{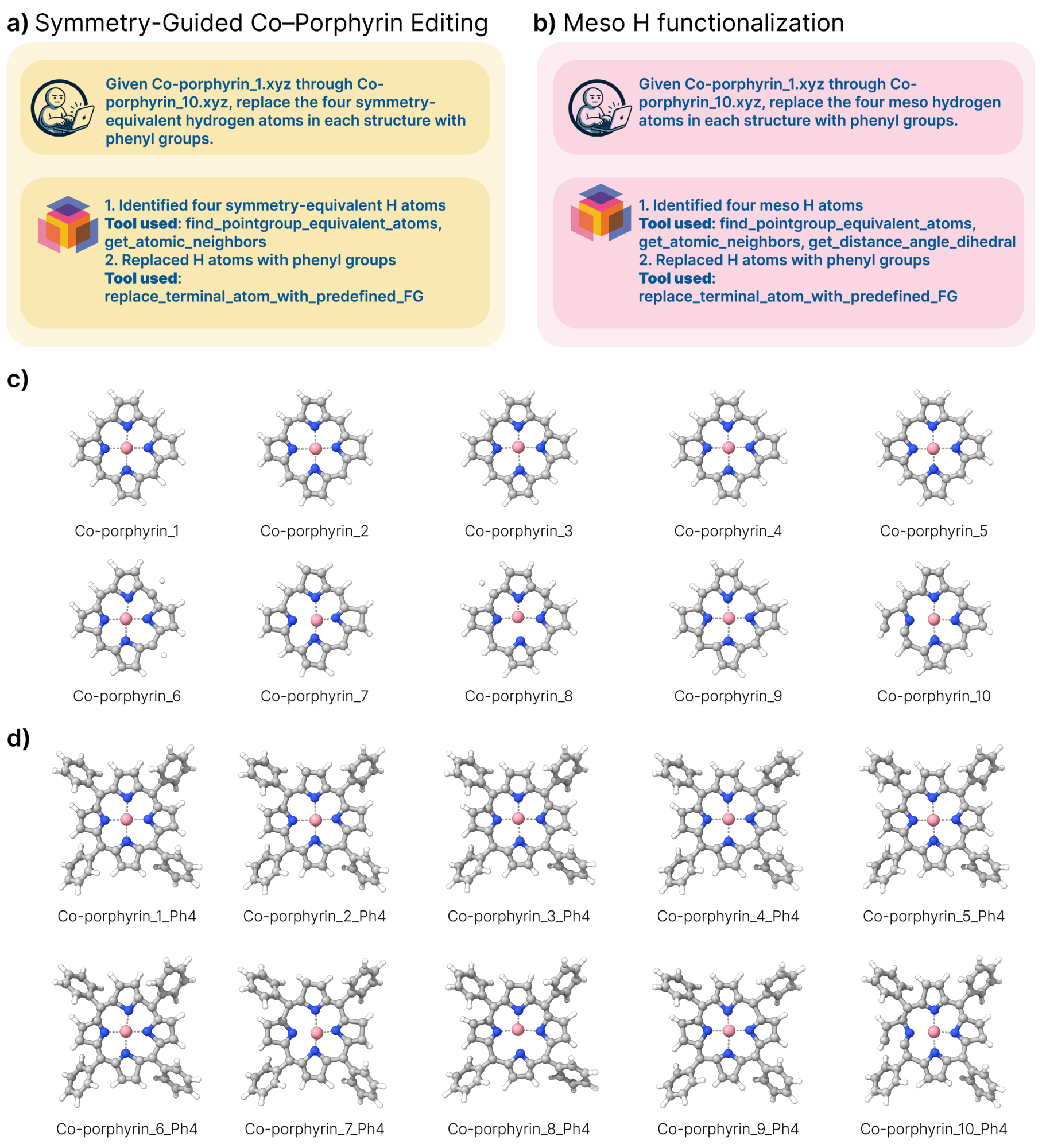}
        \caption{Robustness of \estructural{} to perturbed Co-porphyrin input coordinates. Workflows used to functionalize (a) the four symmetry-equivalent hydrogen atoms and (b) the four meso hydrogen atoms with phenyl groups. (c) Perturbed Co-porphyrin inputs generated with $\sigma = 0.05$~\AA\ (structures 1--5) and $\sigma = 0.10$~\AA\ (structures 6--10). (d) Corresponding tetraphenyl-substituted products. All ten structures were correctly modified in each of five independent runs.
        Atom colour codes: pink, Co; blue, N; grey, C; white, H.
}
        \label{SI:distortion}
\end{figure}

To evaluate robustness against imperfect and nearly symmetric inputs, the original Co-porphyrin structure was randomly perturbed before applying the same phenyl-substitution task (\autoref{SI:distortion}). Five structures (\texttt{Co-porphyrin\_1.xyz}--\texttt{Co-porphyrin\_5.xyz}) were generated using a Cartesian perturbation with $\sigma = 0.05$~\AA, and five additional structures (\texttt{Co-porphyrin\_6.xyz}--\texttt{Co-porphyrin\_10.xyz}) were generated using $\sigma = 0.10$~\AA. The agent was then instructed to replace the four symmetry-equivalent meso hydrogen atoms with phenyl groups.

The smaller perturbations were sufficient to break the exact $D_{4h}$ symmetry of structures 1--5, while the larger perturbations in structures 6--10 produced more pronounced geometric distortions. As a result, some meso C--H, Co--N, and C--C contacts were not recognized as bonds under the default distance-based bond-perception settings used for visualization. To further prevent reliance on fixed atomic indices, the atom ordering was independently randomized for each of the ten perturbed Co-porphyrin structures, and hence, the four meso hydrogen atoms had different atomic indices in each \texttt{Co-porphyrin\_1.xyz}--\texttt{Co-porphyrin\_10.xyz} structure.

\estructural{} initially attempted to recover the target sites using \texttt{find\_pointgroup\_equivalent\_atoms} while progressively increasing the symmetry-tolerance argument. This approach successfully identified the four target sites for some of structures 1--5. For the remaining structures, including the more strongly perturbed cases, the agent instead used \texttt{get\_atomic\_neighbors} and identified the meso positions from their local connectivity. 

Across all five independent runs, \estructural{} identified the four intended meso hydrogen atoms in all ten perturbed structures and replaced them with phenyl groups. The same replacement tool and corresponding arguments were subsequently applied to each identified site, yielding the expected tetraphenyl-substituted structures without residual steric clashes.

\subsubsection{Robustness to varied protonation states}
\begin{figure}[ht!]
        \centering        \includegraphics[width=1.0\textwidth]{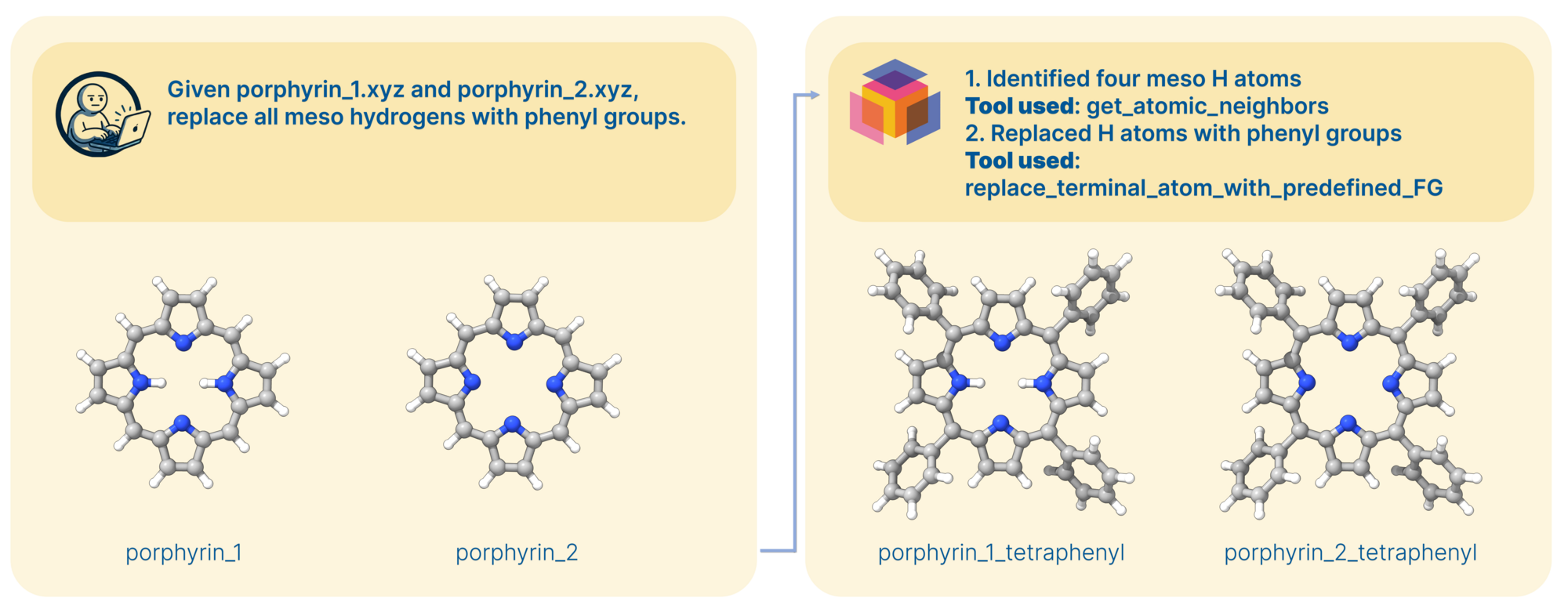}
        \caption{Meso-H functionalization of porphyrins with different protonation states. A neutral porphyrin containing two inner N--H protons and a dianionic porphyrin without inner protons were provided as inputs. In both cases, the four meso hydrogen atoms were identified and replaced with phenyl groups. The expected tetraphenyl-substituted structures were obtained in all five independent runs.
        Atom colour codes: blue, N; grey, C; white, H.
}
        \label{SI:varied_protonation}
\end{figure}

To evaluate robustness with respect to protonation state, two metal-free porphyrin structures were provided: a neutral porphyrin containing two inner N--H protons and a dianionic porphyrin obtained by removing the metal centre from the corresponding metalloporphyrin without adding the two inner protons (\autoref{SI:varied_protonation}). The agent was instructed to replace all meso hydrogen atoms with phenyl groups without being given the corresponding atomic indices.

For both protonation states, the agent used \texttt{get\_atomic\_neighbors} to identify the meso C--H sites from the local connectivity and subsequently replaced the four meso hydrogen atoms with phenyl groups. The different protonation states of the inner nitrogen atoms therefore did not affect identification of the peripheral meso positions. The expected tetraphenyl-substituted structures were obtained in all five independent runs for both input structures.

\subsection{Case study 2: Binding of reaction intermediates and ligands}
This case study employed two sequential prompts. In the first prompt, \estructural{} was instructed to bind CO$_2$, COOH, CO, H, OH, and H$_2$O to the Co centre of \texttt{Co-Pc.xyz}. In the second prompt, \estructural{} was instructed to take both the original Co-Pc structure and all structures generated in the first step and attach one additional axial ligand, selected from thiophene, pyridine, and Cl, to the Co centre.

For all generated complexes, the newly formed metal--ligand distances were within the expected bonding range established by the binding tool, approximately corresponding to the sum of the covalent radii of the Co and the ligand anchoring atom. No steric clashes were observed among the core structure and newly attached fragments according to the post-attachment sanitization criteria described in Section~\ref{SI_Note:editing}. 
For complexes containing two axial ligands, the angle defined by the anchoring atom of the first ligand, Co, and the anchoring atom of the second ligand was approximately $175^\circ$, indicating an almost linear axial arrangement and preservation of the intended octahedral coordination geometry around the Co centre.
Across the five runs, \estructural{} mainly used \texttt{bind\_predefined\_fragment} for ligand attachment, although \texttt{bind\_atom\_ligands\_library} was also employed in some runs.
The higher number of tool calls in Run 1 (\autoref{tab:per-Case2}) primarily resulted from additional structural verification after the binding operations. The binding tool places the two binding atoms at a distance determined by the sum of their covalent radii by default. In Run 1, \estructural{} additionally verified the resulting binding geometries by repeatedly calling \texttt{get\_distance\_angle\_dihedral} to check the distance between the Co centre and the anchoring atom of each newly bound ligand. In particular, Run 1 included 27 such distance-checking calls, which were the primary contributor to its higher tool-call count.
All five independent iterations successfully generated the expected structures and therefore received full credit (\autoref{tab:per-Case2}).

\renewcommand{\arraystretch}{1.05}
\begin{table}[H]
\centering
\caption{Per-iteration evaluation scores for ``Binding of reaction intermediates and ligands''.}
\label{tab:per-Case2}
\small

\begin{tabularx}{\linewidth}{@{}
>{\centering\arraybackslash}p{0.16\linewidth}
>{\centering\arraybackslash}p{0.20\linewidth}
>{\centering\arraybackslash}p{0.20\linewidth}
>{\centering\arraybackslash}p{0.20\linewidth}
@{}}
\toprule
\textbf{Iteration} &
\textbf{Number of tool calls} &
\textbf{Runtime (s)} &
\textbf{Evaluation score} \\
\midrule

Run 1 & 70 & 99 & 100 \\
Run 2 & 37 & 53 & 100 \\
Run 3 & 37 & 79 & 100 \\
Run 4 & 37 & 81 & 100 \\
Run 5 & 37 & 123 & 100 \\

\midrule
\textbf{Average} &
$\mathbf{43.6 \pm 14.8}$ &
$\mathbf{87.0 \pm 26.0}$ &
$\mathbf{100 \pm 0}$ \\
\bottomrule
\end{tabularx}
\end{table}

\begin{figure}[htbp]
        \centering        \includegraphics[width=1.0\textwidth]{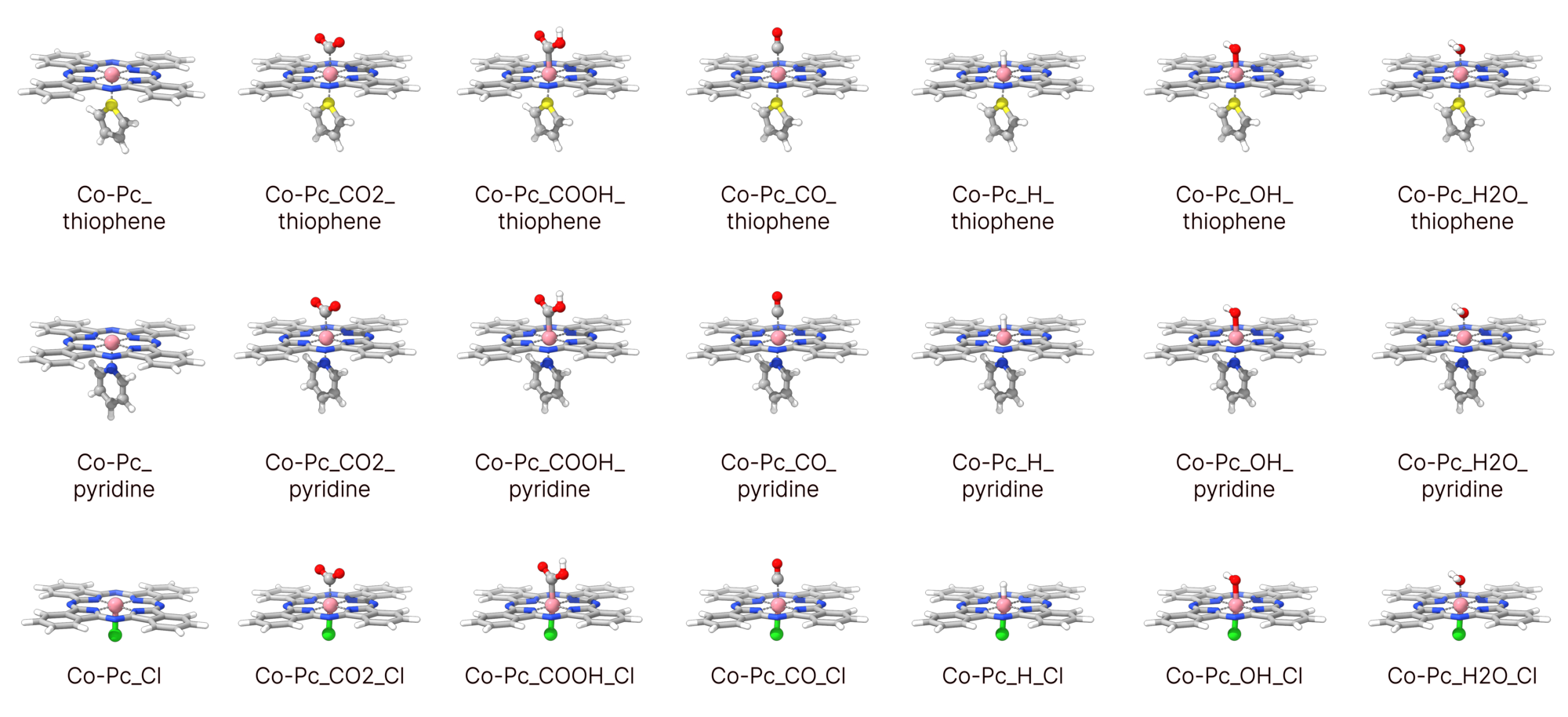}
        \caption{Full list of Co--phthalocyanine (Co--Pc) complexes generated by the agent with different axial ligands (thiophene, pyridine, and Cl). The structure labels shown below each complex correspond to the filenames used by the agent during structure generation, while the \texttt{.xyz} extension is omitted for clarity. 
        The Co centre, together with the bound intermediates (CO$_2$, COOH, CO, H, OH, and H$_2$O), and axial ligands (thiophene, pyridine, or Cl) are highlighted using a ball-and-stick representation, while the remaining atoms are rendered as sticks for clarity.
        Atom colour codes: pink, Co; lime green, Cl; yellow, sulfur; red, O; blue, N; grey, C; white, H.
}
        \label{fig:SI_CoPc}
\end{figure}

\subsubsection{Generic combinatorial generation of axial-ligand variants}

\begin{figure}[t!]
        \centering        \includegraphics[width=1.0\textwidth]{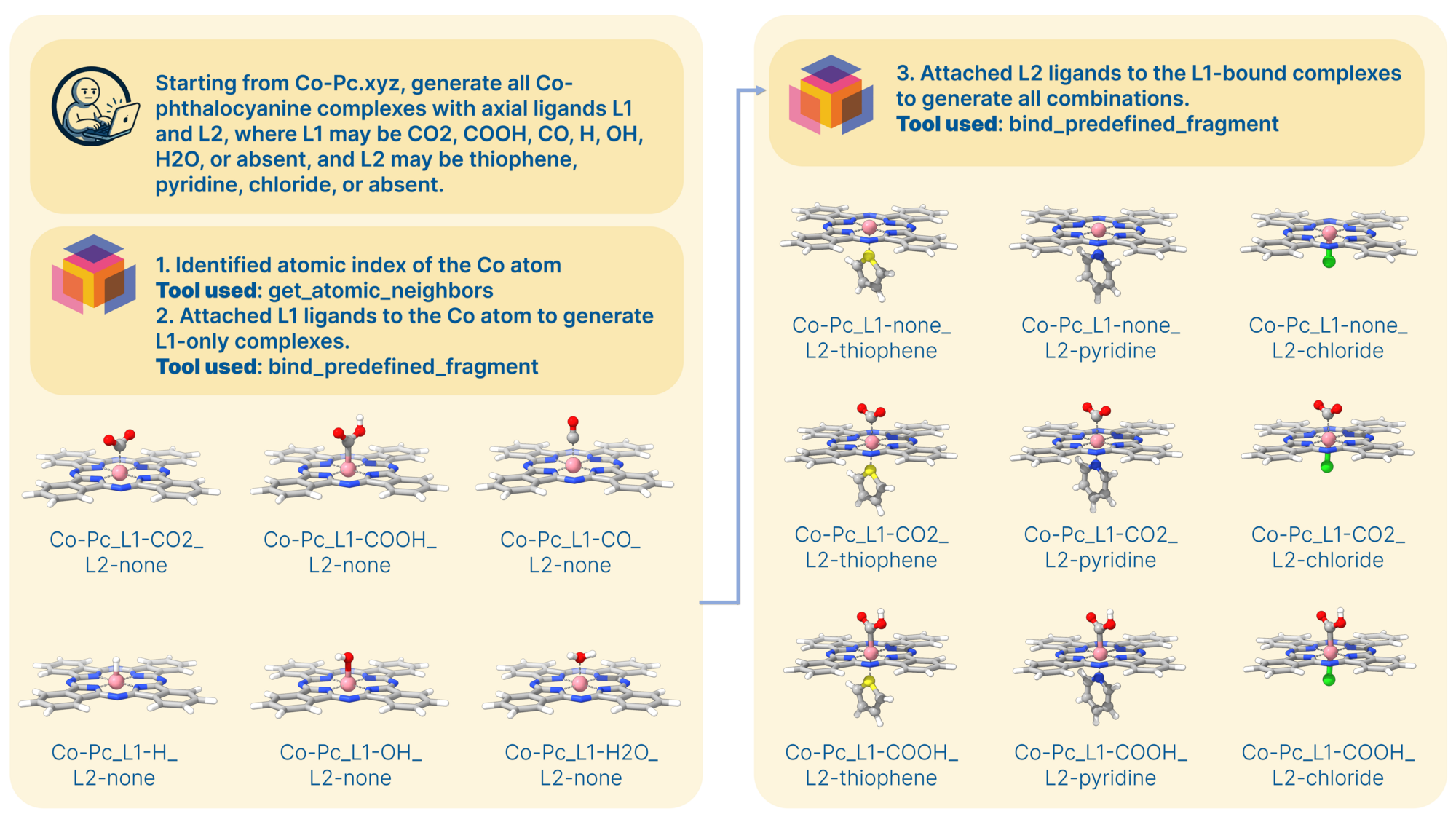}
        \caption{Combinatorial generation of Co-phthalocyanine axial-ligand variants using a single generic prompt. The agent enumerated the specified L1 and L2 combinations, identified the Co binding centre, and generated the corresponding complexes using the binding tool. The resulting complexes had the same molecular compositions and metal--ligand connectivity as those obtained using the sequential-prompt workflow. The full set of axial-ligand–coordinated structures is provided in \autoref{fig:SI_CoPc}.
        Atom colour codes: pink, Co; lime green, Cl; yellow, sulfur; red, O; blue, N; grey, C; white, H.
}
        \label{fig:SI_Case2_generic}
\end{figure}

To evaluate the same task with less procedural guidance, the sequential two-prompt workflow was replaced by a single combinatorial prompt requesting all Co-phthalocyanine complexes formed from the specified L1 and L2 ligand sets. The user provided the allowed ligands for the two axial positions, while the enumeration of all combinations, identification of the Co binding centre, and execution of the required binding operations were left to the agent (\autoref{fig:SI_Case2_generic}).

Across all five iterations, the agent identified the Co atom using \texttt{get\_atomic\_neighbors} and generated the requested ligand combinations using \texttt{bind\_predefined\_fragment}. 
The ligand enumeration workflow varied across the five runs. In Run~1, \estructural{} generated L2-only complexes before attaching L1 ligands to construct the remaining combinations, whereas Runs~2--5 mainly generated L1-only complexes before adding L2 ligands.
The resulting complexes had the same intended molecular compositions and metal--ligand connectivity as those obtained in the sequential-prompt case (\autoref{fig:SI_CoPc}). All generated complexes satisfied the post-attachment sanitization criteria, and all five iterations successfully generated the complete requested set and received full credit (\autoref{tab:per-Case2_generic}).

\renewcommand{\arraystretch}{1.05}
\begin{table}[H]
\centering
\caption{Per-iteration evaluation scores for ``Binding of reaction intermediates and ligands (generic prompt)''.}
\label{tab:per-Case2_generic}
\small

\begin{tabularx}{\linewidth}{@{}
>{\centering\arraybackslash}p{0.16\linewidth}
>{\centering\arraybackslash}p{0.20\linewidth}
>{\centering\arraybackslash}p{0.20\linewidth}
>{\centering\arraybackslash}p{0.20\linewidth}
@{}}
\toprule
\textbf{Iteration} &
\textbf{Number of tool calls} &
\textbf{Runtime (s)} &
\textbf{Evaluation score} \\
\midrule

Run 1 & 51 & 130 & 100 \\
Run 2 & 66 & 122 & 100 \\
Run 3 & 53 & 119 & 100 \\
Run 4 & 52 & 109 & 100 \\
Run 5 & 61 & 125 & 100 \\

\midrule
\textbf{Average} &
$\mathbf{56.6 \pm 6.6}$ &
$\mathbf{121.0 \pm 7.8}$ &
$\mathbf{100 \pm 0}$ \\
\bottomrule
\end{tabularx}
\end{table}

\subsection{Case study 3: Organometallic complex construction with stereochemical and ligand variations}
\subsubsection{Selective isomer generation}
In this case study, the user prompt was: ``Generate the following structures with their respective isomers: (1) cisplatin and transplatin, (2) $\Lambda$-Fe(bpy)$_3$ and $\Delta$-Fe(bpy)$_3$, and (3) \textit{fac}-Ir(ppy)$_3$ and \textit{mer}-Ir(ppy)$_3$.'' The requested complexes and isomer types were explicitly specified by the user, while the corresponding ligand arrangements and template-based construction were determined by the agent. Across all five runs, the agent generated all six structures using the \texttt{build\_tm\_complex\_from\_template} tool.

For cisplatin and transplatin, the generated square-planar structures exhibited the expected approximately $90^\circ$ and $180^\circ$ ligand--Pt--ligand angles, consistent with the requested \textit{cis} and \textit{trans} arrangements. For $\Lambda$- and $\Delta$-Fe(bpy)$_3$, the coordination stereochemistry of the generated structures was visually inspected by the authors and confirmed to correspond to the requested enantiomeric forms. For \textit{fac}-Ir(ppy)$_3$, the three N--Ir--N angles were approximately $98.8^\circ$, consistent with a distorted octahedral facial arrangement. For \textit{mer}-Ir(ppy)$_3$, the corresponding N--Ir--N angles were $98.7^\circ$, $91.3^\circ$, and $167.6^\circ$, showing two \textit{cis}-like and one \textit{trans}-like relationships and therefore the expected meridional coordination geometry.

The higher number of tool calls in Run 3 arose in part from additional
geometric verification of the generated isomers (\autoref{tab:per-Case3A}). After structure generation,
\estructural{} additionally called \texttt{get\_distance\_angle\_dihedral} six
times to verify the characteristic coordination geometries: the Cl--Pt--N angle was checked for each of the \textit{cis} and \textit{trans} Pt complexes, and two N--Ir--N angles were checked for each of the \textit{fac} and \textit{mer} Ir(ppy)$_3$ complexes. These additional angle
checks contributed to the higher tool-call count in Run 3.

All generated structures from all five iterations exhibited heavy-atom RMSDs below $0.5$~\AA\ relative to their corresponding GFN2-xTB-optimized geometries. For the GFN2-xTB optimizations of cisplatin and transplatin, neutral singlet states were used. For $\Lambda$-Fe(bpy)$_3$ and $\Delta$-Fe(bpy)$_3$, a total charge of $+2$ and a singlet spin state were used, while \textit{fac}-Ir(ppy)$_3$ and \textit{mer}-Ir(ppy)$_3$ were optimized as neutral singlets (\autoref{tab:per-Case3A} and \autoref{tab:per-Case3A_rmsd}).

\renewcommand{\arraystretch}{1.05}
\begin{table}[H]
\centering
\caption{Per-iteration evaluation results for ``Selective isomer generation''.}
\label{tab:per-Case3A}
\small

\begin{tabularx}{\linewidth}{@{}
>{\centering\arraybackslash}p{0.20\linewidth}
>{\centering\arraybackslash}p{0.24\linewidth}
>{\centering\arraybackslash}p{0.24\linewidth}
>{\centering\arraybackslash}p{0.24\linewidth}
@{}}
\toprule
\textbf{Iteration} &
\textbf{Number of tool calls} &
\textbf{Runtime (s)} &
\textbf{Evaluation score} \\
\midrule

Run 1 & 9  & 27 & 100 \\
Run 2 & 9  & 33 & 100 \\
Run 3 & 17 & 44 & 100 \\
Run 4 & 9  & 23 & 100 \\
Run 5 & 9  & 20 & 100 \\

\midrule
\textbf{Average} &
$\mathbf{10.6 \pm 3.6}$ &
$\mathbf{29.4 \pm 9.5}$ &
$\mathbf{100 \pm 0}$ \\
\bottomrule
\end{tabularx}
\end{table}

\renewcommand{\arraystretch}{1.05}
\begin{table}[H]
\centering
\caption{Heavy-atom RMSDs (\AA) between the generated coordination-isomer structures from  ``Selective isomer generation'' and their corresponding GFN2-xTB-optimized geometries. The same structures were generated in all five iterations because the same structure-generation tool and arguments were used; therefore, identical RMSD values were obtained across all five runs and are reported once.}
\label{tab:per-Case3A_rmsd}
\small

\begin{tabularx}{\linewidth}{@{}
>{\centering\arraybackslash}X
>{\centering\arraybackslash}X
>{\centering\arraybackslash}X
>{\centering\arraybackslash}X
>{\centering\arraybackslash}X
>{\centering\arraybackslash}X
@{}}
\toprule
\textbf{cisplatin} &
\textbf{transplatin} &
\textbf{$\Delta$-Fe(bpy)$_3$} &
\textbf{$\Lambda$-Fe(bpy)$_3$} &
\textit{fac}-\textbf{Ir(ppy)$_3$} &
\textit{mer}-\textbf{Ir(ppy)$_3$} \\
\midrule

0.13 & 0.15 & 0.33 & 0.33 & 0.21 & 0.26 \\

\bottomrule
\end{tabularx}
\end{table}

\subsubsection{Generation for zirconocene complexes}
In this case study, the user prompt was: ``Generate XYZ files for the following zirconocene complexes: (1) ZrCp$_2$Cl$_2$ and (2) ZrCp$_2$(CH$_3$)($\eta^2$-C$_2$H$_4$)$^+$.'' The user explicitly specified the identities of the two target complexes, while the corresponding coordination arrangements, ligand placement, and structure-generation procedure had to be inferred by the agent.

Across all five iterations, \estructural{} generated both zirconocene
complexes using \texttt{build\_tm\_complex\_from\_template}, followed
by structural inspection using geometric analysis tools.
Because these structures were generated as complete coordination
complexes rather than by local modification of a pre-existing parent
structure, their overall structural quality was additionally assessed
by comparison with the corresponding GFN2-xTB-optimized geometries.
Heavy-atom RMSDs were calculated between the generated and optimized
structures.

All five iterations produced the expected zirconocene structures
and satisfied the structural validation criteria, resulting in
full evaluation scores. 
In Run~3, \estructural{} additionally performed GFN2-xTB geometry
optimization, correctly assigning a neutral singlet state to
ZrCp$_2$Cl$_2$ and a charge of $+1$ with a singlet spin state to
ZrCp$_2$(CH$_3$)($\eta^2$-C$_2$H$_4$)$^+$.
As a result, both optimized structures showed heavy-atom RMSDs of
0.00~\AA\ relative to the corresponding reference geometries,
whereas the structures generated in the other four runs exhibited
RMSDs ranging from 0.33 to 0.38~\AA.
For the reference GFN2-xTB optimizations, ZrCp$_2$Cl$_2$ was treated as a neutral singlet, whereas ZrCp$_2$(CH$_3$)($\eta^2$-C$_2$H$_4$)$^+$ was treated as a singlet with a total charge of $+1$ (\autoref{tab:per-Case3B}).

\renewcommand{\arraystretch}{1.10}
\setlength{\tabcolsep}{3pt}

\begin{table}[H]
\centering
\caption{Per-iteration evaluation results for ``Generation of zirconocene complexes''. Heavy-atom RMSDs were calculated between the generated structures and their corresponding GFN2-xTB-optimized geometries.}
\label{tab:per-Case3B}
\small

\begin{tabularx}{\linewidth}{@{}
>{\centering\arraybackslash}p{0.11\linewidth}
>{\centering\arraybackslash}p{0.14\linewidth}
>{\centering\arraybackslash}p{0.12\linewidth}
>{\centering\arraybackslash}p{0.20\linewidth}
>{\centering\arraybackslash}p{0.25\linewidth}
>{\centering\arraybackslash}p{0.11\linewidth}
@{}}
\toprule
\textbf{Iteration} &
\makecell{\textbf{Number of}\\\textbf{tool calls}} &
\makecell{\textbf{Runtime}\\\textbf{(s)}} &
\makecell{\textbf{ZrCp$_2$Cl$_2$}\\\textbf{RMSD (\AA)}} &
\makecell{
\textbf{ZrCp$_2$(CH$_3$)}\\
\textbf{($\eta^2$-C$_2$H$_4$)$^+$}\\
\textbf{RMSD (\AA)}
} &
\makecell{\textbf{Evaluation}\\\textbf{score}} \\
\midrule

Run 1 & 30 & 73 & 0.33 & 0.33 & 100 \\
Run 2 & 26 & 42 & 0.34 & 0.38 & 100 \\
Run 3\rlap{\textsuperscript{[a]}} & 32 & 87 & 0.00 & 0.00 & 100 \\
Run 4 & 19 & 47 & 0.34 & 0.38 & 100 \\
Run 5 & 17 & 42 & 0.36 & 0.38 & 100 \\

\midrule
\textbf{Average} &
$\mathbf{24.8 \pm 6.6}$ &
$\mathbf{58.2 \pm 20.6}$ &
$\mathbf{0.27 \pm 0.15}$ &
$\mathbf{0.30 \pm 0.17}$ &
$\mathbf{100 \pm 0}$ \\
\bottomrule
\end{tabularx}

\vspace{0.3em}

\begin{minipage}{\linewidth}
\footnotesize
\raggedright
[a] In Run 3, the agent performed GFN2-xTB geometry optimization, resulting in RMSD values of 0.00~\AA. Before optimization, the initial structures had RMSD values of 0.36 and 0.38~\AA\ for ZrCp$_2$Cl$_2$ and ZrCp$_2$(CH$_3$)($\eta^2$-C$_2$H$_4$)$^+$, respectively.
\end{minipage}

\end{table}

\subsubsection{Construction of concave and convex CpRu($\eta^{6}$-sumanene)$^{+}$ complexes}
In this case study, the user prompt was: ``Build concave and convex CpRu($\eta^{6}$-sumanene)$^{+}$ complexes by inspecting the three-dimensional shape of the molecules.'' This task introduced an additional level of structural ambiguity, as the agent had to infer the relevant $\eta^{6}$ coordination site on the curved sumanene framework in addition to distinguishing between concave and convex coordination geometries. The user specified only that both concave and convex coordination geometries should be generated, while the choice between the central and peripheral six-membered rings, the corresponding atomic indices, and the structure-generation workflow had to be inferred by the agent.

The score was divided into planning (30 points) and final output (70 points). The experimentally reported coordination motif corresponds to Ru binding to a peripheral six-membered ring rather than to the central hexagon of sumanene~\cite{amaya2010dynamically}. In Runs~1, 2, 4, and 5, the agent initially selected the central rather than peripheral $\eta^6$ coordination site. Because the issue arose from the initial choice of configuration, 10 points were deducted from the planning score. The intended peripheral structures were subsequently generated after a follow-up prompt specifying peripheral coordination, so full credit was retained for the final output. In Run~3, the agent directly selected the experimentally relevant peripheral $\eta^6$ coordination site and generated both the concave and convex structures without additional user guidance, resulting in a full score.

The overall structural quality of the generated complexes was evaluated by comparison with the corresponding GFN2-xTB-optimized geometries. All generated concave and convex structures showed heavy-atom RMSDs below 0.50~\AA\ across the five runs (\autoref{tab:per-Case3C}), indicating that the generated geometries remained close to their corresponding GFN2-xTB-relaxed structures. Particularly low RMSD values were obtained in Runs~3--5, where geometry sanitization was further assisted by constrained GFN2-xTB optimization using a charge of +1 and uhf = 0. The agent selected Ru--C distance constraints of approximately 2.2~\AA\ for each of the six carbon atoms in the coordinated hexagonal ring. This value is consistent with the experimentally observed Ru--C distances for the concave complex, which range from 2.145 to 2.282~\AA~\cite{amaya2010dynamically}, indicating that \estructural{} selected chemically reasonable constraint values. A force constant of 0.5 or 1.0 was used for these constrained optimizations. As a result, particularly low RMSD values were observed in Runs~3--5.

Accordingly, Runs~1, 2, 4, and 5 received scores of 90, while Run~3 received a score of 100, giving an average evaluation score of $92.0 \pm 4.5$. This scoring therefore distinguishes direct generation of the intended configuration from successful generation achieved after interactive refinement.

\renewcommand{\arraystretch}{1.10}
\setlength{\tabcolsep}{4pt}

\begin{table}[H]
\centering
\caption{Per-iteration evaluation results for ``Construction of concave and convex CpRu($\eta^{6}$-sumanene)$^{+}$ complexes''. Heavy-atom RMSDs were calculated between the generated structures and their corresponding GFN2-xTB-optimized geometries.}
\label{tab:per-Case3C}
\small

\begin{tabularx}{\linewidth}{@{}
>{\centering\arraybackslash}p{0.13\linewidth}
>{\centering\arraybackslash}p{0.15\linewidth}
>{\centering\arraybackslash}p{0.13\linewidth}
>{\centering\arraybackslash}p{0.18\linewidth}
>{\centering\arraybackslash}p{0.18\linewidth}
>{\centering\arraybackslash}p{0.15\linewidth}
@{}}
\toprule
\textbf{Iteration} &
\makecell{\textbf{Number of}\\\textbf{tool calls}} &
\makecell{\textbf{Runtime}\\\textbf{(s)}} &
\makecell{\textbf{Concave}\\\textbf{RMSD (\AA)}} &
\makecell{\textbf{Convex}\\\textbf{RMSD (\AA)}} &
\makecell{\textbf{Evaluation}\\\textbf{score}} \\
\midrule

Run 1 & 37 & 253 & 0.23 & 0.12 & 90 \\
Run 2 & 29 & 176 & 0.29 & 0.09 & 90 \\
Run 3 & 22 & 214 & 0.06 & 0.08 & 100 \\
Run 4 & 35 & 329 & 0.05 & 0.08 & 90 \\
Run 5 & 39 & 274 & 0.05 & 0.08 & 90 \\

\midrule
\textbf{Average} &
$\mathbf{32.4 \pm 6.9}$ &
$\mathbf{249.2 \pm 58.3}$ &
$\mathbf{0.14 \pm 0.12}$ &
$\mathbf{0.09 \pm 0.02}$ &
$\mathbf{92.0 \pm 4.5}$ \\
\bottomrule
\end{tabularx}
\end{table}

\subsection{Case study 4: Fragment-level replacement in organometallic complexes}
\subsubsection{Substitution of PNP tert-butyl groups}
In this case study, the user prompt was: ``Given MoCl$_3$PNP.xyz, find all t-butyl groups and replace them with H, Me, and Ph, creating separate XYZ files for each replacement.'' The user specified the input structure, the functional group to be replaced, and the three requested substituents, while identification of the corresponding tBu groups, their atomic indices, and the required structure-editing operations were left to the agent.

Because this task involved only local substitution of the tBu groups, the MoCl$_3$PNP core and the pre-existing metal--ligand coordination environment were preserved during editing. Accordingly, no additional RMSD-based assessment of the complete structure was performed. Instead, the generated structures were validated by removing the transition-metal centre and converting the remaining fragments into canonical connectivity SMILES. This confirmed the expected replacement of all tBu groups by H, Me, or Ph in the corresponding structures. The target modifications were unambiguous, and all five iterations generated the expected structures without steric clashes; therefore, each run received full credit (\autoref{tab:per-Case4A}).

\renewcommand{\arraystretch}{1.05}
\begin{table}[H]
\centering
\caption{Per-iteration evaluation results for ``Substitution of PNP tert-butyl groups''.}
\label{tab:per-Case4A}
\small

\begin{tabularx}{\linewidth}{@{}
>{\centering\arraybackslash}p{0.20\linewidth}
>{\centering\arraybackslash}p{0.24\linewidth}
>{\centering\arraybackslash}p{0.24\linewidth}
>{\centering\arraybackslash}p{0.24\linewidth}
@{}}
\toprule
\textbf{Iteration} &
\makecell{\textbf{Number of}\\\textbf{tool calls}} &
\makecell{\textbf{Runtime}\\\textbf{(s)}} &
\makecell{\textbf{Evaluation}\\\textbf{score}} \\
\midrule

Run 1 & 14 & 52 & 100 \\
Run 2 & 31 & 70 & 100 \\
Run 3 & 14 & 56 & 100 \\
Run 4 & 15 & 32 & 100 \\
Run 5 & 18 & 60 & 100 \\

\midrule
\textbf{Average} &
$\mathbf{18.4 \pm 7.2}$ &
$\mathbf{54.0 \pm 14.0}$ &
$\mathbf{100 \pm 0}$ \\
\bottomrule
\end{tabularx}
\end{table}

\subsubsection{Ligand exchange in metal complexes}
In this case study, the user prompt was: ``Given Fe\_CO\_6.xyz and Ru\_bpy\_2\_H2O\_2.xyz: (1) For Fe\_CO\_6, replace the two axial CO ligands with water molecules. (2) For Ru\_bpy\_2\_H2O\_2, replace one water ligand with acetonitrile.'' The user specified the input structures, the ligands to be exchanged, and the target coordination positions, while identification of the corresponding atomic indices and execution of the required ligand-exchange operations were handled by the agent.

Because the modifications were localized to the exchanged ligands, the remainder of each parent metal complex was preserved during editing. We therefore evaluated the local coordination geometry rather than performing an RMSD-based comparison of the complete structures. For Fe\_CO\_6, the two axial CO ligands were correctly replaced by H$_2$O in all five runs, with Fe--O distances of $1.98$~\AA. For Ru\_bpy\_2\_H2O\_2, one H$_2$O ligand was correctly replaced by acetonitrile, giving a Ru--N distance of $2.17$~\AA\ in all five runs. These metal--ligand distances were consistent with the bonding distances generated from the corresponding covalent-radius sums used by the binding tool. No steric clashes were observed according to the post-attachment sanitization criteria. All five iterations therefore received full credit (\autoref{tab:per-Case4B}). These examples were intended to demonstrate geometric ligand-exchange
operations rather than to establish the electronic stability of the resulting complexes. The formal oxidation states, spin states, and energetic viability of the generated structures were not evaluated.

\renewcommand{\arraystretch}{1.05}
\begin{table}[H]
\centering
\caption{Per-iteration evaluation results for ``Ligand exchange in metal complexes''.}
\label{tab:per-Case4B}
\small

\begin{tabularx}{\linewidth}{@{}
>{\centering\arraybackslash}p{0.20\linewidth}
>{\centering\arraybackslash}p{0.24\linewidth}
>{\centering\arraybackslash}p{0.24\linewidth}
>{\centering\arraybackslash}p{0.24\linewidth}
@{}}
\toprule
\textbf{Iteration} &
\makecell{\textbf{Number of}\\\textbf{tool calls}} &
\makecell{\textbf{Runtime}\\\textbf{(s)}} &
\makecell{\textbf{Evaluation}\\\textbf{score}} \\
\midrule

Run 1 & 20 & 61 & 100 \\
Run 2 & 21 & 49 & 100 \\
Run 3 & 20 & 72 & 100 \\
Run 4 & 28 & 72 & 100 \\
Run 5 & 18 & 54 & 100 \\

\midrule
\textbf{Average} &
$\mathbf{21.4 \pm 3.8}$ &
$\mathbf{61.6 \pm 10.4}$ &
$\mathbf{100 \pm 0}$ \\
\bottomrule
\end{tabularx}
\end{table}

\subsection{Case study 5: Molecular fragment operations and fragment analysis}
\subsubsection{Isomer and enantiomer conversion}
In this case study, the user prompt was: ``You are given stilbene.xyz and alanine.xyz. (1) Generate a transition-state-like structure and the corresponding photoisomerization product of stilbene.xyz. (2) Generate the enantiomer of alanine.xyz.'' The user specified the input structures and desired transformations, while the specific structural operations required to generate the corresponding isomer and enantiomer were determined by the agent. Because the provided stilbene structure was the \textit{trans} isomer, the expected transition-state-like structure should exhibit a torsion near $90^\circ$, corresponding to an approximately orthogonal arrangement of the two phenyl groups, while the photoisomerization product should adopt the \textit{cis} configuration with a torsion near $0^\circ$.

For alanine, stereochemical SMILES analysis confirmed the correct generation of the enantiomer in all five runs, with the stereochemical label of the chiral carbon changing from \texttt{@@} in the input structure to \texttt{@} in the generated structure.
In Run~2, the H and CH$_3$ were interchanged by rotating them by 180$^\circ$ around the central carbon using the \texttt{rotate\_substituents\_around\_bisector} tool, whereas the other four runs interchanged the NH$_2$ and CH$_3$. In all five runs, the resulting structures were subsequently refined by unconstrained xTB geometry optimization.

In all five runs for stilbene, \estructural{} used \texttt{set\_dihedral\_between\_fragments} to adjust the relevant torsion to approximately $90^\circ$ for the TS-like structure and $0^\circ$ for the \textit{cis} product. The resulting structures were then refined using \texttt{constraint\_xtb}, with the C=C bond length constrained to $1.355$~\AA\ and the corresponding dihedral constrained to $90^\circ$ or $0^\circ$, respectively. After constrained optimization, the TS-like structures exhibited dihedral angles of approximately $87^\circ$, consistent with the expected near-orthogonal geometry along the photoisomerization coordinate. The corresponding \textit{cis} products exhibited dihedral angles ranging from $0.2^\circ$ to $6.2^\circ$ (\autoref{tab:per-Case5A}). Because the alanine enantiomer and the stilbene \textit{trans}-to-\textit{cis} photoisomerization product were correctly generated in all five runs, and the TS-like structures exhibited the expected near-$90^\circ$ torsions, all five iterations received full credit.

\renewcommand{\arraystretch}{1.05}
\setlength{\tabcolsep}{4pt}

\begin{table}[H]
\centering
\caption{Per-iteration evaluation results for ``Isomer and enantiomer conversion''. For stilbene, the TS-like structure and photoisomerization product were evaluated using the C--C=C--C dihedral angle describing the relative orientation of the two benzene rings. Product dihedral angles are reported relative to $0^\circ$ for clarity.}
\label{tab:per-Case5A}
\small

\begin{tabularx}{\linewidth}{@{}
>{\centering\arraybackslash}p{0.12\linewidth}
>{\centering\arraybackslash}p{0.16\linewidth}
>{\centering\arraybackslash}p{0.14\linewidth}
>{\centering\arraybackslash}p{0.18\linewidth}
>{\centering\arraybackslash}p{0.20\linewidth}
>{\centering\arraybackslash}p{0.14\linewidth}
@{}}
\toprule
\textbf{Iteration} &
\makecell{\textbf{Number of}\\\textbf{tool calls}} &
\makecell{\textbf{Runtime}\\\textbf{(s)}} &
\makecell{\textbf{TS-like}\\\textbf{dihedral ($^\circ$)}} &
\makecell{\textbf{Product}\\\textbf{dihedral ($^\circ$)}} &
\makecell{\textbf{Evaluation}\\\textbf{score}} \\
\midrule

Run 1 & 17 & 143 & 87.0 & 0.2 & 100 \\
Run 2 & 20 & 156 & 87.0 & 0.2 & 100 \\
Run 3 & 18 & 151 & 87.0 & 6.2 & 100 \\
Run 4 & 20 & 142 & 87.0 & 6.1 & 100 \\
Run 5 & 19 & 144 & 87.0 & 6.1 & 100 \\

\midrule
\textbf{Average} &
$\mathbf{18.8 \pm 1.3}$ &
$\mathbf{147.2 \pm 6.1}$ &
$\mathbf{87.0 \pm 0.0}$ &
$\mathbf{3.8 \pm 3.2}$ &
$\mathbf{100 \pm 0}$ \\
\bottomrule
\end{tabularx}
\end{table}

\subsubsection{Molecular fragment identification}
In this case study, the user prompt was: ``The provided PDB-code-2FRJ.xyz file contains six distinct fragments. Identify which atoms belong to each fragment (heme group, lysine, histidine, water, NO ligand, and the histidine fragment coordinated to the heme iron). For each fragment, list each atom with its atomic species and atomic index explicitly paired.'' The user specified the identities of the six target fragments, while the corresponding atomic indices and fragment-identification procedure were determined by the agent. The agent analyzed molecular connectivity to identify the atoms belonging to each requested fragment and reported the corresponding atomic species and indices.
In the representative workflow shown in Figure~7b of the main manuscript (Run~1), \estructural{} primarily used Python-based connectivity analysis via \texttt{bash\_kernel}. Tool usage varied across runs, with other runs additionally employing tools such as \texttt{get\_atomic\_neighbors}, \texttt{get\_connected\_subgraph\_indices}, and \texttt{get\_distance\_angle\_dihedral}.

Evaluation was based on whether the atoms belonging to each requested fragment were assigned to the correct atomic indices. All six fragments were correctly identified in all five independent runs, and each run therefore received full credit (\autoref{tab:Case5B}).

\renewcommand{\arraystretch}{1.05}
\begin{table}[H]
\centering
\caption{Per-iteration evaluation results for ``Molecular fragment identification''.}
\label{tab:Case5B}
\small

\begin{tabularx}{\linewidth}{@{}
>{\centering\arraybackslash}p{0.20\linewidth}
>{\centering\arraybackslash}p{0.24\linewidth}
>{\centering\arraybackslash}p{0.24\linewidth}
>{\centering\arraybackslash}p{0.24\linewidth}
@{}}
\toprule
\textbf{Iteration} &
\makecell{\textbf{Number of}\\\textbf{tool calls}} &
\makecell{\textbf{Runtime}\\\textbf{(s)}} &
\makecell{\textbf{Evaluation}\\\textbf{score}} \\
\midrule

Run 1 & 7  & 45 & 100 \\
Run 2 & 30 & 56 & 100 \\
Run 3 & 12 & 61 & 100 \\
Run 4 & 28 & 72 & 100 \\
Run 5 & 11 & 70 & 100 \\

\midrule
\textbf{Average} &
$\mathbf{17.6 \pm 10.6}$ &
$\mathbf{60.8 \pm 11.0}$ &
$\mathbf{100 \pm 0}$ \\
\bottomrule
\end{tabularx}
\end{table}

\subsection{Case study 6: Multimodal reaction mechanism--guided structure generation}
In this case study, the user provided only the reaction scheme shown in \texttt{mechanism.png} together with the prompt: ``Based on the reaction scheme shown in mechanism.png, generate 3D structures corresponding to the reactant, transition-state-like structure, and product.'' Thus, the identities and three-dimensional arrangements of the reactant, TS-like structure, and product, as well as the structure-generation and refinement workflow, were determined by the agent.

For TS-like structure generation, the agent additionally had to identify the reaction centre, determine which interatomic distances and/or angles should be constrained, select physically reasonable constraint values, and assign the appropriate molecular charge and spin multiplicity. This case therefore required not only interpretation of the reaction scheme but also autonomous construction of a chemically meaningful TS initialization strategy.

The generated reactant and product structures were evaluated by heavy-atom RMSD relative to their corresponding GFN2-xTB-optimized geometries. The TS-like structures were instead evaluated by using them as initial geometries for subsequent DFT TS optimization using the B3LYP-D4/def2-SVP level in ORCA 6.0~\cite{beckeDensityfunctionalThermochemistryIII1993,leeDevelopmentColleSalvettiCorrelationenergy1988,stephensInitioCalculationVibrational1994,caldeweyher2017extension_d4correction,caldeweyherGenerallyApplicableAtomiccharge2019,weigendBalancedBasisSets2005,orca6}, and confirming the presence of a single imaginary frequency associated with the resulting transition state.

Across all five runs, the reactant and product structures were first generated using \texttt{build\_tm\_complex\_from\_template}. In Runs~1--4, both structures were subsequently subjected to a short GFN2-xTB optimization as an optional sanitization step, whereas this additional optimization was not performed in Run~5. The TS-like structure was then generated from the reactant geometry by applying appropriate geometric constraints using \texttt{constraint\_xtb}; thus, the optimized reactant geometry was used as the starting point in Runs~1--4, while the unoptimized reactant geometry was used in Run~5. Depending on the run, the agent inspected the generated structures using tools for analyzing atomic connectivity
and geometric parameters, supplemented by visual inspection in some runs.

The application of geometric constraints during these refinement steps varied between runs. The structures illustrated in Figure~8 correspond to Run~2, in which both the reactant and product underwent xTB refinement, with geometric constraints applied to the reactant-like structure but not to the product. All xTB optimizations in this case study were performed with a total charge of $+1$ and singlet state.

Accordingly, Runs~1--4 showed only limited structural relaxation, with heavy-atom RMSDs of $0.15$~\AA\ or less for both the reactant and product. Run~5 exhibited somewhat larger RMSDs of $0.53$~\AA\ and $0.66$~\AA\ for the reactant and product, respectively. These differences reflect modest conformational relaxation associated with the absence of the optional sanitization step rather than a change in molecular connectivity, and the structures were readily relaxed to their corresponding optimized geometries by a short GFN2-xTB optimization.

As summarized in \autoref{tab:gpt54_constraints}, \estructural{} applied a distance
constraint to the forming C1--C2 bond in every run, with target
values ranging from 2.10 to 2.33~\AA{}. In addition, constraints were frequently applied to other interatomic distances associated with the bond rearrangements shown in the reaction schematic, including elongation of the ethylene C0--C1 bond and the Zr--C2 bond to the migrating methyl group, as well as the Zr--C0 interaction involving the ethylene carbon that remains coordinated to Zr. Thus, although the exact set and values of the constraints varied between runs, the key forming C1--C2 bond was consistently constrained, while the additional constraints generally targeted bonds being formed, elongated, or reorganized during ethylene insertion.

The TS-like structures generated in all five runs successfully converged during subsequent DFT transition-state optimization to structures exhibiting a single imaginary frequency of approximately $-300$~cm$^{-1}$ (\autoref{tab:Case6}).
Due to the aforementioned less-sanitized reactant and product geometries, Run~5 received only a minor deduction, while full credit was retained for the TS-like structure.

\renewcommand{\arraystretch}{1.10}
\begin{table}[H]
\centering
\caption{Interatomic distance constraints selected to construct the TS-like geometry in the multimodal reaction-mechanism case. The atomic labels C0, C1, and C2 are illustrated in \autoref{fig:SI_ethylene}: C0 and C1 are the two ethylene carbon atoms, while C2 is the methyl carbon atom. During ethylene insertion, C0 remains coordinated to Zr, while formation of the C1--C2 bond produces the propyl ligand. Constrained distances are given in \AA, and force constants are given in Hartree/Bohr$^2$.}
\label{tab:gpt54_constraints}
\small

\begin{tabularx}{\linewidth}{@{}
>{\raggedright\arraybackslash}p{0.24\linewidth}
*{5}{>{\centering\arraybackslash}X}
@{}}
\toprule
Constrained distance &
Run 1 &
Run 2 &
Run 3 &
Run 4 &
Run 5 \\
\midrule

Zr--C0 & 2.30 & 2.35 & 2.35 & 2.35 & -- \\
Zr--C1 & -- & -- & 2.65 & -- & 2.563 \\
Zr--C2 & 2.35 & 2.25 & 2.70 & 2.40 & 2.475 \\
C0--C1 & 1.40 & 1.42 & 1.40 & 1.42 & 1.335 \\
C0--C2 & -- & -- & -- & -- & -- \\
C1--C2 & 2.10 & 2.20 & 2.20 & 2.20 & 2.33 \\
Force constant & 0.50 & 1.00 & 1.00 & 1.00 & 0.50 \\

\bottomrule
\end{tabularx}
\end{table}

\renewcommand{\arraystretch}{1.05}
\setlength{\tabcolsep}{3pt}

\begin{table}[H]
\centering
\caption{Per-iteration evaluation results for ``Multimodal reaction mechanism--guided structure generation''. Reactant and product RMSDs were calculated for heavy atoms between the generated structures and their corresponding GFN2-xTB-optimized geometries. For the generated initial TS structure, the imaginary frequency obtained after DFT TS optimization is reported.}
\label{tab:Case6}
\small

\begin{tabularx}{\linewidth}{@{}
>{\centering\arraybackslash}p{0.10\linewidth}
>{\centering\arraybackslash}p{0.12\linewidth}
>{\centering\arraybackslash}p{0.11\linewidth}
>{\centering\arraybackslash}p{0.15\linewidth}
>{\centering\arraybackslash}p{0.18\linewidth}
>{\centering\arraybackslash}p{0.15\linewidth}
>{\centering\arraybackslash}p{0.13\linewidth}
@{}}
\toprule

\textbf{Iteration} &
\makecell{\textbf{Number of}\\\textbf{tool calls}} &
\makecell{\textbf{Runtime}\\\textbf{(s)}} &
\makecell{\textbf{Reactant}\\\textbf{RMSD (\AA)}} &
\makecell{\textbf{TS imaginary}\\\textbf{frequency (cm$^{-1}$)}} &
\makecell{\textbf{Product}\\\textbf{RMSD (\AA)}} &
\makecell{\textbf{Evaluation}\\\textbf{score}} \\
\midrule

Run 1 & 67 & 230 & 0.00 & $-301.2$ & 0.00 & 100 \\
Run 2 & 47 & 208 & 0.15 & $-296.1$ & 0.00 & 100 \\
Run 3 & 40 & 288 & 0.15 & $-300.8$ & 0.14 & 100 \\
Run 4 & 37 & 150 & 0.00 & $-301.1$ & 0.00 & 100 \\
Run 5 & 68 & 252 & 0.53 & $-297.2$ & 0.66 & 90 \\

\midrule
\textbf{Average} &
$\mathbf{51.8 \pm 14.8}$ &
$\mathbf{225.6 \pm 51.6}$ &
$\mathbf{0.17 \pm 0.22}$ &
$\mathbf{-299.3 \pm 2.4}$ &
$\mathbf{0.16 \pm 0.29}$ &
$\mathbf{98.0 \pm 4.5}$ \\
\bottomrule
\end{tabularx}
\end{table}

\subsection{Case study 7: Mechanism reasoning-driven geometry construction}\label{SIsec:case7}
In this case study, the user supplied the reaction diagram in \texttt{TS.png} and
specified methyl substitution at the ring carbon alpha to sulfur, hydrogen at the
remaining peripheral ring positions, and methyl as the N-alkyl group. The prompt
further stated that the exchangeable N--H proton could equilibrate between the two
ring nitrogen atoms before N-alkylation, thereby giving two neutral,
compositionally identical starting structures and two possible reaction pathways.
The agent was asked to infer both N-methyl TDSO substrates and their corresponding
pyrazole products, construct their three-dimensional geometries, generate a
transition-state guess for each pathway by directly editing the corresponding
substrate, refine each guess using Sella optTS at the xTB level, and verify the
result by frequency analysis.

The exact full prompt supplied to benchmark \estructural{} agent performance for case 7:

\begin{quote}
\small
Work autonomously and complete every requested subtask. Save all requested structures and reports as clearly named files in the working directory. Do not ask for confirmation. Using \texttt{TS.png}, set the substituent on the ring carbon alpha to sulfur to Me, all other peripheral ring substituents to H, and the $N$-alkyl group to Me. The exchangeable N--H can equilibrate between the two ring nitrogens before $N$-alkylation, producing two neutral, compositionally identical starting structures and two possible reaction pathways. Infer both starting structures, the transformation involved, and the corresponding products. For each pathway, build the substrate and product geometries, generate a transition-state guess by directly editing the substrate geometry, then refine it using Sella \texttt{optTS} at xTB level and verify it by frequency analysis. Report consistent relative energies for both pathways. Save all structures, calculations, energy data, and a concise mechanistic report. Use the following file structure:
\begin{verbatim}
ts-study/
|-- structures/
|   |-- pathway_A_substrate.xyz
|   |-- pathway_A_TS_Initial_Guess.xyz
|   |-- pathway_A_TS_optimized.xyz
|   |-- pathway_A_product.xyz
|   |-- pathway_B_substrate.xyz
|   |-- pathway_B_TS_Initial_Guess.xyz
|   |-- pathway_B_TS_optimized.xyz
|   `-- pathway_B_product.xyz
|-- calculations/
|   |-- pathway_A/
|   `-- pathway_B/
|-- energies.csv
`-- report.md
\end{verbatim}
\end{quote}

Five independent runs were performed using GPT-5.4 with maximum reasoning effort.

This task required several coupled forms of chemical reasoning. The agent first had
to interpret the six-membered TDSO scaffold and its substitution pattern from the
reaction image, distinguish the two N-methyl regioisomers produced after proton
exchange, and map these substrates onto the corresponding 1,5- and
1,3-dimethylpyrazole products. It then had to identify the concerted SO-extrusion
coordinate, edit the substrate geometry toward simultaneous C--N bond formation and
S--C/S--N bond cleavage, and determine whether the Sella-refined structures were
first-order saddle points. The reported reaction barriers were not used in the
evaluation; scoring emphasized the identities and geometries of the substrates and
products, the chemical relevance of the proposed TS structures, and frequency-based
TS verification.

All deposited structures were re-examined directly from the \texttt{xyz} files
(connectivity from interatomic distances, heavy-atom RMSD after exhaustive atom
mapping with mirror images allowed) and from the raw xTB/ASE frequency outputs in
the run directories, rather than from the agents' own reports. The two Sella-refined
saddle points of Run~1 serve as the reference TS geometries.

Run~1 produced a fully correct result. Both substrate and product pairs were assigned correctly, the TS guesses were built by relaxing the substrate under four xTB distance constraints (forming N--C 1.9~\AA, breaking S--N and S--C 2.1~\AA{} each, and retained S--O 1.48~\AA) and lay within 0.1--0.2~\AA{} of the final saddles, and the two optimized TS structures exhibited one significant imaginary frequency ($-433.0$ and $-389.3$~cm$^{-1}$) whose displacement followed the intended ring-contraction coordinate.

Run~2 built both substrates correctly, constructed TS guesses by the same constrained
relaxation strategy, and located the same two saddle points as Run~1 (heavy-atom RMSD
0.02 and 0.04~\AA; $-429.5$ and $-379.2$~cm$^{-1}$). It failed at the product stage:
its report argued that the 1,3- and 1,5-dimethylpyrazole regioisomers are
constitutionally equivalent, and the geometry it saved for both pathways is in fact
1,4-dimethylpyrazole, which is the product of neither pathway. Run~3 failed at the
reaction-identification stage. It replaced the TDSO substrates with methylpyrazole
tautomers and diazomethane and modelled N-methylation with \ce{N2} extrusion;
consequently, its substrate, product, and TS structures all described a different
reaction and had the wrong elemental composition (\ce{C5H8N4}).

Runs~4 and~5 inferred the intended reaction and constructed all four endpoints
correctly, but both failed when translating the mechanism into a TS guess by direct
editing of the substrate. In Run~4, the pathway-A guess used chemically reasonable
target distances (N--C 1.94, S--N 2.30, S--C 2.37~\AA) but displaced the S(O) unit
toward the two methyl groups rather than away from the ring, leaving sulfur 2.3~\AA{}
from both methyl carbons (heavy-atom RMSD 0.98~\AA{} from the reference saddle). The
subsequent constrained relaxation collapsed this geometry (S--CH$_3$ 1.68~\AA{} and an
\ce{H2} molecule at 0.78~\AA), and Sella then located a proton-transfer saddle rather
than the ring-contraction saddle: in the deposited TS the S--N bond is fully broken
(3.38~\AA), the transferring hydrogen sits 1.48~\AA{} from both carbon and sulfur, and
the $-1270.4$~cm$^{-1}$ mode is dominated by this hydrogen. The pathway-B guess had
plausible reactive distances but a misoriented SO unit (RMSD 0.52~\AA); Sella
recovered the reference saddle ($-381.5$~cm$^{-1}$, RMSD 0.03~\AA). The report
nevertheless described both structures as the intended first-order saddles, so the
presence of one imaginary frequency did not by itself establish mechanistic
correctness.

In Run~5, the agent performed a constrained relaxation with appropriate targets
(N--C 1.95, S--N 2.10, S--C 2.10~\AA) that converged to a sound guess, but did not
use it. Instead, it saved and passed to Sella a hand-edited geometry in which the ring
atoms N3 and C8 had been pushed together without their methyl substituents, so the
spectator N--Me and ring-C--Me bonds were stretched from 1.44 to 1.77~\AA{} and from
1.49 to 1.81~\AA{} (pathway~A; ring-C--Me from 1.48 to 1.78~\AA{} in pathway~B),
giving guesses 1.02 and 0.52~\AA{} from the reference saddles. Because the system is
small and rigid, Sella still recovered both reference saddle points (RMSD 0.02 and
0.04~\AA{} from the Run~1 structures; electronic energies within
0.15~kcal\,mol$^{-1}$), albeit at a residual force of 0.08--0.10~eV\,\AA$^{-1}$
against the requested 0.05. The frequency verification then failed for a
software-handling reason: the agent's ASE script exported the frequencies by casting
complex values to real, which silently set the imaginary modes to zero. The agent read
this as ``0 imaginary frequencies'' and reported that no transition state had been
validated, although the ASE summary written in the same directory lists $427.5i$ and
$385.9i$~cm$^{-1}$. The deposited TS geometries are therefore correct, but the TS
guesses, the verification, and the mechanistic conclusion are not.

Scoring follows the rubric in \autoref{tab:case7-rubric}, which is result-oriented
on a 0--100 scale and assigns points per deliverable so that every entry can be
checked against the deposited files. Permutation of the pathway labels is allowed.
Tool-call count and runtime are reported as efficiency diagnostics but do not
contribute to the score, and the reported reaction barriers were not scored. Per-run
results are given in \autoref{tab:case7-evaluation}.

\renewcommand{\arraystretch}{1.05}
\setlength{\tabcolsep}{3pt}

\begin{table}[H]
\centering
\caption{Scoring rubric for ``Mechanism reasoning-driven geometry construction''.
RMSD denotes the heavy-atom RMSD after optimal atom mapping (mirror images allowed).
The reference saddles are the Run~1 Sella-refined TS structures, whose imaginary modes
follow the SO-extrusion coordinate.}
\label{tab:case7-rubric}
\small
\begin{tabularx}{\linewidth}{@{}
>{\raggedright\arraybackslash}p{0.32\linewidth}
>{\centering\arraybackslash}p{0.10\linewidth}
>{\raggedright\arraybackslash}X
@{}}
\toprule
\textbf{Item} & \textbf{Points} & \textbf{Criterion} \\
\midrule
Substrate connectivity, pathways A and B & $2\times10$ &
Bond graph matches the N-methyl TDSO regioisomer (\ce{C5H8N2OS}; N--Me adjacent to or
distal from S) \\
Product connectivity, pathways A and B & $2\times10$ &
1,5- and 1,3-dimethylpyrazole, respectively; duplicates or other regioisomers receive
no credit \\
TS initial guess, pathways A and B & $2\times5$ &
RMSD $<0.3$~\AA{} to the reference saddle and no spectator bond changed by more than
0.1~\AA{} relative to the optimized substrate \\
Sella-refined TS, pathways A and B & $2\times10$ &
RMSD $<0.1$~\AA{} to the reference saddle \\
Frequency verification, pathways A and B & $2\times10$ &
Exactly one significant imaginary frequency for that saddle is both computed and
correctly reported \\
Mechanistic report & 10 &
Assignments consistent with the deposited structures (10); one major conclusion
contradicted by the structures (5); wrong reaction (0) \\
\midrule
\textbf{Total} & \textbf{100} & \\
\bottomrule
\end{tabularx}
\end{table}

\begin{table}[H]
\centering
\caption{Per-iteration evaluation results for ``Mechanism reasoning-driven geometry
construction'' under the rubric in \autoref{tab:case7-rubric}. Guess RMSD is the
heavy-atom RMSD of the deposited TS initial guess to the reference saddle. Imaginary
frequencies are those in the xTB/ASE frequency outputs of the deposited TS candidates.
Run~4 pathway~A is a first-order saddle for an unintended proton-transfer coordinate.
Run~5 located both reference saddles, but its report states zero imaginary frequencies
because its frequency export discarded the imaginary components. ``Wrong reaction''
indicates that the run modelled N-methylation rather than the requested SO-extrusion
pathways.}
\label{tab:case7-evaluation}
\small

\begin{tabularx}{\linewidth}{@{}
>{\centering\arraybackslash}p{0.09\linewidth}
>{\centering\arraybackslash}p{0.10\linewidth}
>{\centering\arraybackslash}p{0.09\linewidth}
>{\centering\arraybackslash}p{0.14\linewidth}
>{\centering\arraybackslash}p{0.17\linewidth}
>{\centering\arraybackslash}p{0.17\linewidth}
>{\centering\arraybackslash}p{0.11\linewidth}
@{}}
\toprule

\textbf{Iteration} &
\makecell{\textbf{Number of}\\\textbf{tool calls}} &
\makecell{\textbf{Runtime}\\\textbf{(s)}} &
\makecell{\textbf{Guess RMSD}\\\textbf{A / B (\AA)}} &
\makecell{\textbf{Pathway-A TS}\\\textbf{imaginary frequency}\\\textbf{(cm$^{-1}$)}} &
\makecell{\textbf{Pathway-B TS}\\\textbf{imaginary frequency}\\\textbf{(cm$^{-1}$)}} &
\makecell{\textbf{Evaluation}\\\textbf{score}} \\
\midrule

Run 1 & 82  & 2316 & 0.09 / 0.19 & $-433.0$  & $-389.3$ & 100 \\
Run 2 & 121 & 2695 & 0.11 / 0.18 & $-429.5$  & $-379.2$ & 75 \\
Run 3 & 68  & 2209 & \multicolumn{3}{c}{wrong reaction} & 0 \\
Run 4 & 123 & 2623 & 0.98 / 0.52 & $-1270.4$ & $-381.5$ & 65 \\
Run 5 & 120 & 2985 & 1.02 / 0.52 &
\makecell{$-427.5$\\(reported as 0)} & \makecell{$-385.9$\\(reported as 0)} & 65 \\

\midrule
\textbf{Average} &
$\mathbf{102.8 \pm 26}$ &
$\mathbf{2565.6 \pm 311}$ &
\multicolumn{3}{c}{\textbf{not averaged across failed outcomes}} &
$\mathbf{61.0 \pm 37.0}$ \\
\bottomrule
\end{tabularx}
\end{table}

\subsubsection{Endpoint-only vision baselines}\label{SIsec:case7_endpoint}
To isolate multimodal scaffold interpretation from transition-state optimization,
we additionally evaluated Claude Opus~4.5 (high reasoning effort) using an
endpoint-only version of the task. These runs
received the same reaction image and substitution instructions but were asked to
construct only the two substrates and two products; Sella optimization, frequency
analysis, and energy calculations were omitted. Each of the four reference endpoint
structures contributed 25 points to the evaluation score. Because the pathway labels
were not defined independently of the structures in the user prompt, permutation of
the A/B labels was allowed, but duplicate or constitutionally different structures
did not receive credit.

The exact prompt supplied to the endpoint-only baseline model was:

\begin{quote}
\small
Work autonomously and complete every requested subtask. Save all requested
structures and reports as clearly named files in the working directory. Do not ask
for confirmation. Using \texttt{TS.png}, set the substituent on the ring carbon
alpha to sulfur to Me, all other peripheral ring substituents to H, and the N-alkyl
group to Me. The exchangeable N--H can equilibrate between the two ring nitrogens
before N-alkylation, producing two neutral, compositionally identical starting
structures and two possible reaction pathways. Infer both starting structures, the
transformation involved, and the corresponding products. For each pathway, build
the substrate and product geometries. Save a concise mechanistic report and the
structures as \texttt{ts-study/structures/pathway\_A\_substrate.xyz},
\texttt{ts-study/structures/pathway\_A\_product.xyz},
\texttt{ts-study/structures/pathway\_B\_substrate.xyz}, and
\texttt{ts-study/structures/pathway\_B\_product.xyz}. The dependency files are
already available in your working directory: \texttt{TS.png}.
\end{quote}

Opus~4.5 completed all five runs, but none produced either correct substrate. The
reference substrates are six-membered TDSO structures with composition
\ce{C5H8N2OS}; instead, four runs produced \ce{C4}-containing five-membered
TDSO-like rings and Run~4 produced a \ce{C3}-containing analogue. Only Run~4
generated the correct pair of dimethylpyrazole product regioisomers, with the pathway
labels interchanged. Runs~1, 2, and~5 recovered one of the two required product
connectivities, together with either a duplicate (Runs~1 and~5) or the wrong
1,4-dimethylpyrazole regioisomer (Run~2). Run~3 saved 1,4-dimethylpyrazole for both
pathways, although its report labels them 1,3- and 1,5-dimethylpyrazole, and therefore
received no product credit.

\renewcommand{\arraystretch}{1.05}
\setlength{\tabcolsep}{3pt}
\begin{table}[H]
\centering
\caption{Endpoint-only Case~7 results for Claude Opus~4.5. ``Correct pair''
denotes recovery of both reference product connectivities after allowing permutation
of the pathway labels; product identities were verified from the deposited
\texttt{xyz} connectivity rather than from the agent's report.}
\label{tab:case7-opus45-endpoints}
\small
\begin{tabularx}{\linewidth}{@{}
>{\centering\arraybackslash}p{0.10\linewidth}
>{\centering\arraybackslash}p{0.14\linewidth}
>{\centering\arraybackslash}p{0.13\linewidth}
>{\centering\arraybackslash}p{0.21\linewidth}
>{\centering\arraybackslash}p{0.21\linewidth}
>{\centering\arraybackslash}p{0.13\linewidth}
@{}}
\toprule
\textbf{Iteration} &
\makecell{\textbf{Number of}\\\textbf{tool calls}} &
\makecell{\textbf{Runtime}\\\textbf{(s)}} &
\makecell{\textbf{Substrate}\\\textbf{outcome}} &
\makecell{\textbf{Product}\\\textbf{outcome}} &
\makecell{\textbf{Evaluation}\\\textbf{score}} \\
\midrule
Run 1 & 37 & 655 & Incorrect (\ce{C4}) & One of two & 25 \\
Run 2 & 31 & 765 & Incorrect (\ce{C4}) & One of two & 25 \\
Run 3 & 40 & 557 & Incorrect (\ce{C4}) & Neither (1,4-isomer twice) & 0 \\
Run 4 & 47 & 949 & Incorrect (\ce{C3}) & Correct pair & 50 \\
Run 5 & 44 & 833 & Incorrect (\ce{C4}) & One of two & 25 \\
\midrule
\textbf{Average} &
$\mathbf{39.8 \pm 6.2}$ &
$\mathbf{752.0 \pm 152.4}$ &
\multicolumn{2}{c}{\textbf{0/5 correct substrate pairs}} &
$\mathbf{25.0 \pm 17.7}$ \\
\bottomrule
\end{tabularx}
\end{table}

Inspection of the agent traces identified image interpretation as the dominant
failure mode. Opus~4.5 consistently described the substrate immediately after
viewing \texttt{TS.png} as a five-membered 1,2,3-thiadiazole oxide with an
\ce{S-N-N-C-C} ring. All subsequent SMILES generation and geometry editing therefore
started from a scaffold missing one carbon, even though the model generally inferred
SO extrusion and pyrazole formation correctly. Because these failures occurred in the
endpoint-only experiment, before any TS search was requested, they are attributable
primarily to incorrect visual parsing and chemical identification of the substrate
rather than to Sella optimization. This behaviour is consistent with the broader
model-dependent trend in the vision-only spatial benchmark
(Section~\ref{sec3:VLM} and \autoref{fig:VLM_Bench}), in which Opus~4.5
becomes less reliable as the amount of molecular visual information increases. The
two evaluations probe different inputs---rendered three-dimensional structures in
the spatial benchmark and a two-dimensional reaction diagram here---but both expose
visual molecular parsing as a bottleneck that precedes downstream geometry
operations. The substantially more size-robust performance of GPT-5.4 in the spatial
benchmark further suggests that continuing improvements in the underlying VLM may
directly reduce this class of scaffold-identification failure.

\subsubsection{DFT refinement of the Run~1 pathways}\label{Case7_refinement}
In the representative run (Run~1), \estructural{} reported the relative electronic energies of the reactants, transition states, and products along both reaction pathways at the GFN2-xTB level (\autoref{tab:case7-energies}). For the separated products (pyrazole and free SO), the agent evaluated and reported both singlet and triplet states of SO, with singlet SO predicted to be lower in energy. In its final report, \estructural{} acknowledged that isolated SO experimentally has a triplet ground state but used singlet SO for the reported reaction profiles to maintain consistency with the closed-shell singlet surface employed in the TS searches.

This discrepancy in the SO spin-state energetics is consistent with a known limitation of conventional GFN2-xTB, whose spin-independent energy expression systematically favours lower-spin states over higher-spin states~\cite{neugebauer2023high}. Although a spin-polarized variant (spGFN2-xTB) has been developed~\cite{neugebauer2023high}, it was not implemented in our \texttt{constraint\_xtb} tool, which was primarily designed for geometry refinement. Consequently, although \estructural{} successfully executed the requested calculations, the resulting GFN2-xTB reaction profiles placed the SO-containing products at higher relative energies.

The structures generated by \estructural{} were subsequently used as initial geometries for DFT geometry and TS optimizations at the B3LYP-D4/def2-SVP level. The reactant and product structures converged to local minima with no imaginary frequencies, while the optimized transition states exhibited a single imaginary frequency of $-434.9$ and $-404.3$~cm$^{-1}$ for Paths~A and B, respectively. The resulting DFT Gibbs free energies were used to construct the reaction energy profiles shown in Figure~9f of the main manuscript, and the corresponding electronic and Gibbs free energies are summarized in \autoref{tab:case7-energies}.

\begin{table}[H]
\centering
\caption{Relative energies (kcal~mol$^{-1}$) of the reactants, transition states, and products along the Run~1 reaction pathways, with the Path~A reactant as the common energy reference. GFN2-xTB electronic energies were reported by \estructural{}, whereas DFT energies were obtained from subsequent calculations at the B3LYP-D4/def2-SVP level. For the GFN2-xTB product energies, singlet and triplet SO are reported separately. DFT product energies include triplet SO. Gibbs free energies were calculated at 298.15~K and 1~atm.}
\label{tab:case7-energies}

\small
\setlength{\tabcolsep}{10pt}
\renewcommand{\arraystretch}{1.15}

\begin{tabular}{lccc}
\toprule
Species
& $\Delta E$ (GFN2-xTB)
& $\Delta E$ (DFT)
& $\Delta G$ (DFT) \\
\midrule
Path A reactant             & 0.0  & 0.0   & 0.0   \\
Path A TS                   & 23.0 & 32.8  & 31.5  \\
Path A product (singlet SO) & 16.7 & --   & --   \\
Path A product (triplet SO) & 18.4 & $-16.0$ & $-28.8$ \\
\addlinespace[3pt]
Path B reactant             & 10.1 & 12.8  & 13.0  \\
Path B TS                   & 20.5 & 29.3  & 28.2  \\
Path B product (singlet SO) & 17.5 & --   & --   \\
Path B product (triplet SO) & 19.1 & $-16.0$ & $-29.1$ \\
\bottomrule
\end{tabular}
\end{table}

\renewcommand{\arraystretch}{0.95}
\begin{table}[H]
\centering
\caption{Average evaluation scores (out of 100) across five independent runs for all case studies. Values are reported as mean $\pm$ standard deviation. Per-run results are provided in Supporting Information Section~2.}
\label{tab:all_case_scores}
\small

\begin{tabularx}{\linewidth}{@{}
>{\raggedright\arraybackslash}X|
>{\centering\arraybackslash}p{0.22\linewidth}
@{}}
\toprule
\textbf{Case study} &
\textbf{Average score} \\
\midrule

Selective amine protection in spermidine
& $100 \pm 0$ \\

Symmetry-guided Co-porphyrin editing
& $100 \pm 0$ \\

Binding of reaction intermediates and ligands
& $100 \pm 0$ \\

Selective isomer generation
& $100 \pm 0$ \\

Generation of zirconocene complexes
& $100 \pm 0$ \\

Construction of concave and convex CpRu($\eta^{6}$-sumanene)$^{+}$ complexes
& $92.0 \pm 4.5$ \\

Substitution of PNP tert-butyl groups
& $100 \pm 0$ \\

Ligand exchange in metal complexes
& $100 \pm 0$ \\

Isomer and enantiomer conversion
& $100 \pm 0$ \\

Molecular fragment identification
& $100 \pm 0$ \\

Multimodal reaction mechanism--guided structure generation
& $98.0 \pm 4.5$ \\

Mechanism reasoning-driven geometry construction
& $61.0 \pm 37.0$ \\

\bottomrule
\end{tabularx}
\end{table}

\clearpage
\subsection{Failure modes and cases requiring user guidance}
\begin{figure}[!ht]
        \centering        \includegraphics[width=1.0\textwidth]{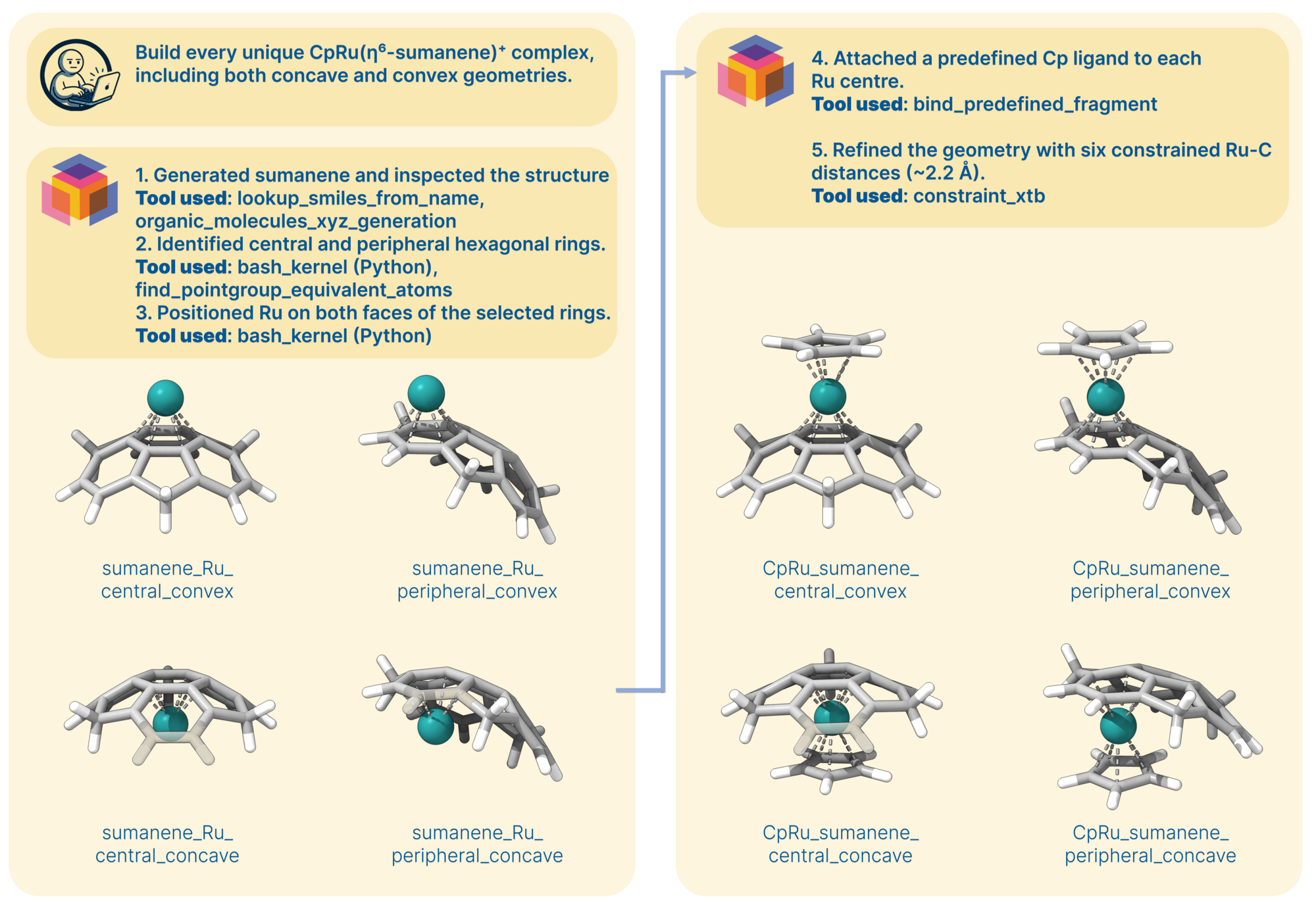}
        \caption{Generation of all unique CpRu($\eta^{6}$-sumanene)$^{+}$ configurations using an explicit configuration-coverage prompt. The agent identified both central and peripheral $\eta^{6}$ coordination sites and generated concave and convex geometries for each, yielding four distinct complexes. All four configurations were successfully generated in each of five independent runs. Atom colour codes: teal, Ru; grey, C; white, H.}        \label{fig:SI_sumanene_generic}
\end{figure}

\renewcommand{\arraystretch}{1.05}
\setlength{\tabcolsep}{4pt}

\begin{table}[H]
\centering
\caption{Per-iteration heavy-atom RMSDs for the four CpRu($\eta^{6}$-sumanene)$^{+}$ configurations generated using the prompt ``Build every unique CpRu($\eta^{6}$-sumanene)$^{+}$ complex, including both concave and convex geometries.'' RMSDs were calculated between the generated structures and their corresponding GFN2-xTB-optimized geometries.}
\label{tab:sumanene_every}
\small

\begin{tabularx}{\linewidth}{@{}
>{\centering\arraybackslash}p{0.14\linewidth}
>{\centering\arraybackslash}p{0.20\linewidth}
>{\centering\arraybackslash}p{0.20\linewidth}
>{\centering\arraybackslash}p{0.20\linewidth}
>{\centering\arraybackslash}p{0.20\linewidth}
@{}}
\toprule

\textbf{Iteration} &
\makecell{\textbf{Central concave}\\\textbf{RMSD (\AA)}} &
\makecell{\textbf{Central convex}\\\textbf{RMSD (\AA)}} &
\makecell{\textbf{Peripheral concave}\\\textbf{RMSD (\AA)}} &
\makecell{\textbf{Peripheral convex}\\\textbf{RMSD (\AA)}} \\
\midrule

Run 1 & 0.03 & 0.04 & 0.06 & 0.05 \\
Run 2 & 0.07 & 0.07 & 0.11 & 0.08 \\
Run 3 & 0.07 & 0.07 & 0.11 & 0.08 \\
Run 4 & 0.14 & 0.10 & 0.30 & 0.12 \\
Run 5 & 0.14 & 0.10 & 0.29 & 0.12 \\

\midrule
\textbf{Average} &
$\mathbf{0.09 \pm 0.05}$ &
$\mathbf{0.08 \pm 0.03}$ &
$\mathbf{0.17 \pm 0.11}$ &
$\mathbf{0.09 \pm 0.03}$ \\
\bottomrule
\end{tabularx}
\end{table}

To further examine the configuration-selection failure mode, we tested a more explicit but still generic prompt: ``Build every unique CpRu($\eta^{6}$-sumanene)$^{+}$ complex, including both concave and convex geometries.'' In this formulation, the user did not specify the individual coordination sites or atomic indices, but explicitly requested complete coverage of the unique configurations. The agent therefore had to infer that both the central and peripheral six-membered rings should be considered, together with concave and convex coordination at each site.

Across all five runs, \estructural{} successfully generated all four configurations: central-concave, central-convex, peripheral-concave, and peripheral-convex (\autoref{fig:SI_sumanene_generic}).
In Run~1, illustrated in \autoref{fig:SI_sumanene_generic}, the agent first positioned Ru atoms above the selected rings and subsequently attached Cp ligands. In Runs~2--5, the agent instead constructed CpRu fragments and positioned them on the selected rings, following the approach used in the original case study. 
In Runs~1--3, additional geometry refinement was performed using \texttt{constraint\_xtb} with a charge of $+1$ and \texttt{uhf=0}, imposing Ru--C distance constraints on the coordinated C atoms.
Heavy-atom RMSDs relative to the corresponding GFN2-xTB-optimized structures remained below $0.5$~\AA\ for all generated configurations and all five runs (\autoref{tab:sumanene_every}).

These results suggest that the primary difficulty in the initial task was not the underlying structure-generation capability, but the ambiguity in selecting among multiple chemically plausible coordination configurations. When the prompt explicitly requested every unique configuration, the agent consistently enumerated and generated the complete set without requiring further human correction.

\begin{figure}[ht]
        \centering        \includegraphics[width=1.0\textwidth]{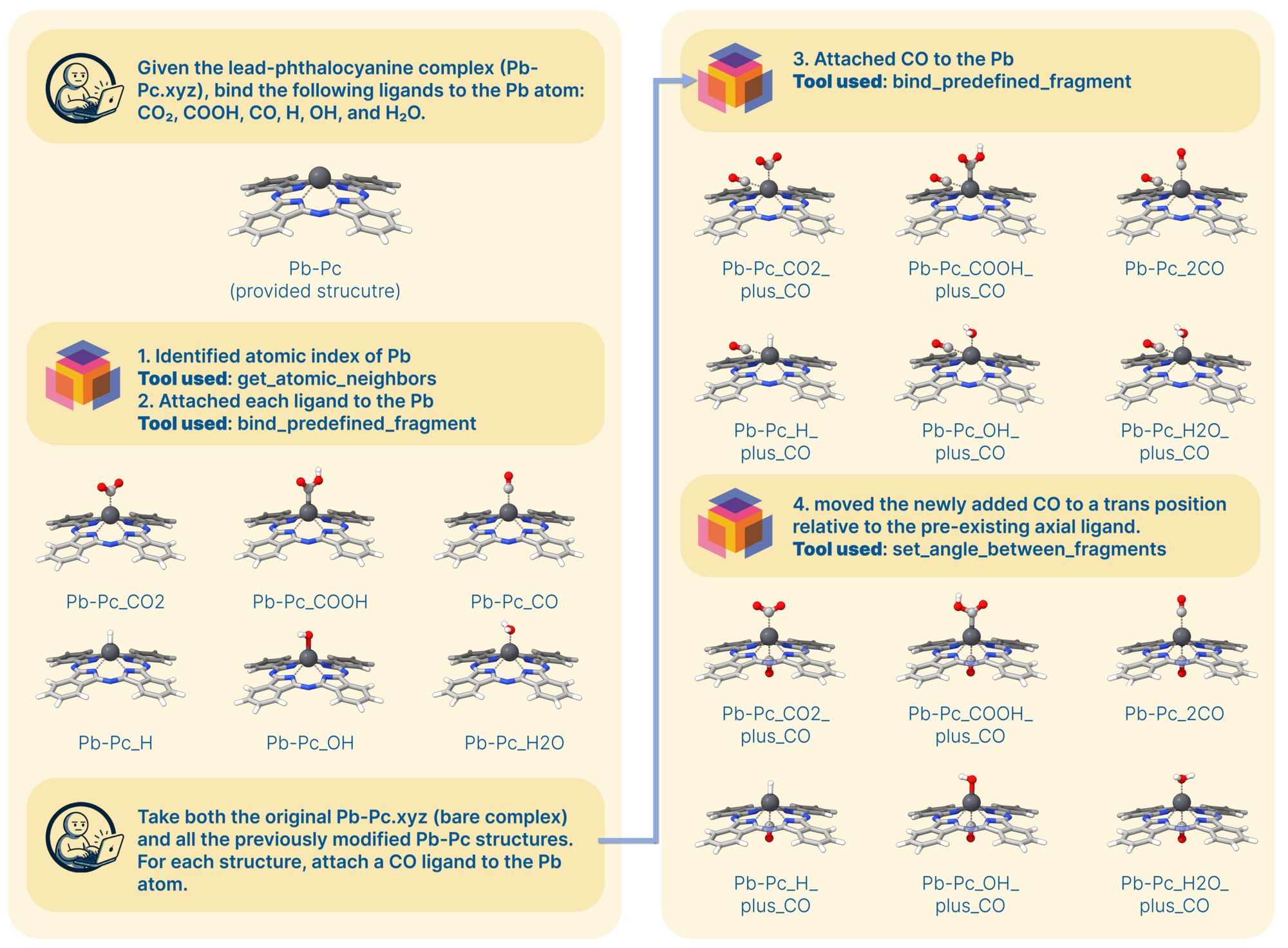}
        \caption{Configuration-dependent axial ligand binding to nonplanar Pb-phthalocyanine. The first ligand was consistently placed on the convex side according to the geometry-selection rules of the binding tool. Upon sequential CO binding, the initial placement was identical across runs, but in two of five runs the agent further moved the newly added CO to an approximately trans position on the opposite, concave side. This example illustrates the increased configuration-selection ambiguity encountered for nonplanar coordination environments. Atom colour codes: dark grey, Pb;  red, O; blue, N; grey, C; white, H.}        \label{fig:SI_PbPc}
\end{figure}

A second configuration-selection challenge was observed for ligand binding to the nonplanar Pb-phthalocyanine (Pb-Pc) complex. Unlike planar Co-Pc, Pb-Pc presents inequivalent concave and convex sides. The ligand-binding tool first determines a binding direction according to the local VSEPR-like coordination environment; when no predefined geometry is applicable, a spherical search is used to identify the least sterically congested direction (Section \ref{SI_Note:editing}). As a result, the first axial ligand was consistently placed on the convex side of Pb-Pc in all runs.

When an additional CO ligand was subsequently introduced, the binding tool again placed the ligand in the least sterically congested accessible direction, which consistently corresponded to the convex side of Pb-Pc across all five runs. However, the agent differed in how it interpreted the resulting two-ligand configuration. In three of the five runs, the initially generated geometry was retained, whereas in the remaining two runs the agent further adjusted the newly added CO using \texttt{set\_angle\_between\_fragments} to place it approximately trans to the pre-existing axial ligand, resulting in coordination on the opposite, concave side of Pb-Pc (\autoref{fig:SI_PbPc}).

This behaviour illustrates a common failure mode in tasks with multiple chemically plausible configurations. The underlying binding operation itself remained reproducible, but the subsequent selection or refinement of the preferred configuration depended on the agent's interpretation of the coordination environment. In such cases, additional user guidance or a more explicit specification of the desired configuration can improve consistency across independent runs.

\subsection{Tool ablation study}\label{SIsec:ablation}
To evaluate the contribution of the dedicated molecular manipulation tools, we repeated representative Co-porphyrin editing tasks with the structure-editing tools disabled. The agent retained access to \texttt{bash\_kernel}, through which it could execute Python code using commonly available cheminformatics and atomistic libraries, including RDKit~\cite{rdkit}, Open Babel~\cite{obabel}, ASE~\cite{ase-paper}, pymatgen~\cite{pymatgen}, and NumPy, to attempt the requested modifications.

\begin{figure}[t!]
        \centering        \includegraphics[width=1.0\textwidth]{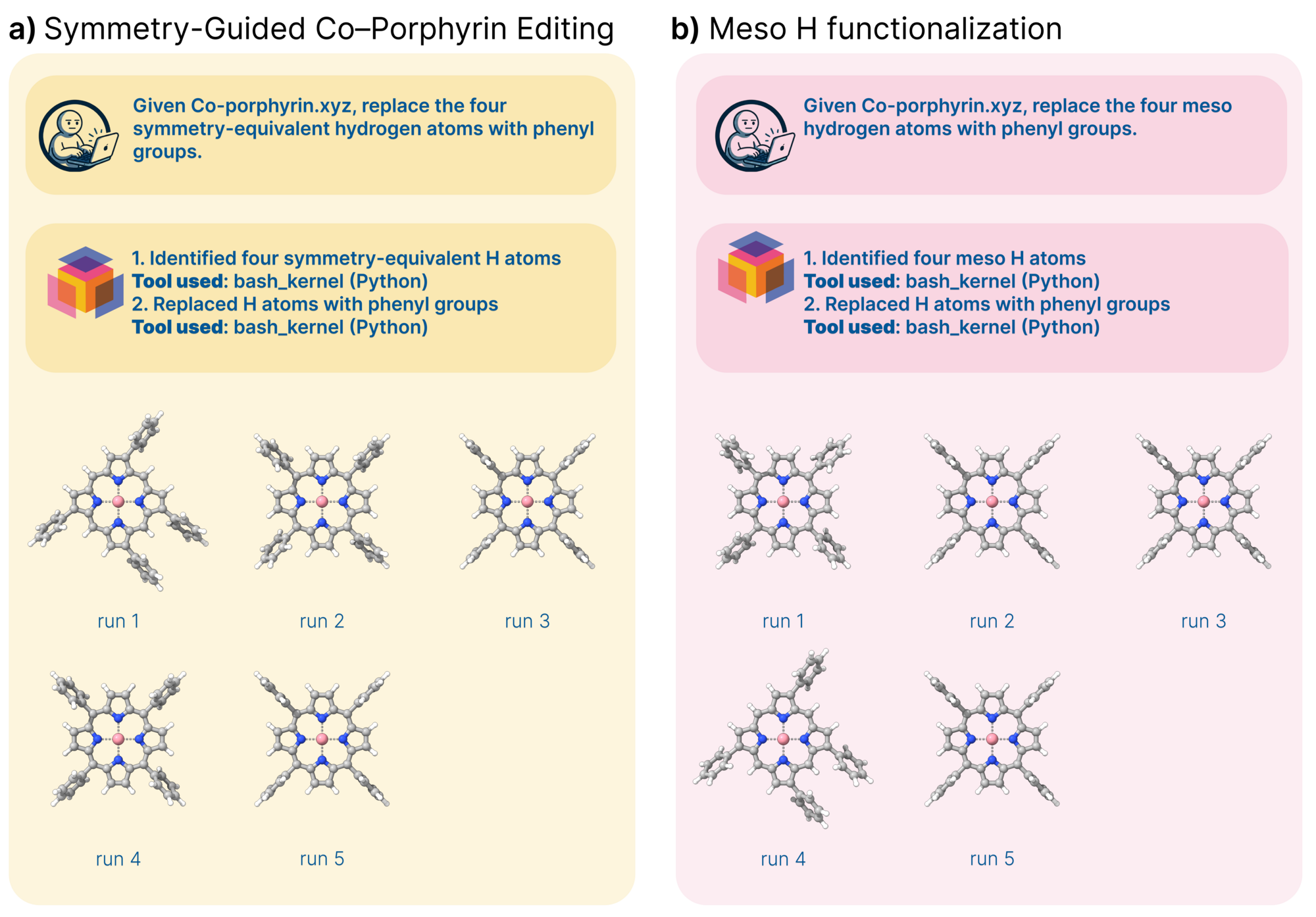}
        \caption{Tool-ablation results for Co-porphyrin phenyl substitution. 
(a) Replacement of four symmetry-equivalent hydrogen atoms and 
(b) replacement of four meso hydrogen atoms with phenyl groups using only Python-based operations without the dedicated molecular analysis and editing tools. Although phenyl fragments were successfully introduced in most cases, incorrect target-site selection was observed in individual runs (run 1 in panel (a) and run 4 in panel (b)), illustrating reduced editing reliability in the ablated setting. 
The two steps denote distinct reasoning stages, although both are performed through \texttt{bash\_kernel} using Python in the tool-ablation setting. Atom colour codes: pink, Co; blue, N; grey, C; white, H.}        \label{fig:SI_ablation_editing}
\end{figure}

To examine the contributions of both the structural-analysis and editing tools, we selected the Co-porphyrin functionalization case, in which these capabilities are both central to the workflow. We evaluated two tasks: replacement of the four symmetry-equivalent hydrogen atoms with phenyl groups and replacement of the four meso hydrogen atoms with phenyl groups, using the same prompts as in the original case study (\autoref{fig:SI_ablation_editing}). In the tool-ablation setting, the agent generally succeeded in constructing and attaching the phenyl fragments, but target-site identification was less reliable. In each task, one of the five runs produced a structure in which the intended meso positions were not correctly selected, despite successful phenyl attachment itself.

The absence of dedicated tools also increased the execution time. For the symmetry-guided prompt, the average runtime increased from $15.2 \pm 3.2$~s with the dedicated tools to $91.6 \pm 41.9$~s without them. For the meso-H prompt, the corresponding runtime increased from $40.6 \pm 16.6$~s to $65.6 \pm 20.4$~s (\autoref{tab:ablation_CoPor}). Thus, although the agent could partially reproduce the editing workflow by generating Python code directly, the dedicated tools provided both more reliable site-specific editing and substantially greater execution efficiency.

\renewcommand{\arraystretch}{1.05}
\begin{table}[H]
\centering
\caption{Runtime comparison between the tool-ablation and tool-enabled workflows for Co-porphyrin phenyl substitution. Values are reported in seconds.}
\label{tab:ablation_CoPor}
\small

\begin{tabular*}{\linewidth}{@{\extracolsep{\fill}}ccccc@{}}
\toprule
&
\multicolumn{2}{c}{\textbf{Symmetry-guided}} &
\multicolumn{2}{c}{\textbf{Meso-H}} \\
\cmidrule(lr){2-3}
\cmidrule(lr){4-5}

\textbf{Iteration} &
\makecell{\textbf{No dedicated}\\\textbf{tools}} &
\makecell{\textbf{With dedicated}\\\textbf{tools}} &
\makecell{\textbf{No dedicated}\\\textbf{tools}} &
\makecell{\textbf{With dedicated}\\\textbf{tools}} \\
\midrule

Run 1 & 111 & 19 & 70 & 31 \\
Run 2 & 61  & 11 & 55 & 62 \\
Run 3 & 38  & 13 & 41 & 28 \\
Run 4 & 105 & 16 & 96 & 55 \\
Run 5 & 143 & 17 & 66 & 27 \\

\midrule
\textbf{Average} &
$\mathbf{91.6 \pm 41.9}$ &
$\mathbf{15.2 \pm 3.2}$ &
$\mathbf{65.6 \pm 20.4}$ &
$\mathbf{40.6 \pm 16.6}$ \\
\bottomrule
\end{tabular*}
\end{table}

\begin{figure}[!htbp]
        \centering        \includegraphics[width=1.0\textwidth]{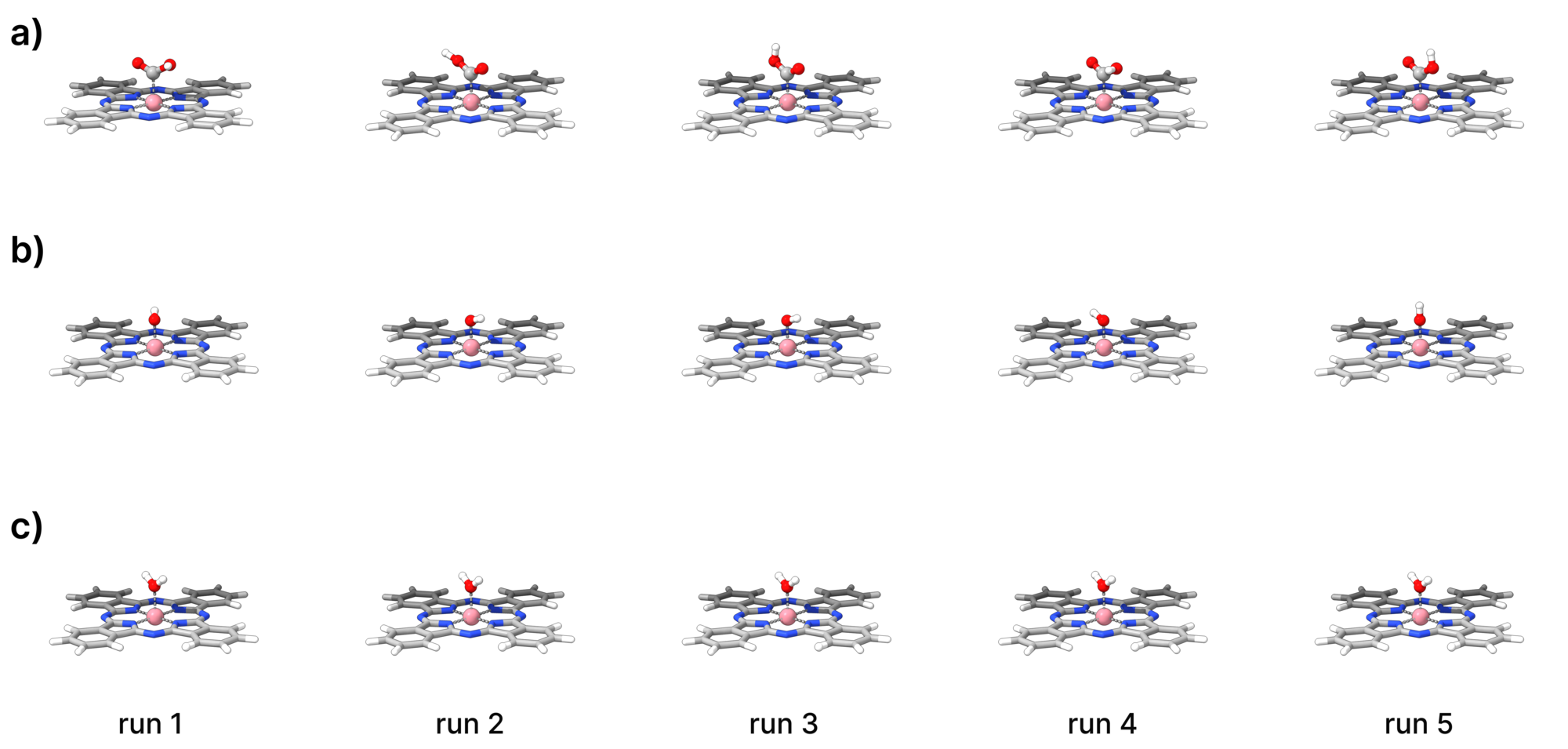}
        \caption{Tool-ablation results for ligand binding to Co-phthalocyanine. 
(a) COOH-, (b) OH-, and (c) H$_2$O-bound structures generated without the dedicated molecular binding tools over five independent runs. 
Although the intended ligands were attached to the Co centre, geometric issues were observed in each case, including nonplanar COOH groups in (a), a nearly linear Co–O–H geometry in Run 5 of (b), and approximately trigonal-planar local geometries around the coordinated water O atom in all five runs of (c). Atom colour codes: pink, Co; red, O; blue, N; grey, C; white, H. 
}     \label{fig:SI_ablation_binding}
\end{figure}

\begin{figure}[t]
        \centering        \includegraphics[width=1.0\textwidth]{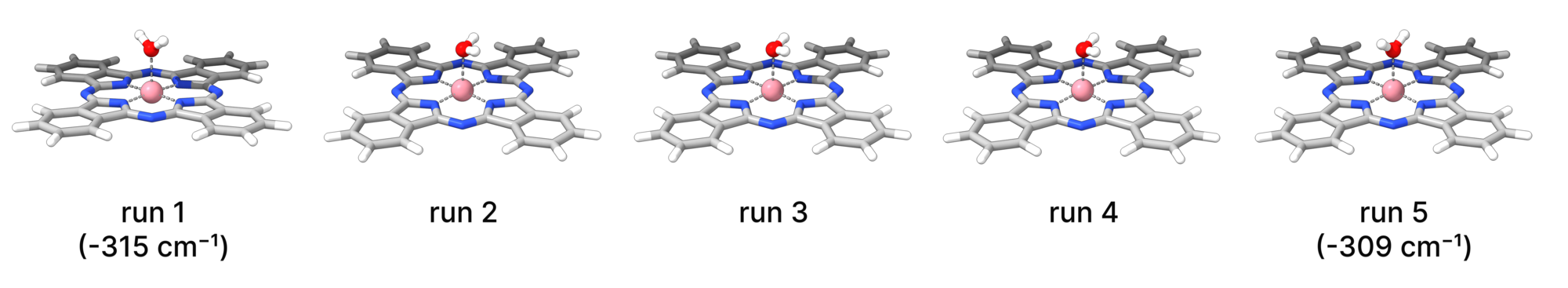}
        \caption{     
        DFT geometry optimization and frequency analysis of the five
H$_2$O-bound structures generated in the tool-ablation test. Geometry
optimizations and frequency calculations were performed at the
B3LYP-D4/def2-SVP level. Runs 2--4 relaxed to trigonal-pyramidal-like local geometries around the coordinated water O atom with no imaginary frequencies. Runs 1 and 5 retained approximately planar geometries and exhibited imaginary frequencies of $-315$ and $-309$~cm$^{-1}$, respectively; the corresponding imaginary modes
are directed toward pyramidalization. Atom colour codes: pink, Co; red, O; blue, N; grey, C; white, H.
}     \label{fig:SI_water}
\end{figure}

\renewcommand{\arraystretch}{1.05}
\begin{table}[H]
\centering
\caption{Runtime comparison between the tool-ablation and tool-enabled workflows for ligand binding to Co-phthalocyanine. Values are reported in seconds.}
\label{tab:ablation_Case2}
\small

\begin{tabular*}{\linewidth}{@{\extracolsep{\fill}}ccc@{}}
\toprule
\textbf{Iteration} &
\textbf{No dedicated tools} &
\textbf{With dedicated tools} \\
\midrule

Run 1 & 48 & 54 \\
Run 2 & 30 & 26 \\
Run 3 & 50 & 38 \\
Run 4 & 36 & 40 \\
Run 5 & 37 & 33 \\

\midrule
\textbf{Average} &
$\mathbf{40.2 \pm 8.5}$ &
$\mathbf{38.2 \pm 10.4}$ \\
\bottomrule
\end{tabular*}
\end{table}

To examine the contributions of the dedicated ligand-binding tool and predefined fragment library, we repeated the Co-Pc ligand-binding task in the same tool-ablation setting, while keeping the original prompt unchanged. For this analysis, we used only the first prompt from the original sequential ligand-binding workflow, which requests attachment of the initial set of axial ligands to Co; the subsequent prompt introducing additional axial ligands was not included.

The tool-ablation workflow was able to attach the requested ligands to the Co centre, but the resulting local geometries were less consistently sanitized (\autoref{fig:SI_ablation_binding}). 
For the coordinated COOH ligand, the O=C--O--H dihedral angle varied between runs, with some generated structures deviating from a planar O=C--O--H arrangement. Geometric issues were also observed for the OH and
H$_2$O ligands. For OH, run 5 produced a nearly linear Co--O--H
arrangement. For H$_2$O, all five runs produced similar local geometries in
which the Co atom and the two H atoms were arranged approximately in the same
plane around the coordinated water O atom, resulting in an approximately
trigonal-planar local geometry around the coordinated water O atom.

To assess the stability of these trigonal-planar H$_2$O-bound geometries, we performed geometry optimization and frequency calculations at the B3LYP-D4/def2-SVP level for the neutral system in the doublet state, starting from the five structures shown in \autoref{fig:SI_ablation_binding}c (\autoref{fig:SI_water}). Runs 2--4 were relaxed to trigonal-pyramidal-like local geometries around the coordinated water O atom, with no imaginary frequencies. Runs 1 and 5 retained
planar geometries but exhibited imaginary frequencies of $-315$ and $-309$~cm$^{-1}$, respectively. Analysis of the corresponding imaginary modes showed displacement toward pyramidalization. These results indicate that the approximately trigonal-planar geometries generated in the tool-ablation
test do not correspond to stable local minima.

These results show that correct ligand attachment does not necessarily lead to a reasonable local geometry. The dedicated binding tools include predefined geometry and post-attachment sanitization procedures that help preserve realistic ligand conformations and metal--ligand geometries. Without these tools, Python-based manipulation could reproduce the intended connectivity, but the local geometry was less consistent.

The runtime difference between the two settings was comparatively small. The average runtime was $40.2 \pm 8.5$~s without the dedicated tools and $38.2 \pm 10.4$~s with the tools (\autoref{tab:ablation_Case2}). Thus, in this example, the principal benefit of the dedicated tools was not execution speed but improved geometric reliability and sanitization of the resulting metal--ligand structures.

\begin{figure}[!t]
        \centering        \includegraphics[width=1.0\textwidth]{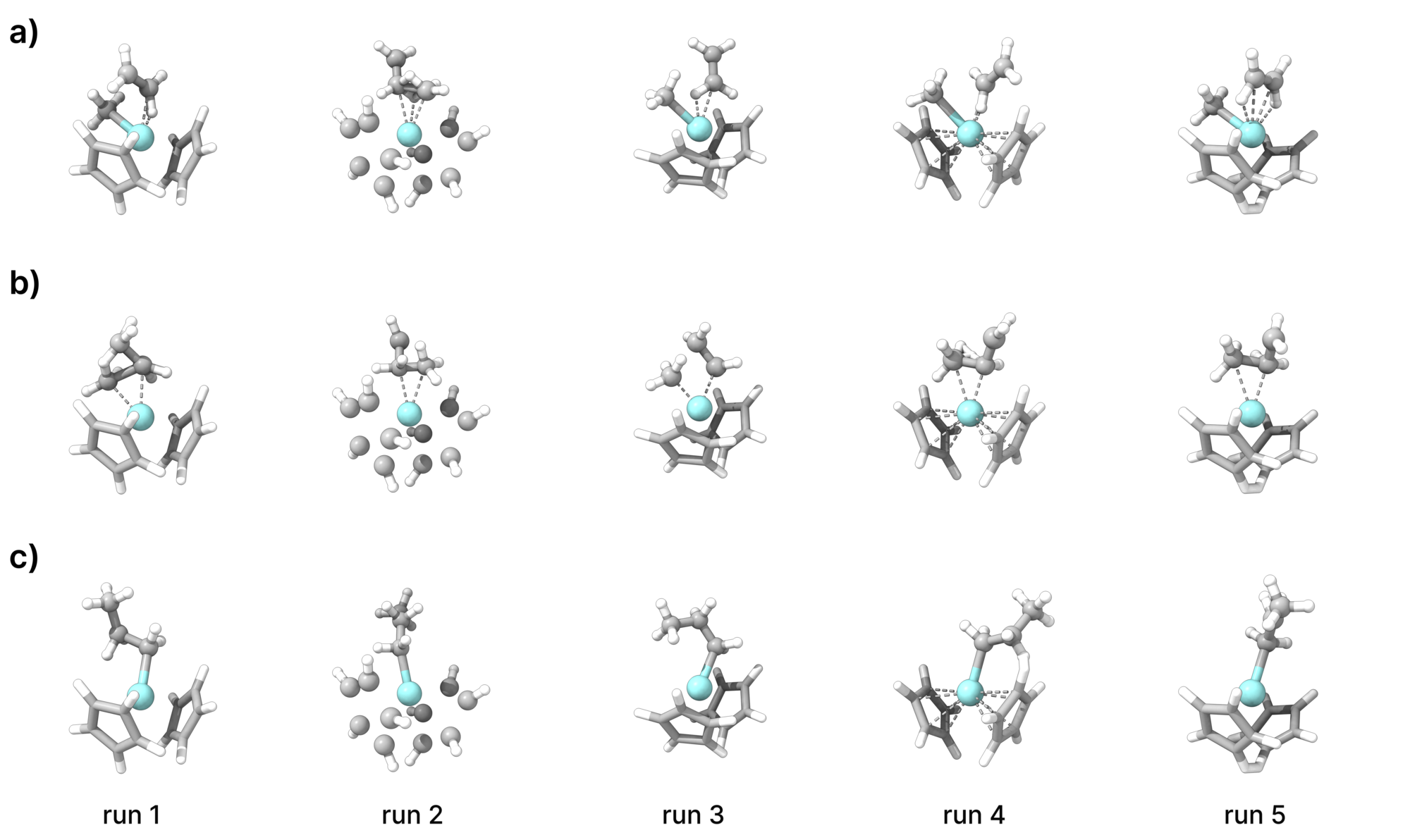}
        \caption{Tool-ablation results for multimodal reaction mechanism--guided structure generation. 
(a) Reactant, (b) initial TS-like structure, and (c) product generated without the dedicated molecular structure tools for five independent runs. The agent attempted to construct the structures using Python-based coordinate manipulation, but multiple runs exhibited severe geometric artifacts, including steric clashes, atomic overlaps, disconnected fragments, and distorted coordination environments. Atom colour codes: sky blue, Zr; grey, C;
white, H.}        \label{fig:SI_ablation_Case6}
\end{figure}

To examine the contributions of \texttt{build\_tm\_complex\_from\_template} for coordination-complex construction and \texttt{constraint\_xtb} for TS-like geometry generation, we repeated the multimodal reaction mechanism--guided structure generation case in the same tool-ablation setting, with only \texttt{bash\_kernel} available for executing Python code using general-purpose scientific libraries. The agent attempted to reconstruct the reactant, TS-like structure, and product directly from the reaction scheme by writing its own coordinate-generation and editing code.

As shown in \autoref{fig:SI_ablation_Case6}, the resulting structures frequently contained severe geometric artifacts, including steric clashes, atomic overlaps, disconnected fragments, and distorted coordination environments. These issues appeared in the reactant, TS-like, and product structures across multiple runs. Although the agent was still able to infer the intended chemical transformation from the multimodal input, the absence of dedicated structure-generation, editing, and constrained-geometry tools made it difficult to translate this interpretation into physically reasonable three-dimensional geometries.

This ablation therefore highlights a distinction between mechanistic reasoning and reliable molecular construction. General Python-based code generation could partially reproduce the intended workflow, but it did not consistently enforce chemically meaningful connectivity, coordination geometry, or steric constraints. In contrast, the dedicated tools provide controlled geometric operations and sanitization procedures that substantially improve the robustness of the generated structures.

\begin{figure}[!t]
        \centering        \includegraphics[width=1.0\textwidth]{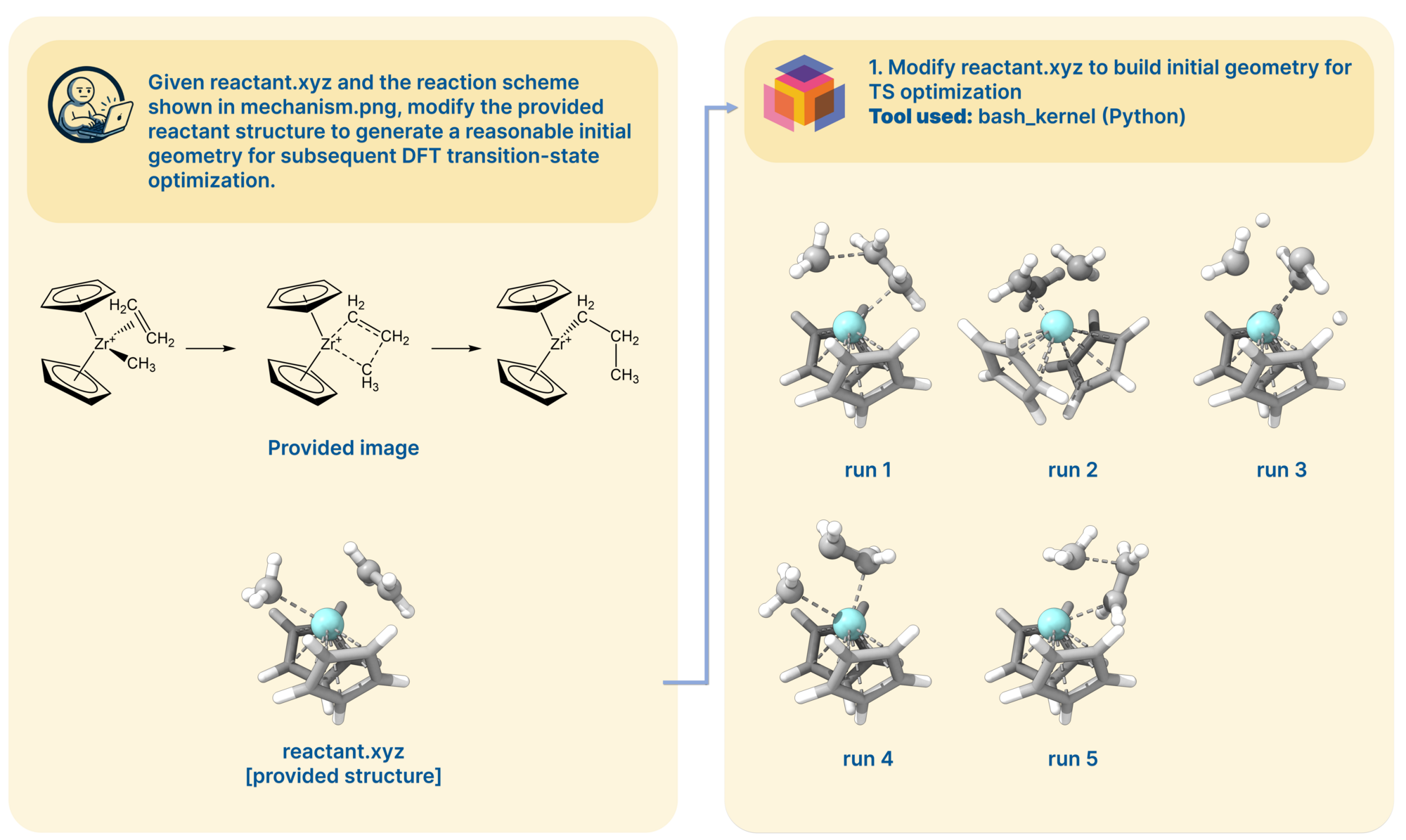}
        \caption{Tool-ablation results for TS-like geometry generation from a provided reactant structure and reaction scheme. The reactant \texttt{xyz} structure was given directly, and the agent was asked to generate a reasonable initial geometry for subsequent DFT transition-state optimization using only Python-based operations. The five generated TS guesses (Runs 1--5) all exhibited chemically unreasonable geometries, and all corresponding DFT transition-state optimizations failed to converge. In contrast, analogous TS-like structures generated with the dedicated molecular editing tools yielded successful transition-state optimizations with a single imaginary frequency near $-300$~cm$^{-1}$.  Atom colour codes: sky blue, Zr; grey, C;
white, H.}        \label{fig:SI_ablation_TS}
\end{figure}

Because the full tool-ablation workflow produced severe geometric artifacts already at the reactant-generation stage (\autoref{fig:SI_ablation_Case6}), it was unclear whether the subsequent failure of TS-like structure generation arose primarily from the absence of \texttt{constraint\_xtb} or simply from starting with a poorly constructed reactant geometry. We therefore performed an additional test to isolate the TS-initialization step more directly. In this test, the correctly constructed reactant structure was provided explicitly as \texttt{reactant.xyz}, together with the reaction scheme image, and the agent was asked only to generate a reasonable initial geometry for subsequent DFT transition-state optimization. By removing the need to generate the reactant and product de novo, this setup allowed the effect of the TS-like geometry construction step itself to be evaluated independently (\autoref{fig:SI_ablation_TS}).

In the absence of \texttt{constraint\_xtb}, the agent attempted to construct the TS-like geometry using Python-based coordinate manipulation. As shown in \autoref{fig:SI_ablation_TS}, the resulting structures remained chemically unreasonable across all five runs, with distorted coordination environments, poor relative placement of the reacting fragments, and geometries that did not provide suitable starting points for TS optimization. DFT transition-state optimization initiated from these five TS guesses failed to converge in every case.

By contrast, when the same task was performed using \estructural{} with the dedicated toolset available, the generated TS-like geometries led to successful DFT transition-state optimizations and yielded a single imaginary frequency near $-300$~cm$^{-1}$, consistent with the main benchmark results. This comparison shows that the failures observed in the ablated workflow were not caused solely by poor reactant generation. Even when a reasonable reactant structure was supplied directly, the absence of a dedicated constrained-geometry tool substantially reduced the reliability of TS-like initial-structure generation and the subsequent transition-state search. These results therefore indicate that \texttt{constraint\_xtb} plays an important role in preparing suitable initial geometries for downstream DFT transition-state optimization.

\subsubsection{Model dependence in multimodal TS construction}
Because multimodal transition-state construction requires interpretation of the reaction scheme in addition to geometric manipulation, we further examined the dependence of the workflow on the underlying LLM. In particular, the agent needs to identify the relevant reaction centre, determine which atoms participate in bond formation or cleavage, select suitable geometric constraints, and infer physically reasonable constraint values before generating a TS-like structure.

\begin{figure}[htbp!]
        \centering        \includegraphics[width=1.0\textwidth]{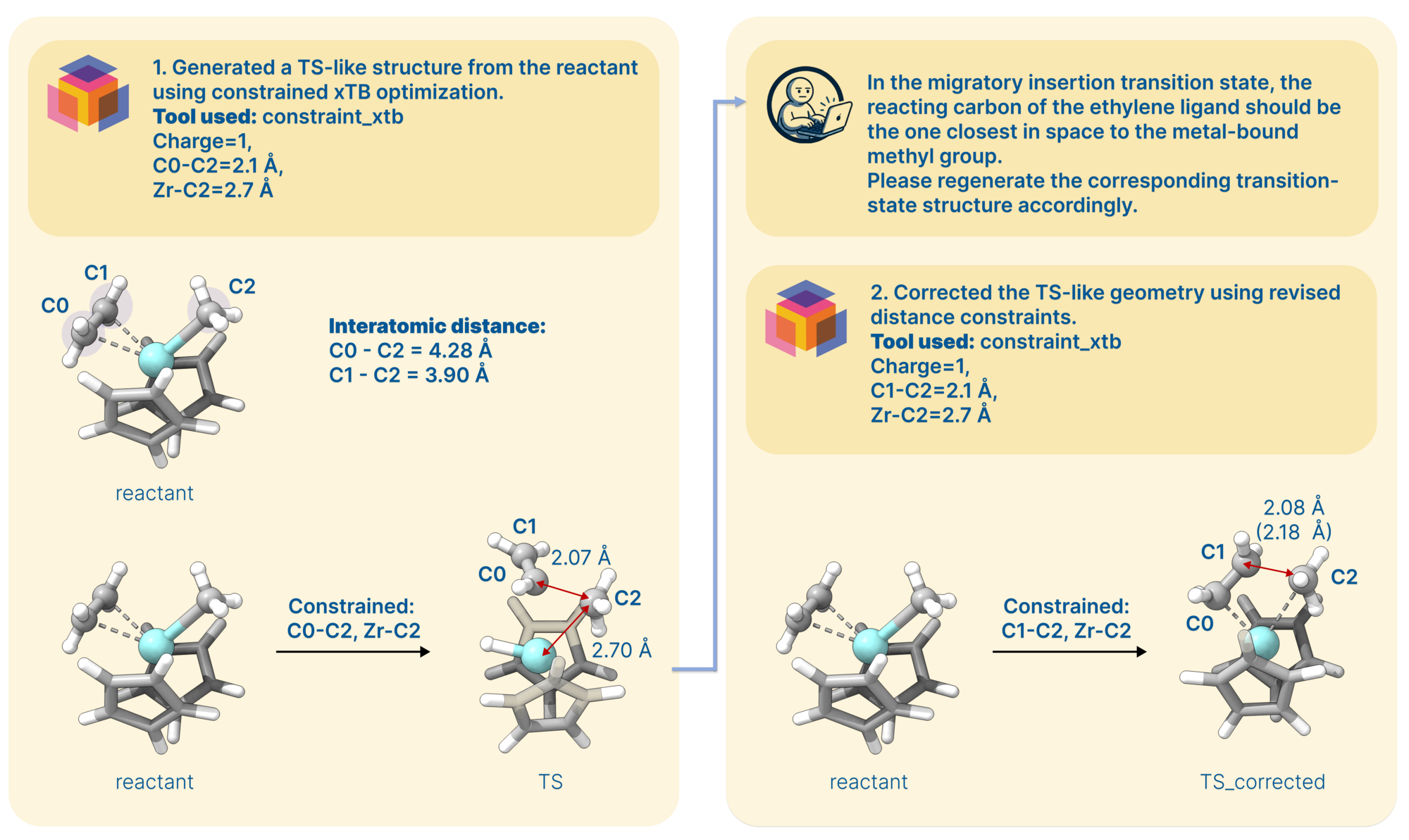}
        \caption{An unwanted structure obtained when an arbitrary carbon atom of the ethylene ligand was selected for constraint application (left) and a chemically reasonable TS-like geometry obtained after selecting the ethylene carbon atom closest to the metal-bound methyl group through the second prompt. The results shown are from Run~1 using Claude Opus~4.5. Key interatomic distances obtained from constrained xTB optimization are highlighted, with the corresponding DFT reference distance shown in parentheses. The Zr--Cp coordination bonds are omitted for clarity. Atom colour codes: sky blue, Zr; grey, C; white, H.}        \label{fig:SI_ethylene}
\end{figure}

We therefore repeated the same multimodal TS-generation task using Claude Opus 4.5, GPT-5.2, Claude Opus 4.8, and GPT-5.4, while keeping the system prompt, agent architecture, available toolset, and other workflow settings unchanged. Thus, only the underlying LLM was varied in this comparison. For most runs, the models independently identified the appropriate reacting atoms and generated TS-like structures that subsequently converged during DFT transition-state optimization to structures with a principal imaginary frequency near $-300$~cm$^{-1}$ (\autoref{tab:LLMs}). However, human intervention was required in Opus~4.5 Run~1 and GPT-5.2 Runs~4 and~5. In these cases, the model initially applied the distance constraint to the more distant ethylene carbon (C0--C2) rather than to the carbon spatially closest to the metal-bound methyl group (C1--C2) as shown in \autoref{fig:SI_ethylene}. After a follow-up instruction clarifying the intended reacting carbon, the agent regenerated the TS-like structure using the corrected constraint, and the subsequent DFT TS optimization successfully recovered the expected transition state.

As shown in \autoref{tab:gpt54_constraints} and \autoref{tab:model_constraints_comparison}, the selection of geometric constraints also showed model-dependent behaviour. 
After any necessary human correction, all models consistently constrained the forming C1--C2 bond, with target values ranging from 2.10 to 2.33~\AA{}, mostly between 2.10 and 2.20~\AA{}. However, GPT-5.2 and GPT-5.4 generally applied additional constraints to other key interatomic distances involved in the bond rearrangement, including Zr--C0, Zr--C1, Zr--C2, and the elongating ethylene C0--C1 bond. In contrast, Opus 4.5 and Opus 4.8 tended to use a more minimal constraint set, most commonly focusing on the forming C1--C2 bond and, when included, the Zr--C2 interaction involving the methyl carbon. These results indicate that, while the key forming C--C bond was ultimately constrained in all runs, the extent to which additional reaction-centre distances were constrained depended on the underlying LLM.

\renewcommand{\arraystretch}{1.05}
\begin{table}[t]
\centering
\caption{Principal imaginary frequencies obtained after DFT transition-state optimization of TS-like structures generated using the same agentic workflow with different underlying LLMs. Values are reported in cm$^{-1}$.}
\label{tab:LLMs}
\small

\begin{tabular*}{\linewidth}{@{\extracolsep{\fill}}ccccc@{}}
\toprule
\textbf{Iteration} &
\textbf{Opus 4.5} &
\textbf{GPT-5.2} &
\textbf{Opus 4.8} &
\textbf{GPT-5.4} \\
\midrule

Run 1 & $-301$\rlap{\textsuperscript{[a]}} & $-299$ & $-301$ & $-301$ \\
Run 2 & $-298$ & $-297$ & $-296$ & $-296$ \\
Run 3 & $-296$ & $-300$ & $-301$ & $-301$ \\
Run 4 & $-296$ & $-297$\rlap{\textsuperscript{[a,b]}} & $-296$ & $-301$ \\
Run 5 & $-301$ & $-300$\rlap{\textsuperscript{[a]}} & $-296$ & $-297$ \\

\midrule
\textbf{Average} &
$\mathbf{-299 \pm 3}$ &
$\mathbf{-299 \pm 2}$ &
$\mathbf{-298 \pm 3}$ &
$\mathbf{-299 \pm 2}$ \\
\bottomrule
\end{tabular*}

\vspace{0.3em}

\begin{minipage}{\linewidth}
\footnotesize
\raggedright
[a] Human intervention was required to correct the initially selected reacting carbon before regeneration of the TS-like structure.

[b] A small additional imaginary frequency of $-4$~cm$^{-1}$ was also obtained.
\end{minipage}
\end{table}

\clearpage
\makeatletter
\setlength{\@fptop}{0pt}
\makeatother

\begin{table}[p!]
\renewcommand{\arraystretch}{1.10}

\centering
\caption{Interatomic distance constraints selected by different underlying LLMs to construct the TS-like geometry in the multimodal reaction-mechanism case. The atomic labels C0, C1, and C2 are defined in \autoref{fig:SI_ethylene}. Constrained distances are given in \AA, and force constants are given in Hartree/Bohr$^2$.}
\label{tab:model_constraints_comparison}
\small

\begin{tabularx}{\linewidth}{@{}
>{\raggedright\arraybackslash}p{0.24\linewidth}
*{5}{>{\centering\arraybackslash}X}
@{}}

\toprule
\multicolumn{6}{c}{\textbf{Model = GPT-5.2}} \\
\midrule
Constrained distance &
Run 1 &
Run 2 &
Run 3 &
Run 4 &
Run 5 \\
\midrule

Zr--C0 & 2.40 & 2.45 & 2.40 & 2.50 & 2.55 \\
Zr--C1 & -- & 2.55 & -- & 2.50 & 2.45 \\
Zr--C2 & 2.70 & 2.65 & 2.60 & 2.60 & 2.65 \\
C0--C1 & -- & 1.40 & 1.40 & 1.38 & 1.42 \\
C0--C2 & -- & -- & -- & -- & 3.20 \\
C1--C2 & 2.10 & 2.15 & 2.20 & 2.10 & 2.10 \\
Force constant & 0.50 & 0.70 & 0.70 & 0.80 & 0.80 \\

\midrule
\multicolumn{6}{c}{\textbf{Model = Opus 4.5}} \\
\midrule
Constrained distance &
Run 1 &
Run 2 &
Run 3 &
Run 4 &
Run 5 \\
\midrule

Zr--C0 & -- & -- & -- & -- & -- \\
Zr--C1 & -- & -- & -- & -- & -- \\
Zr--C2 & 2.70 & -- & 2.60 & 2.70 & 2.60 \\
C0--C1 & -- & -- & -- & -- & 1.45 \\
C0--C2 & -- & -- & -- & -- & -- \\
C1--C2 & 2.10 & 2.10 & 2.10 & 2.20 & 2.10 \\
Force constant & 0.80 & 0.80 & 0.80 & 0.80 & 0.70 \\

\midrule
\multicolumn{6}{c}{\textbf{Model = Opus 4.8}} \\
\midrule
Constrained distance &
Run 1 &
Run 2 &
Run 3 &
Run 4 &
Run 5 \\
\midrule

Zr--C0 & 2.45 & 2.50 & 2.40 & 2.40 & -- \\
Zr--C1 & -- & -- & -- & -- & -- \\
Zr--C2 & 2.45 & 2.35 & 2.45 & 2.45 & -- \\
C0--C1 & -- & 1.40 & -- & -- & -- \\
C0--C2 & -- & -- & -- & -- & -- \\
C1--C2 & 2.15 & 2.15 & 2.10 & 2.10 & 2.15 \\
Force constant & 0.70 & 1.00 & 1.00 & 1.00 & 1.00 \\

\bottomrule
\end{tabularx}
\end{table}

\clearpage

\makeatletter
\setlength{\@fptop}{0pt plus 1fil}
\makeatother

\clearpage
\section{Limitations and roadmap}\label{SIsec:roadmap}

\begin{xltabular}{\textwidth}{
>{\raggedright\arraybackslash}p{0.16\textwidth}
>{\raggedright\arraybackslash}p{0.20\textwidth}
>{\raggedright\arraybackslash}X
>{\raggedright\arraybackslash}p{0.25\textwidth}
}

\caption{Practical scope and recommended user checks for \estructural{}, with representative successful applications.}
\label{tab:user_scope}\\

\toprule
\textbf{Task category} &
\textbf{Representative successful examples} &
\textbf{Current scope and limitations} &
\textbf{Recommended user guidance and checks} \\
\midrule
\endfirsthead

\multicolumn{4}{l}{\small\textit{Table \thetable\ continued}}\\
\toprule
\textbf{Task category} &
\textbf{Representative successful examples} &
\textbf{Current scope and limitations} &
\textbf{Recommended user guidance and checks} \\
\midrule
\endhead

\midrule
\multicolumn{4}{r}{\small\textit{Continued on next page}}\\
\endfoot

\bottomrule
\endlastfoot

\textbf{Organic molecule generation} &
Spermidine; Boc-protected spermidine &
Molecular structures are generally generated from SMILES. \estructural{} can infer SMILES directly or retrieve them from common chemical names using the PubChemPy-based lookup tool. This provides reliable connectivity for common compounds, but uncommon or ambiguous names may not correspond to an appropriate PubChem entry. In addition, SMILES does not uniquely specify all aspects of three-dimensional conformation, so the generated geometry may not represent a particular desired conformer. &
For uncommon or ambiguous names, confirm that the intended molecular identity and connectivity were obtained. If a specific three-dimensional conformer is important, inspect the generated geometry and perform conformer search or geometry optimization as appropriate.
\\
\addlinespace[1.0em]

\textbf{Molecular functionalization and editing} &
Co-porphyrin and MoCl$_3$PNP functionalization; selective Boc protection; ligand substitution &
Dedicated editing tools set new bond lengths based on covalent radii and apply post-editing sanitization to avoid severe steric clashes between the core structure and newly attached fragments. These procedures are intended to produce chemically reasonable starting geometries, but they do not systematically minimize steric hindrance, identify the lowest-energy conformer, or produce a fully optimized structure. &
Perform a conformer search and subsequent geometry optimization when a specific or lowest-energy conformation is required.
\\
\addlinespace[1.0em]

\textbf{Coordination-complex generation} &
cis/transplatin; $\Lambda/\Delta$-Fe(bpy)$_3$; \textit{fac}/\textit{mer}-Ir(ppy)$_3$; zirconocene complexes &
The predefined coordination-complex construction tools currently cover common coordination environments and geometries, including the tested geometries up to six-coordinate octahedral complexes. More unusual coordination numbers, geometries, or bonding motifs may fall outside the currently implemented templates. &
Complexes outside the currently supported coordination environments may require additional user guidance.
\\
\addlinespace[1.0em]

\textbf{Ligand binding and coordination} &
Co--Pc axial-ligand binding; Pb--Pc axial-ligand binding; CpRu($\eta^{6}$-sumanene)$^{+}$ &
The binding tools place ligands according to predefined VSEPR-based coordination geometries when applicable, or otherwise along sterically accessible directions. When multiple chemically plausible binding configurations exist, the generated configuration may depend on the agent's interpretation. &
For systems that do not closely match the predefined VSEPR-based geometries or that allow multiple plausible binding configurations, additional user guidance may be helpful.
\\
\addlinespace[1.0em]

\textbf{Stereochemical and configurational selection} &
cis/transplatin; $\Lambda/\Delta$-Fe(bpy)$_3$; \textit{fac}/\textit{mer}-Ir(ppy)$_3$; alanine enantiomerization &
When the target stereochemistry or configuration is explicitly specified, the agent generated the requested structures in the cases examined here. However, the present benchmark covers a limited set of stereochemical and configurational transformations. &
Verify the intended stereochemistry or configuration when it is important.
\\
\addlinespace[1.0em]

\textbf{Electronic-structure settings} &
Charge and uhf assignment for xTB-assisted organometallic structure refinement &
The agent selected chemically appropriate charge and uhf settings for the systems examined in this work. Systems with multiple plausible spin states, such as high-spin and low-spin configurations, were not included in the present benchmark. &
For systems with multiple plausible spin states, additional user guidance or verification of the charge and spin state may be needed.
\\
\addlinespace[1.0em]

\textbf{TS-like structure generation} &
Multimodal zirconocene migratory-insertion case &
The \texttt{constraint\_xtb} workflow can generate chemically reasonable starting geometries for subsequent TS optimization. However, identification of the reaction centre, selection of constrained atoms, and choice of constraint values depend partly on LLM reasoning and showed model-dependent variation in our tests. The resulting structure is an initial TS-like geometry rather than a validated saddle point. &
For a new reaction class or LLM, additional testing on a representative case may be needed before broader application. Subsequent quantum-chemical TS optimization and frequency analysis are required to validate the transition state.
\\
\addlinespace[1.0em]

\textbf{Multimodal structure interpretation} &
Multimodal zirconocene migratory-insertion case; mechanism-reasoning-driven geometry construction &
\estructural{} can identify the main molecular components, coordination environment, and reaction-centre geometry from chemical diagrams. However, subtle three-dimensional details depicted schematically, such as the relative orientation of Cp rings, may not be reproduced unless they are explicitly required. In addition, the vision-only benchmark showed decreasing spatial-perception accuracy as molecular size increased. &
When subtle three-dimensional details are important, explicitly specifying or guiding the desired geometry may improve the generated structure.
\\
\end{xltabular}

\subsection{Integration with multi-agent quantum chemistry (Stage 1)}
At present (Stage 0), \estructural{} is designed as a dedicated agent specialized in molecular structure generation, editing, and geometric manipulation. Although it supports xTB-based geometry refinement, including constrained optimization for generating TS-like structures, it does not perform systematic quantum-chemical calculations
such as DFT optimization, frequency analysis, or energy profiling. An important next step (Stage 1) is to integrate \estructural{} into our multi-agent quantum chemistry architecture, \quntur{}. Within this framework, \estructural{} can serve as a specialized geometry-manipulation agent that collaborates with other agents responsible for tasks such as a high-level planner, input-file generation, and DFT calculations. This integration enables \quntur{} to operate on user-demand structures across a substantially broader range of chemical systems, including stereochemically sensitive molecules, organometallic complexes, weakly bound intermediates, and catalytic reaction pathways involving intermediates and transition states.

\subsection{Data-driven structure retrieval and editing (Stage 2)}
Complementary to the construction of molecules from scratch, Stage 2 focuses on data-driven core-structure retrieval and editing. As illustrated in the case studies (Figure 6), editing an existing, chemically similar structure is often more reliable than full \textit{de novo} construction for large or highly coordinated organometallic systems.
Future versions of \estructural{} will therefore incorporate data querying and web-search capabilities, enabling retrieval of relevant structures from the literature or curated databases---such as the Cambridge Structural Database (CSD)~\cite{CSD} and the Open Molecules 2025 (OMol25) dataset~\cite{omol25}---followed by targeted, geometry-preserving modifications. This hybrid strategy would further enhance the robustness and practicality of \estructural{} for real-world organometallic chemistry.

\subsection{Interactive user interfaces (Stage 3)}
Although the current implementation of \estructural{} supports human-in-the-loop interaction through iterative natural language dialogue---allowing users to guide, correct, or refine the agent's decisions---structural editing is currently mediated exclusively through language-based instructions and agent-executed operations. 
In practice, however, certain geometry-editing tasks can be more efficiently performed through direct user manipulation than through detailed natural language descriptions. For example, when a user intends to adjust the distance or angle between a specific pair of atoms within a large or complex molecule, directly selecting those atoms and applying the modification can be more intuitive and efficient than specifying the operation textually.

User interaction could also be beneficial when a request allows multiple distinct structural alternatives. In the current framework, \estructural{} autonomously selects a structural solution based on its reasoning, which the user can subsequently refine through dialogue. In future implementations, \estructural{} could instead recognize such ambiguity and ask the user to choose among plausible alternatives before proceeding. Alternatively, the system could generate multiple candidate structures for the user to select from. This would provide users with greater control when alternative binding configurations, isomers, conformations, or other structurally distinct possibilities exist.

Accordingly, an interactive user interface (UI) represents an important future direction for \estructural{}. Providing a visual editing environment in which users can directly select atoms and invoke existing structure-editing operations---such as substitution, distance tuning, or angle control---while also enabling interactive clarification and selection among alternative structures could seamlessly combine LLM-driven automation with familiar direct-manipulation workflows found in conventional molecular viewers. Such a hybrid interface would enhance usability and flexibility, enabling users to choose the most efficient interaction mode depending on the task while complementing the agent's autonomous capabilities in complex modelling workflows.

\subsection{Agentic model training on molecular manipulation (Stage 4)}
Geometry manipulation tasks are clearly out of distribution for current vision–language models (VLMs). Two fundamental challenges remain. From a perception standpoint, VLMs lack the ability to reliably observe and analyze molecular systems, as discussed in \ref{sec3:VLM}. From a reasoning standpoint, existing systems have limited capacity to integrate chemical reasoning with spatial reasoning. Targeted agentic training will therefore be necessary to advance \estructural{} to the next level. Promising directions include synthetic data and task generation, as well as targeted post-training approaches such as reinforcement learning with verifiable rewards (RLVR)~\cite{guo_deepseek-r1_2025} and supervised fine-tuning (SFT)~\cite{ouyang2022training}. 

\subsection{Advanced organometallic structure construction (Stage 5)}
While \estructural{} currently supports commonly used coordination geometries and ligand types, systems with very high coordination numbers (e.g., greater than six, as commonly encountered in lanthanide and actinide chemistry), multinuclear metal centres, or multiple polydentate ligands remain difficult to construct reliably using predefined templates alone.
Promising directions include expanding the coordination-template library, developing more flexible rule-based construction strategies, and integrating external frameworks such as Architector~\cite{taylor_architector_2023} and molSimplify~\cite{molSimplify2} as complementary tools to support the construction of more complex organometallic structures.

\subsection{Extension to slab models and solid-state systems (Stage 6)}
The current implementation of \estructural{} is limited to molecular and coordination chemistry and does not yet support slab models or extended solid-state systems. Extending the framework to solids and surfaces represents Stage 6 of the roadmap.
Supporting slab models, surface adsorption, and heterogeneous catalytic reactions will require explicit treatment of periodic boundary conditions, identification of adsorption sites (e.g., atop, bridge, and hollow), and extended coordination environments.

\subsection{Reaction pathway generation (Stage 7)}
Extending \estructural{} toward systematic reaction pathway generation represents Stage~7 of the roadmap, building on the mechanism-guided case studies presented in this work. By generalizing image- and instruction-guided structure construction to more complex reaction networks, \estructural{} could support the generation of consistent sets of reaction intermediates and TS-like structures, thereby providing structured initial geometries for downstream optimization, transition-state searches, and energy profiling. Furthermore, a future closed-loop workflow could couple agent-proposed mechanistic hypotheses, geometry construction for each hypothesis, and computational validation through quantum chemistry. Such a mechanism–geometry–validation loop could enable scalable, hypothesis-driven exploration of reaction space, allowing chemistry agents to tackle broader domains.

\subsection{Enhanced multimodal understanding and model improvement (Stage 8)}
Many types of chemically relevant information are commonly communicated through schematic images rather than textual descriptions. In addition to reaction mechanisms, a series of functionalized derivatives derived from a common core structure is often presented visually. Such depictions can involve complex abstraction, including Markush structures or substitution patterns, which are difficult to encode solely with textual representations. Although \estructural{} already supports image-guided interpretation of reaction mechanisms, its ability to parse and reason over more abstract chemical diagrams remains limited. Enhancing multimodal understanding of these visual representations, therefore, represents an important direction for future development.

Beyond visual input, Stage 8 also envisions richer multimodal interaction modalities, including voice-based instructions~\cite{darvish2025organa,raucci2021voice} and immersive interfaces such as virtual or augmented reality~\cite{aspuru2018matter,taylor2025optimising,o2018sampling,baaden2025virtual,marti2009haptic}. Supporting these interaction modes would enable more intuitive and user-friendly control over molecular structure generation and editing, particularly for exploratory tasks where precise textual descriptions are cumbersome. Such interfaces can support education and training~\cite{fombona2022vr,bennie2019teaching,taylor2024breaking,sohail2025vr} by enabling users to interactively explore molecular structures, reaction pathways, and stereochemical relationships in a more natural and accessible manner.


\end{document}